\documentstyle[twocolumn,pra,aps,epsf]{revtex}     

\begin{document}

\title{OPTICAL DIPOLE TRAPS FOR NEUTRAL ATOMS}

\author{Rudolf Grimm and Matthias Weidem\"uller}
\address{Max-Planck-Institut f\"ur Kernphysik, 69029 Heidelberg, Germany}
\author{Yurii B. Ovchinnikov}
\address{National Institute of Standards and Technology, PHY B167, 
Gaithersburg, MD 20899, USA}


\maketitle

\tableofcontents

\section{Introduction} 

Methods for storage and trapping of charged and neutral particles 
have very often served as the experimental key to great scientific advances, 
covering physics in the vast energy range
from elementary particles to ultracold atomic quantum matter.
The ultralow-energy region became experimentally accessible as a result 
of the dramatic developments in the field of laser cooling and trapping, 
which have taken place over the last two decades 
(Stenholm, 1986; Minogin and Letokhov, 1988; Arimondo {\it et al.}, 1992; 
Metcalf and van der Straten, 1994; Chu, 1998; Cohen-Tannoudji, 1998; 
Phillips, 1998). 

For {\em charged particles},  
the strong Coulomb interaction can be used for trapping in electric or 
electromagnetic fields (Bergstr\"om {\it et al.}, 1995; Ghosh, 1995). 
At the very low temperatures reached by laser cooling, 
single ions show a variety of interesting quantum effects
(Wineland {\it et al.}, 1995), 
and ensembles form a crystalline ordered state (Walther, 1993).
Laser cooling in ion traps has opened up completely new possibilities 
for ultrahigh precision spectroscopy and related fundamental applications
(Thompson, 1993).
It is a very important feature of ion traps that 
the confining mechanism does not rely on the internal structure
of the ion, which is therefore accessible for all kinds of experiments.

For {\em neutral atoms}, 
it has become experimental routine to produce ensembles 
in the mikrokelvin region, and many experiments are 
being performed with such laser-cooled ultracold gases.
It is thus possible to trap the atoms by
much weaker mechanisms as compared to the Coulomb interaction.
Traps for neutral atoms can be realized on the basis of three different 
interactions, 
each class having specific properties and offering particular 
advantages:

\vspace{-2.5mm}
\begin{itemize}

\item
{\em Radiation-pressure traps} operating with near-resonant light 
(Pritchard {\it et al.}, 1986; Raab {\it et al.}, 1987) 
have a typical depth of a few Kelvin, and because of very strong 
dissipation they allow to 
capture and accumulate atoms even from a thermal gas. 
In these traps, the atomic ensemble can be cooled 
down to temperatures in the range of a few 10\,$\mu$K.
The trap performance, however, is limited in several ways by the 
strong optical excitation:
The attainable temperature is limited by the photon recoil, the achievable 
density is limited by radiation trapping and light-assisted inelastic 
collisions, 
and the internal dynamics is strongly perturbed by resonant processes
on a time scale in the order of a microsecond.

\item
{\em Magnetic traps} (Migdall {\it et al.}, 1986; Bergeman {\it et al.}, 1987)
are based on the state-dependent force on the 
magnetic dipole moment in an inhomogeneous field. They represent ideal 
conservative traps with typical depths in the order of 100\,mK,  
and are excellent tools for evaporative cooling and Bose-Einstein 
condensation. 
For further applications, a fundamental restriction is imposed
by the fact that the trapping mechanism relies on the internal 
atomic state. 
This means that experiments concerning the internal dynamics
are limited to a few special cases. Furthermore, 
possible trapping geometries 
are restricted by the necessity to use arrangements of coils or permanent
magnets.

\item
{\em Optical dipole traps} 
rely on the electric dipole interaction with far-detuned light, which 
is much weaker than all mechanisms discussed above. 
Typical trap depths are in the range below one millikelvin.
The optical excitation can be kept extremely low, so that such a trap is not
limited by the light-induced mechanisms present in radiation-pressure traps.
Under appropriate conditions, the trapping mechanism is independent of the 
particular sub-level of the electronic ground state. The 
internal ground-state dynamics can thus be fully exploited for 
experiments, which is possible on a time scale of many seconds.
Moreover, a great variety of different trapping geometries can be realized 
as, e.g., highly anisotropic or multi-well potentials.

\end{itemize}

\noindent
The subject of this review are atom traps of the last described class
along with their unique features as storage devices at ultralow energies. 

Historically, the optical dipole force, acting as confining mechanism in
a dipole trap, was first considered by Askar'yan
(1962) in connection with plasmas as well as neutral atoms.
The possibility of trapping atoms with this force was considered by
Letokhov (1968) who suggested that atoms might be one-dimensionally 
confined at the nodes or antinodes of a standing wave tuned far below or
above the atomic transition frequency. Ashkin (1970) demonstrated
the trapping of micron-sized particles in laser light based on the
combined action of radiation pressure and the dipole force. Later he
suggested three-dimensional traps for neutral atoms (1978). 
The dipole force on neutral atoms was demontrated by Bjorkholm {\it et al.}\ 
(1978)
by focusing an atomic beam by means of a focused laser beam.
As a great breakthrough, Chu {\it et al.}\ (1986) exploited this force to 
realize 
the first optical trap for neutral atoms. 
After this demonstration, enormous progress in laser cooling and trapping was 
made in many different directions, and
much colder and denser atomic samples became available for the 
efficient loading of shallow dipole traps. In the early '90s  
optical dipole forces began to attract rapidly increasing interest not
only for atom trapping, but also in the emerging field of atom optics 
(Adams {\it et al.}, 1994).

In this review, we focus on dipole 
traps realized with {\em far-detuned} light. 
In these traps an ultracold ensemble of atoms is confined in a 
nearly conservative potential well
with very weak influence from spontaneous photon scattering. 
The basic physics of the dipole interaction 
is discussed in Sec.~\ref{background1}. 
The experimental background of dipole trapping experiments is 
then explained in Sec.~\ref{background2}. Specific trapping schemes and 
experiments are presented in Secs.~\ref{red} and \ref{blue}, 
where we explore the wide range of applications of dipole traps considering
particular examples.

\section{Optical dipole potential} \label{background1}

Here we introduce the basic concepts of atom trapping in optical dipole
potentials that result from the interaction with {\em far-detuned} light. 
In this case of particular interest, the optical excitation is very low and 
the 
radiation force due to photon scattering is negligible 
as compared to the dipole force.
In Sec.~\ref{oscillator}, we first consider the atom as 
a simple classical or quantum-mechanical oscillator
to derive the main equations for the optical dipole interaction.
We then  
discuss the case of real multi-level atoms in Sec.~\ref{multilevel}, 
in particular of alkali atoms as used in the great majority of experiments.
discussed

\subsection{Oscillator model} \label{oscillator}

The optical dipole force arises from the dispersive interaction of the 
induced atomic dipole moment with the intensity gradient of the light field
(Askar'yan, 1962; Kazantsev, 1973; Cook, 1979; Gordon and Ashkin, 1980).
Because of its conservative character, the force can be derived from
a potential, the minima of which can be used for atom trapping.
The absorptive part of the dipole interaction in far-detuned light
leads to residual photon scattering of the trapping light, 
which sets limits to the performance of dipole traps. 
In the following, we derive
the basic equations for the dipole potential and the scattering rate
by considering the atom as a simple oscillator subject to the classical
radiation field.

 \subsubsection{Interaction of induced dipole with driving field}

When an atom is placed into laser light, the
electric field ${\bf E}$ induces an atomic dipole moment ${\bf p}$ 
that oscillates at the driving frequency $\omega$. 
In the usual complex notation
${\bf E}({\bf r},t) = {\bf \hat{e}} \, \tilde{E}({\bf r}) 
\exp(-i \omega t) + c.c.$
and
${\bf p}({\bf r},t) = {\bf \hat{e}} \, \tilde{p}({\bf r}) 
\exp(-i \omega t) + c.c.$, 
where ${\bf \hat{e}}$ is the unit polarization vector,
the amplitude $\tilde{p}$ of the dipole moment
is simply related to the field amplitude $\tilde{E}$ by 
\begin{equation}
\tilde{p} = \alpha \, \tilde{E} \, .
\label{palphaE}
\end{equation}
Here $\alpha$ is the {\em complex polarizability}, which depends
on the driving frequency $\omega$.

The {\em interaction potential} of the induced dipole moment ${\bf p}$ in
the driving field ${\bf E}$ is given by
\begin{equation}
U_{\rm dip} = - \frac{1}{2} \, \langle {\bf p}{\bf E}\rangle 
= - \frac{1}{2 \epsilon_0 c} \, {\rm Re}(\alpha) \, I \, ,
\label{URealpha}
\end{equation}
where the angular brackets denote the time average over the rapid 
oscillating terms, the field intensity is 
$I= 2 \epsilon_0 c |\tilde{E}|^2$, and the factor $\frac{1}{2}$ takes into
account that the dipole moment is an induced, not a permanent one.
The potential energy of the atom
in the field is thus proportional to the intensity $I$ and
to the real part of the polarizability, 
which describes the in-phase component of the dipole oscillation
being responsible for the dispersive properties of the interaction. 
The {\em dipole force} results from the gradient of the interaction 
potential
\begin{equation}
{\bf F}_{\rm dip}({\bf r}) = - \nabla U_{\rm dip}({\bf r}) 
= \frac{1}{2 \epsilon_0 c} \, {\rm Re}(\alpha) \, \nabla I({\bf r}) \, .
\label{FgradU}
\end{equation}
It is thus a conservative force, proportional to the intensity gradient
of the driving field. 

The power absorbed by the oscillator from the driving field 
(and re-emitted as dipole radiation) is given by  
\begin{equation}
P_{\rm abs} = \langle {\bf \dot{p}}{\bf E}\rangle 
= 2 \omega \, {\rm Im}(\tilde{p}\tilde{E}^*) 
= \frac{\omega}{\epsilon_0 c} \,{\rm Im}(\alpha) \, I \,.
\label{Pabs}
\end{equation}
The absorption results from the imaginary part of the 
polarizability, which describes the out-of-phase component 
of the dipole oscillation.
Considering the light as a stream of photons $\hbar \omega$, 
the absorption can be interpreted in terms of photon scattering
in cycles of absorption and subsequent spontaneous reemisson
processes. The corresponding {\em scattering rate} is
\begin{equation}
\Gamma_{\rm sc}({\bf r}) = \frac{P_{\rm abs}}{\hbar \omega}
= \frac{1}{\hbar \epsilon_0 c} \,{\rm Im}(\alpha) \, I({\bf r}) \,.
\label{GscImalpha}
\end{equation}
 
We have now expressed the two main quantities of interest 
for dipole traps, 
the interaction potential and the scattered radiation power,
in terms of the position-dependent field intensity $I({\bf r})$  
and the polarizability $\alpha(\omega)$.
We point out that these expressions are 
valid for any polarizable neutral particle in an oscillating 
electric field. This can be an atom in a near-resonant or far 
off-resonant laser field, or even a molecule in an optical or 
microwave field.

\subsubsection{Atomic polarizability}

In order to calculate its polarizability $\alpha$, 
we first consider the atom
in Lorentz's model of a classical oscillator.
In this simple and very useful picture,
an electron (mass $m_e$, elementary charge $e$) is considered
to be bound elastically to the core with an oscillation
eigenfrequency $\omega_0$ corresponding to the optical 
transition frequency.
Damping results from the dipole radiation of the oscillating electron 
according to Larmor's well-known formula (see, e.g., Jackson, 1962)
for the power radiated by an accelerated charge.

It is straightforward to calculate the polarizability
by integration of the equation of motion 
$\ddot{x} + \Gamma_{\omega} \dot{x} + \omega_0^2 x = - e E(t) /m_e$ 
for the driven oscillation of the electron with the result
\begin{equation}
\alpha = \frac{e^2}{m_e} \, 
\frac{1}{\omega_0^2 - \omega^2 - {\rm i} \omega \Gamma_{\omega}} \,
\label{alphaclass}
\end{equation}
Here
\begin{equation}
\Gamma_{\omega} = \frac{e^2 \omega^2}{6 \pi \epsilon_0 m_e c^3}\,
\label{Gclass}
\end{equation}
is the classical damping rate due to the radiative energy loss.
Substituting 
$e^2/m_e = 6 \pi \epsilon_0 c^3 \Gamma_{\omega}/\omega^2$  
and introducing the on-resonance damping rate 
$\Gamma \equiv \Gamma_{\omega_0} = (\omega_0/\omega)^2 \Gamma_{\omega}$, 
we put Eq.~\ref{alphaclass} into the form
\begin{equation}
\alpha = 6 \pi \epsilon_0 c^3 \, 
\frac{\Gamma/\omega_0^2}
{\omega_0^2 - \omega^2 - {\rm i} \, (\omega^3/\omega_0^2) \, \Gamma} \, .
\label{alpha}
\end{equation}

In a {\em semiclassical approach}, described in many textbooks,
the atomic polarizability can be calculated by considering the 
{atom as a two-level quantum system} interacting 
with the classical radiation field.
One finds that, when saturation effects can be neglected, 
the semi-classical calculation 
yields exactly the same result as the classical calculation with only one 
modification:
In general, the damping rate $\Gamma$ (corresponding to the spontaneous
decay rate of the excited level)
can no longer be calculated from Larmor's formula, but it is determined 
by the dipole matrix element between ground and excited state,
\begin{equation}
\Gamma = \frac{\omega_0^3}
{3 \pi \epsilon_0 \hbar c^3} \, |\langle e | \mu |g \rangle|^2 \, .
\label{Gquant}
\end{equation}
For many atoms with a strong dipole-allowed 
transition starting from its ground state, the classical formula
Eq.~\ref{Gclass} nevertheless provides a good approximation to the 
spontaneous decay rate. 
For the $D$ lines of the alkali atoms Na, K, Rb, and Cs, the classical
result agrees with the true decay rate to within a few percent.

An important difference between the quantum-mechanical and
the classical oscillator is the possible occurence of saturation. 
At too high intensities of the driving field, 
the excited state gets strongly populated and the above result 
(Eq.~\ref{alpha}) is no longer valid. 
For dipole trapping, however, we are essentially interested in 
the far-detuned case with 
very low saturation and thus very low scattering rates 
($\Gamma_{\rm sc} \ll \Gamma$). 
We can thus use expression Eq.~\ref{alpha} also as an excellent 
approximation
for the quantum-mechanical oscillator.

\subsubsection{Dipole potential and scattering rate}

With the above expression for the polarizability of the atomic
oscillator the following explicit expressions are 
derived
for the dipole potential and the scattering rate in the relevant 
case of large detunings and negligible saturation: 
\begin{equation}
U_{\rm dip}({\bf r})  
=  - \frac{3 \pi c^2}{2 \omega_0^3} \,
\left( \frac{\Gamma}{\omega_0 -\omega} + 
\frac{\Gamma}{\omega_0 +\omega} \right) \, I({\bf r}) \, ,
\label{UdipTWAgen}
\end{equation}
\begin{equation}
\Gamma_{\rm sc}({\bf r})
=  \frac{3 \pi c^2}{2 \hbar \omega_0^3} \,
\left(\frac{\omega}{\omega_0}\right)^3 \,
\left( \frac{\Gamma}{\omega_0 -\omega} + 
\frac{\Gamma}{\omega_0 +\omega} \right)^2 \,
I({\bf r}) \, . 
\label{GscTWAgen}
\end{equation}
These general expressions are valid for any driving frequency $\omega$ 
and show two resonant contributions: Besides the
usually considered resonance at $\omega = \omega_0$, 
there is also the so-called
counter-rotating term resonant at $\omega = - \omega_0$. 

In most experiments, 
the laser is tuned relatively close to the 
resonance at $\omega_0$ such that the detuning 
$\Delta \equiv \omega - \omega_0$ fulfills $|\Delta| \ll \omega_0$. 
In this case, the counter-rotating term can be neglected in the
well-known {\it rotating-wave approximation} 
(see, e.g., Allen and Eberly, 1972), and one can set 
$\omega/\omega_0 \approx 1$. 
This approximation will be made throughout this article with a few
exceptions discussed in Chapter \ref{red}.

In this case of main practical interest, 
the general expressions for the dipole potential 
and the scattering rate 
simplify to
\begin{equation}
U_{\rm dip}({\bf r})  
=  \frac{3 \pi c^2}{2 \omega_0^3} \,
\frac{\Gamma}{\Delta} \, I({\bf r}) \, ,
\label{UdipTWA}
\end{equation}
\begin{equation}
\Gamma_{\rm sc}({\bf r})
=  \frac{3 \pi c^2}{2 \hbar \omega_0^3} \,
\left( \frac{\Gamma}{\Delta} \right)^2 \,
I({\bf r}) \, .
\label{GscTWA}
\end{equation}
The basic physics of dipole trapping in far-detuned laser fields 
can be understood on the basis of these two equations.
Obviously, a simple relation exists between the scattering rate and
the dipole potential,
\begin{equation}
\hbar \Gamma_{\rm sc} = \frac{\Gamma}{\Delta} \, U_{\rm dip} \, ,
\label{KKrelation}
\end{equation}
which is a direct consequence of the fundamental relation between the 
absorptive and dispersive response of the oscillator.
Moreover, these equations show two very essential points for dipole 
trapping:
%
%
\begin{itemize}
\item
{\it Sign of detuning:} Below an atomic resonance
(``red'' detuning, $\Delta <0$) the dipole potential is negative and 
the interaction thus attracts atoms into the light field.
Potential minima are therefore found at positions with maximum 
intensity. Above resonance (``blue'' detuning, $\Delta >0$) 
the dipole interaction repels atoms out of the
field, and potential minima correspond to minima of the intensity.
According to this distinction, dipole traps can be divided into two
main classes, red-detuned traps (Sec.~\ref{red}) 
and blue-detuned traps (Sec.~\ref{blue}).
\item
{\it Scaling with intensity and detuning:} The dipole potential scales as 
$I/\Delta$, whereas the scattering rate scales as $I/\Delta^2$. 
Therefore, optical dipole traps usually use large detunings and high
intensities to keep the scattering rate as low as possible 
at a certain potential depth.
\end{itemize}

\subsection{Multi-level atoms} \label{multilevel}

In real atoms used for dipole trapping experiments, the electronic
transition has a complex sub-structure.
The main consequence is that the dipole potential in general depends
on the particular sub-state of the atom.
This can lead to some quantitative modifications and also interesting
new effects. 
In terms of the oscillator model discussed before, multi-level atoms 
can be described
by state-dependent atomic polarizabilities. Here 
we use an alternative picture
which provides very intuitive insight into the motion of 
multi-level atoms in far-detuned laser fields: 
the concept of state-dependent ground-state potentials 
(Dalibard and Cohen-Tannoudji, 1985, 1989).
Alkali atoms are discussed as a situation of great practical importance, 
in order to clarify the role of fine-structure, 
hyperfine-structure, and magnetic sub-structure. 

\subsubsection{Ground-state light shifts and optical potentials} 
\label{LSpotentials}

The effect of far-detuned laser light on the atomic levels can be treated
as a perturbation in second order of the electric field, 
i.e.\ linear in terms of the field intensity.
As a general result of 
second-order time-independent perturbation theory for non-degenerate states, 
an interaction (Hamiltonian ${\cal H}_1$) leads to
an energy shift of the $i$-th state (unperturbed energy ${\cal E}_i$)
that is given by 
\begin{equation}
\Delta E_i = 
\sum_{j \ne i}\frac{|\langle j|{\cal H}_1|i\rangle|^2}
{{\cal E}_i - {\cal E}_j} \, .
\label{PTgen}
\end{equation}
For an atom interacting with laser light, the interaction Hamiltonian
is ${\cal H}_1 = -\hat{\mu}{\bf E}$ with $\hat{\mu}=-e {\bf r}$ representing
the electric dipole operator. For the  relevant energies ${\cal E}_i$,
one has to apply a `dressed state' view (Cohen-Tannoudji {\it et al.}, 
1992),
considering the combined system `atom plus field'. 
In its ground state the atom has zero internal energy and the field
energy is $n \hbar \omega$ according to the number $n$ of photons. 
This yields a total energy ${\cal E}_i = n \hbar \omega$ for the unperturbed
state. When the atom is put into an excited state by absorbing a photon, 
the sum of its internal energy $\hbar \omega_0$ and the field energy 
$(n-1) \hbar \omega$ becomes
${\cal E}_j  = \hbar \omega_0 + (n-1) \hbar \omega 
= -\hbar \Delta_{ij} + n \hbar \omega$. Thus the denominator in
Eq.~\ref{PTgen} becomes ${\cal E}_i - {\cal E}_j = \hbar \Delta_{ij}$

\begin{figure}
\vspace{0.5cm}
\epsfxsize=8.6cm   
  \centerline{\epsffile{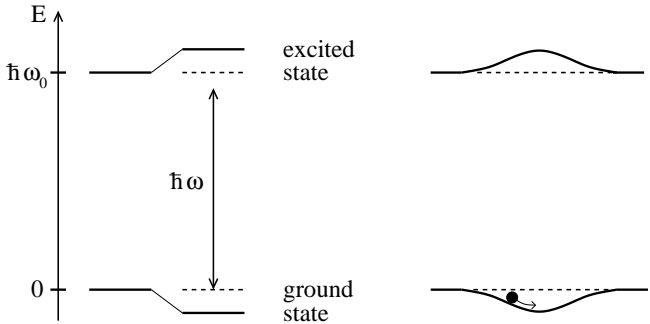}}
\vspace{0.5cm}
\caption{\it
Light shifts for a two-level atom. Left-hand side, red-detuned light 
($\Delta < 0$) shifts the ground state down and the excited state up by same
amounts.
Right-hand side, a spatially inhomogeneous field like a Gaussian laser 
beam produces a ground-state potential well, in which an atom can be
trapped.}
\label{lspotential}
\end{figure}

For a {\em two-level atom}, 
the interaction Hamiltonian 
is ${\cal H}_1 = -{\mu}{E}$ and Eq.~\ref{PTgen} simplifies to 
\begin{equation}
\Delta E = 
\pm \, \frac{|\langle e|\mu|g \rangle|^2}{\Delta} \,|E|^2 =
\pm \frac{3 \pi c^2}{2 \omega_0^3} \,
\frac{\Gamma}{\Delta} \, I  \,
\label{PTTWA}
\end{equation}
for the ground and excited state (upper and lower sign, respectively); 
we have used the relation 
$I = 2 \epsilon_0 c |\tilde{E}|^2$, and 
Eq.~\ref{Gquant} to substitute the dipole matrix element with
the decay rate~$\Gamma$.
This perturbative result 
obtained for the energy shifts reveals a very 
interesting and important fact: The optically induced shift (known as  
`light shift' or `ac Stark shift')
of the ground-state exactly corresponds to the dipole potential 
for the two-level atom (Eq.~\ref{UdipTWA}); 
the excited state shows the opposite shift. 
In the interesting case of low saturation, the atom resides
most of its time in the ground state and
we can interpret the {\em light-shifted ground state}  
as the relevant potential for the motion of atoms.
This situation is illustrated in Fig.~\ref{lspotential}.

For applying Eq.~\ref{PTgen} to a {\em multi-level atom} with transition 
substructure\footnote{Perturbation theory for non-degenerate
states can be applied in the absence of any coupling between degenerate 
ground states. 
This is the case for pure linear $\pi$ or circular $\sigma^{\pm}$ 
polarization, but not for mixed polarizations where Raman couplings between 
different magnetic sub-states become important; see, e.g., Deutsch and 
Jessen (1997).},
one has to know the dipole matrix elements 
$\mu_{ij} = \langle e_i |\mu | g_i  \rangle$ between specific electronic 
ground states $| g_i \rangle$ and excited states $| e_j \rangle$.
It is well-known in atomic physics (see, e.g., Sobelman, 1979) that 
a specific transition matrix element
\begin{equation}
\mu_{ij} = c_{ij} \, \| \mu \| \, ,
\end{equation}
can be written as a product of a reduced matrix element $\| \mu \|$
and a real transition coefficient $c_{ij}$.
The fully reduced matrix element 
depends on the electronic orbital wavefunctions only and  
is directly related to the spontaneous decay rate $\Gamma$ 
according to Eq.~\ref{Gquant}. The coefficients $c_{ij}$, which 
take into account the coupling strength between specific sub-levels 
$i$ and $j$ of the electronic ground and excited state, 
depend on the laser polarization and 
the electronic and nuclear angular momenta involved. 
They can be calculated
in the formalim of irreducible tensor operators 
or can be found in corresponding tables.

With this reduction of the matrix elements, we can now write the
energy shift of an electronic ground state $| g_i \rangle$ in the form
\begin{equation}
\Delta E_{i}  =  
\frac{3 \pi c^2 \, \Gamma}{2 \omega_0^3} \, I \times
\sum_{j}\frac{c^{2}_{ij}}{\Delta_{ij}} \, ,
\label{PTmulti}
\end{equation}
where the summation is carried out over all electronically excited 
states $| e_j \rangle$ .
This means, for a calculation of the state-dependent ground-state 
dipole potential $U_{{\rm dip,}\, i} = \Delta {\cal E}_{i}$, 
one has to sum up the contributions of all
coupled excited states, taking into account the relevant 
line strengths $c^{2}_{ij}$ and detunings $\Delta_{ij}$.

\subsubsection{Alkali atoms} \label{alkalidiscuss}

\begin{figure}
\vspace{0.5cm}
\epsfxsize=8.6cm
\centerline{\epsffile{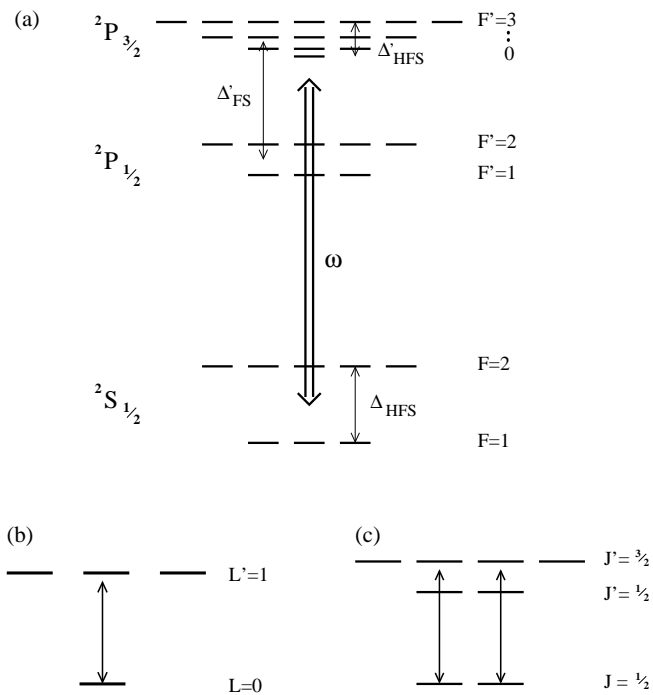}}
\vspace{0.5cm}
\caption{\it Level scheme of an alkali atom: 
(a) full substructure for a nuclear spin $I=\frac{3}{2}$, 
(b) reduced scheme for very large detunings exceeding
the fine-structure splitting ($|\Delta| \gg \Delta'_{\rm FS}$), and
(c) reduced scheme for large detunings in the range
$\Delta'_{\rm FS} \protect\gtrsim |\Delta| 
\gg \Delta_{\rm HFS}, \Delta'_{\rm HFS}$.}
\label{alkali}
\end{figure}

Most experiments in laser cooling and trapping are performed with
alkali atoms because of their closed optical transitions lying in a 
convenient spectral range. The main properties of alkali atoms of 
relevance for dipole trapping are summarized in Table~\ref{alkalitable}.
As an example, the full level scheme of 
the relevant $n\,s \rightarrow n\,p$ transition is shown 
in Fig.~\ref{alkali}(a) for a nuclear spin 
$I=\frac{3}{2}$, as in the case of 
$^7$Li, $^{23}$Na, $^{39,\,41}$K, and $^{87}$Rb.
Spin-orbit coupling in the excited state 
(energy splitting $\hbar \Delta'_{FS}$) leads to 
the well-known $D$ line doublet 
$^2S_{1/2}\rightarrow ^2P_{1/2}, ^2P_{3/2}$. 
The coupling to the nuclear spin then produces the hyperfine structure
of both ground and excited states with energies $\hbar \Delta_{\rm HFS}$ and 
$\hbar \Delta'_{\rm HFS}$, respectively. 
The splitting energies, obeying 
$\Delta'_{\rm FS} \gg \Delta_{\rm HFS} \gg \Delta'_{\rm HFS}$, 
represent the three relevant atomic energy scales.

On the basis of Eq.~\ref{PTmulti}, one can derive a general result 
for the potential of a ground state with total angular momentum $F$ 
and magnetic quantum number $m_F$, which is valid for both linear 
and circular polarization as long as all optical
detunings stay large \footnote{The assumption of unresolved excited-state 
hyperfine structure greatly simplifies the calculation according to 
Eq.~\ref{PTmulti}
because of existing sum rules for the line strength coefficients $c^2_{ij}$;
see also Deutsch and Jessen (1997).} compared with the excited-state 
hyperfine
splitting $\Delta'_{\rm HFS}$: 
\begin{equation}
U_{\rm dip}({\bf r}) = \frac{\pi c^2 \, \Gamma}{2 \omega_0^3} \, 
\left( \frac{2 + {\cal P} g_F m_F}{\Delta_{2,\,F} } +
   \frac{1 - {\cal P} g_F m_F}{\Delta_{1,\,F} }  \right) \, I({\bf r}) \,.
\label{alkaligeneral}
\end{equation} 
Here $g_F$ is the well-known Land\'e factor and ${\cal P}$ characterizes the
laser polarization (${\cal P}=0, \pm 1$ for linearly and 
circularly $\sigma^{\pm}$ polarized light).
The detunings
$\Delta_{2,\,F}$ and $\Delta_{1,\,F}$ refer to the energy splitting between 
the particular ground state $^2$S$_{1/2}, F$ 
and the center of the hyperfine-split 
$^2$P$_{3/2}$ and $^2$P$_{1/2}$ excited states, respectively. 
The two terms in brackets of Eq.~\ref{alkaligeneral} thus represent the 
contributions of the $D_2$ and the $D_1$ line to the total dipole 
potential.

In order to discuss this result, 
let us first consider the case of very large detunings
greatly exceeding the fine-structure splitting 
($|\Delta_{F, \,1}|, |\Delta_{F, \,2}| \gg \Delta'_{\rm FS}$), 
in which we can completely neglect the 
even much smaller hyperfine splitting.
Introducing a detuning $\Delta$ with respect to the center of
the $D$-line dublett, we can linearly expand Eq.~\ref{alkaligeneral}
in terms of the small parameter $\Delta'_{\rm FS}/\Delta$:
\begin{equation}
U_{\rm dip}({\bf r})  
=  \frac{3 \pi c^2}{2 \omega_0^3} \,
\frac{\Gamma}{\Delta} \, 
\left( \, 1 \, + \, \frac{1}{3} \, {\cal P} g_F m_F \, 
\frac{\Delta'_{\rm FS}}{\Delta} \, \right)
\, I({\bf r}) \, .
\label{alkaliexpand}
\end{equation}
While the first-order term describes a small residual dependence on
the polarization ${\cal P}$ and the magnetic sub-state $m_F$, the
dominating zero-order term is just the result obtained for a two-level atom 
(Eq.~\ref{UdipTWA}).
The latter fact can be understood by a simple argument:
If the fine-structure is not resolved, then
the detuning represents the leading term in the total Hamiltonian
and the atomic sub-structure can be ignored in a first perturbative step
by reducing the
atom to a very simple $s \rightarrow p$ transition; see 
Fig.~\ref{alkali}(b).
Such a transition behaves like a two-level atom with the full line strength 
for any laser polarization, and 
the ground-state light shift is thus equal to the one of a two-level atom. 
This single ground state couples to the electronic and nuclear spin
in exactly the same way as it would do without the light.
In this simple case, all resulting hyperfine and magnetic substates states 
directly acquire the light shift of the initial atomic $s$ state.

In the more general case of a resolved fine-structure, but unresolved 
hyperfine structure 
($ \Delta'_{\rm FS} \gtrsim 
|\Delta_{F, \,1}|, |\Delta_{F, \,2}| \gg \Delta_{\rm HFS}$), 
one may first consider the atom in spin-orbit coupling, 
neglecting the coupling to the nuclear spin. 
The interaction with the laser field can thus be considered in the 
electronic angular momentum configuration of the two $D$ lines, 
$J = \frac{1}{2} \rightarrow J'= \frac{1}{2}, \frac{3}{2}$.
In this situation, illustrated in Fig.~\ref{alkali}(c), 
one can first calculate the light shifts of the two electronic
ground states $m_J=\pm\frac{1}{2}$, and in a later step consider their
coupling to the nuclear spin. In this situation, it is important to 
distinguish between linearly and circularly polarized light:

For {\em linear} polarization,
both electronic ground states ($m_J=\pm\frac{1}{2}$) 
are shifted by the same amount because of simple symmetry reasons.
After coupling to the nuclear spin, the resulting 
$F, m_F$ states have to remain degenerate like the two original 
$m_J$ states. Consequently, all magnetic sublevels show the same light 
shifts according to the line strength factors
of $2/3$ for the $D_2$ line and $1/3$ for the $D_1$ line. 

For {\em circular} polarization ($\sigma^{\pm}$), 
the light lifts the degeneracy of the two magnetic sublevels of the
electronic $^2$S$_{\frac{1}{2}}$ ground state, and the 
situation gets more complicated.
The relevant line strength factors are then given by
$\frac{2}{3}(1 \pm m_J)$ for the $D_2$ line 
and $\frac{1}{3}(1 \mp 2 m_J)$ for the $D_1$ line.
The lifted degeneracy of the two ground states can be interpreted
in terms of a `fictitous magnetic field' (Cohen-Tannoudji 
and Dupont-Roc, 1972;  Cho, 1997; Zielonkowski {\it et al.}, 1998a), which 
is very useful to understand how the lifted 
$m_J$ degeneracy affects the $F,m_F$ levels after coupling to the nuclear 
spin. According to the usual theory of the linear Zeeman effect in weak 
magnetic fields, coupling to the nuclear spin affects the
magnetic sub-structure such that one has to 
replace $g_J m_J$ by $g_F m_F$, with
$g_J$ and $g_F$ denoting the ground state Land\'e factors.
Using this analogy and $g_J =2$ for alkali atoms, 
we can substitute $m_J$ by 
$\frac{1}{2} g_F m_F$ to calculate the relevant line strength factors
$\frac{1}{3}(2 \pm g_F m_F)$ and $\frac{1}{3}(1 \mp g_F m_F)$ for the 
$D_2$ and the $D_1$ line, respectively. These factors lead to
the $m_F$ dependent shifts for circularly polarized light
in Eq.~\ref{alkaligeneral}. 
One finds that this result stays valid 
as long as the excited-state hyperfine splitting remains unresolved.


For the {\em photon scattering rate} $\Gamma_{\rm sc}$ of a multi-level 
atom,
the same line strength factors are relevant as for the dipole potential, 
since absorption and light shifts are determined by
the same transititon matrix elements. 
For linear polarization, 
in the most general case $\Delta \gg \Delta'_{\rm HFS}$ considered here, 
one thus explicitely obtains
\begin{equation}
\Gamma_{\rm sc}({\bf r}) =
\frac{\pi c^2 \, \Gamma^2}{2 \hbar \omega_0^3} \,
\left( \frac{2}{\Delta^{\,2}_{2,F}} + \frac{1}{\Delta^{\,2}_{1,F}} \right) 
\, 
I({\bf r})
\label{Gsclin}
\end{equation}
This result is independent of $m_F$, but in general depends on the hyperfine 
state $F$ via the detunings. 
For linearly polarized light, optical pumping tends to equally distribute
the population among the different $m_F$ states, and thus leads to
complete depolarization. If the detuning is large
compared to the ground-state hyperfine splitting, then all sub-states 
$F, m_F$ are equally populated by redistribution via photon scattering.
For circular polarization, Zeeman pumping effects become very important,
which depend on the particular detuning regime and which we 
do not want to discuss in more detail here.
It is interesting to note that 
the general relation between dipole potential and scattering rate takes the 
simple form of Eq.~\ref{KKrelation} either if the contribution of one
of the two $D$ lines dominates for rather small detunings or
if the detuning is large compared to the fine-structure splitting.

Our discussion on multi-level alkali atoms shows that 
{\em linearly polarized light} is usually the right choice for 
a dipole trap\footnote{an interesting exception is the work by
Corwin {\it et al.}\ (1997) on trapping in circularly polarized light; see 
also Sec.~\ref{sec:spinrelaxation}.}, 
because the magnetic sub-levels $m_F$ 
of a certain hyperfine ground-state $F$ are shifted by same amounts. 
This only requires a detuning 
large compared to the excited-state hyperfine splitting, which is
usually very well fulfilled in dipole trapping experiments.
If the detuning also exceeds the ground-state hyperfine splitting
then all sub-states of the electronic ground state are equally shifted, 
and the dipole potential becomes completely independent of $m_F$ and $F$.
For circularly polarized light there is a $m_F$-dependent
contribution, which leads to a splitting analogous to a magnetic field.
This term vanishes only if the optical detuning greatly exceeds the 
fine-structure splitting.

\section{Experimental issues} \label{background2}

Here we discuss several issues of practical importance for experiments 
on dipole trapping. Cooling and heating in the trap is considered
in Sec.~\ref{coolheat}, followed by a summary of the typical experimental
procedures in Sec.~\ref{experimental}. Finally, the particular role of 
collisions is discussed in Sec.~\ref{collisions}.

\subsection{Cooling and heating} \label{coolheat}

Atom trapping in dipole potentials requires cooling to load the trap
and eventually also to counteract heating in the dipole trap. 
We briefly review the various available cooling methods and their 
specific features with respect to dipole trapping. 
Then we discuss sources of heating, and we derive explicite
expressions for the heating rate in the case of thermal equilibrium 
in a dipole trap. This allows for a direct comparison between dipole
traps operating with red and blue detuning.

\subsubsection{Cooling methods} \label{coolingmethods}

Efficient cooling techniques are an essential requirement to load
atoms into a dipole trap since the attainable trap depths are
generally below 1\,mK. Once trapped, further cooling can be applied to
achieve high phase-space densities and to compensate possible heating
mechanisms (see Sec.\ \ref{heating}) which would otherwise boil the
atoms out of the trap. The development of cooling methods for neutral
atoms has proceeded at breathtaking speed during the last decade, and
numerous excellent reviews have been written illuminating these
developments (Foot, 1991; Arimondo {\it et al.}, 1992; Metcalf and van
der Straten, 1994; Sengstock and Ertmer, 1995; Ketterle and van
Druten, 1996; Adams and Riis, 1997; Chu, 1998; Cohen-Tannoudji, 1998;
Phillips, 1998). In this section, we briefly discuss methods which are
of relevance for cooling atoms in dipole traps. Chapters \ref{red} and
\ref{blue} describe the experimental implementations of the cooling
schemes to particular trap configurations.

\paragraph*{Doppler cooling.} 
Doppler cooling is based on cycles of near-resonant absorption of a
photon and subsequent spontaneous emission resulting in a net atomic
momentum change per cycle of one photon momentum $\hbar k$ with $k =
2\pi/\lambda$ denoting the wavenumber of the absorbed photon.  Cooling
is counteracted by heating due to the momentum fluctuations by the
recoil of the spontaneously emitted photons (Minogin and Letokhov,
1987).  Equilibrium between cooling and heating determines the lowest
achievable temperature. For Doppler cooling of two-level atoms in
standing waves (``optical molasses''), the minimum temperature is
given by the Doppler temperature $k_B T_D = \hbar \Gamma/2$. Typical
values of the Doppler temperature are around $100\,\mu$K, which is
just sufficiently low to load atoms into a deep dipole trap.  The
first demonstration of dipole trapping (Chu {\it et al.}, 1986) used
Doppler cooling for loading the trap and keeping the atoms from being
boiled out of the trap.  With the discovery of methods reaching much
lower temperatures, Doppler cooling has lost its importance for the
direct application to dipole traps.

\paragraph*{Polarization-gradient cooling.}
The Doppler temperature is a somewhat artificial limit since it is
based on the simplifying assumption of a two-level atom. It was soon
discovered that atoms with a more complex level structure can be
cooled below $T_D$ in standing waves with spatially varying
polarizations (Lett {\it et al.}, 1988). The cooling mechanisms are
based on optical pumping between ground state Zeeman sublevels. The
friction force providing cooling results either from unbalanced
radiation pressures through motion-induced atomic orientation, or from
a redistribution among the photons in the standing wave (Dalibard and
Cohen-Tannoudji, 1989). In the latter case, the cooling force can be
illustratively
explained by the so-called Sisyphus effect for a moving atom.  The atom
looses kinetic energy by climbing up the dipole potential induced by
the standing wave of the trapping light. When reaching the top of this
potential hill, the atom is optically pumped back into the bottom of
the next potential valley from where again it starts to climb
(Dalibard and Cohen-Tannoudji, 1985).  Polarization-gradient cooling
can be achieved in standing waves at frequencies below an atomic
resonance ({\em red-detuned molasses}) (Lett {\it et al.}, 1988;
Salomon {\it et al.}, 1990) as well as above an atomic resonance ({\em
  blue-detuned molasses}) (Boiron {\it et al.}, 1995; Hemmerich {\it
  et al.}, 1995).  In blue-detuned molasses, atoms primarily populate
states which are decoupled from the light field resulting in a
reduction of photon scattering, but also in a smaller cooling rate as
compared to red-detuned molasses (Boiron {\it et al.}, 1996).

With polarization-gradient molasses one can prepare free-space atomic
samples at temperatures on the order of $\sim 10 T_{\rm rec}$ with the
recoil temperature
\begin{equation}
T_{\rm rec} = \frac{\hbar^2 k^2}{m}
\label{Trec}
\end{equation}
being defined as the temperature associated with the kinetic energy
gain by emission of one photon. For the alkali atoms, recoil
temperatures are given in Table \ref{alkalitable}.  The achievable
temperatures are much below the typical depth of a dipole trap.
Polarization-gradient cooling therefore allows efficient loading of
dipole traps (see Sec.\ \ref{loading}), either by cooling inside a
magneto-optical trap (Steane and Foot, 1991) or by cooling in a short
molasses phase before transfer into the dipole trap.

Besides enhancing the loading efficiency, polarization-gradient
cooling was directly applied to atoms trapped in a dipole
potential by subjecting them to near-resonant standing waves with
polarization gradients (Boiron {\it et al.}, 1998; Winoto {\it et
  al.}, 1998).  Since the cooling mechanism relies on the modification
of the ground-state sublevels by the cooling light, a necessary
condition for efficient cooling is the independency of the trapping
potential from the Zeeman substate which, in contrast to magnetic
traps, can easily be fulfilled in dipole traps as explained in Sec.\ 
\ref{alkalidiscuss}.

\paragraph*{Raman cooling.}
The recoil temperature marks the limit for cooling methods based on
the repeated absorption and emission of photons such as Doppler
cooling and polarization-gradient cooling. To overcome this limit,
different routes were explored to decouple the cold atoms from the
resonant laser excitation once they have reached small velocities.
All approaches are based upon transitions with a narrow line width
which are thus extremely velocity-selective such as dark-state
resonances (Aspect {\it et al.}, 1988) or Raman transition between
ground state sublevels (Kasevich and Chu, 1992).  With free atomic
samples, the interaction time, and thus the spectral resolution
determining the final temperature, was limited by the time the
thermally expanding atomic cloud spent in the laser field (Davidson
{\it et al.}, 1994). Taking advantage of the long storage times in
dipole traps, Raman cooling was shown to efficiently work for trapped
atomic samples (Lee {\it et al.}, 1994; Lee {\it et al.}, 1996; Kuhn
et al., 1996).

The basic principle of Raman cooling is the following. Raman pulses
from two counter-propagating laser beams transfer atoms from one
ground state $|1\rangle$ to another ground state $|2\rangle$
transferring $2\hbar k$ momentum \footnote{The two states are, e.g.,
  the two hyperfine ground states of an alkali atom.}. Using sequences
of Raman pulses with varying frequency width, detuning, and
propagation direction, pulses can be tailored which excite all atoms
except those with a velocity near $v=0$ (Kasevich {\it et al.}, 1992;
Reichel {\it et al.}, 1995; Kuhn {\it et al.}, 1996). The cooling
cycle is completed by a light pulse resonantly exciting atoms in state
$|2\rangle$ in order to optically pump the atoms back to the
$|1\rangle$ state through spontaneous emission.  Each spontaneous
emission randomizes the velocity distribution so that a fraction of
atoms will acquire a velocity $v \approx 0$. By repeating many
sequences of Raman pulses followed by optical pumping pulses, atoms
are accumulated in a small velocity interval around $v=0$. The final
width of the velocity distribution, i.e. the temperature, is
determined by the spectral resolution of the Raman pulses. In dipole
traps, high resolution can be achieved by the long storage times. In
addition, motional coupling of the degrees of freedom through the trap
potential allowed to cool the atomic motion in all three dimensions
with Raman pulses applied along only one spatial direction (Lee {\it
  et al.}, 1996; Kuhn {\it et al.}, 1996).

\paragraph*{Resolved-sideband Raman cooling.}
Cooling with well-resolved motional sidebands, a technique well
known from laser cooling in ion traps (Ghosh, 1995), was very recently
also applied to optical dipole traps (Hamann {\it et al.}, 1998; Perrin et
al., 1998; Vuletic {\it et al.}, 1998; Bouchoule {\it et al.}; 1998).  For
resolved-sideband cooling, atoms must be tightly confined along (at
least) one spatial dimension with oscillation frequencies $\omega_{\rm
  osc}$ large enough to be resolved by Raman transitions between two
ground state levels. In contrast to Raman cooling explained in the
preceeding paragraph, the confining potential of the trap is therefore
a necessary prerequisite for the application of sideband cooling.

Atomic motion in the tightly confining potential is described by a
wavepacket formed by the superposition of vibrational states $|n_{\rm
  osc}\rangle$. In the Lamb-Dicke regime, where the rms size of the
wavepacket is small compared to the wavelength of the cooling
transition, an absorption-spontaneous emission cycle almost
exclusively returns to the same vibrational state it started from
($\Delta n_{\rm osc} = 0$).  The Lamb-Dicke regime is reached by
trapping atoms in dipole potentials formed in the interference pattern
of far-detuned laser beams (see Secs.\ \ref{sec:SWT} and
\ref{sec:CBT}).  Sideband cooling consists of repeated cycles of Raman
pulses which are tuned to excite transitions with $\Delta n_{\rm osc}
= -1$ followed by an optical pumping pulse involving spontaneous
emission back to the initial state with $\Delta n_{\rm osc} = 0$.  In
this way, the motional ground state $n_{\rm osc} = 0$ is selectively
populated since it is the only state not interacting with the Raman
pulses. Achievable temperatures are only limited by the separation and
the width of the Raman sidebands determining the suppression of
off-resonant excitation of the $n_{\rm osc} = 0$ state.

A particularly elegant realization of sideband cooling was
accomplished by using the trapping light itself to drive the Raman
transition instead of applying additional laser fields (Hamann et al,
1998; Vuletic {\it et al.}, 1998). For this purpose, a small magnetic
field was applied shifting the energy of two adjacent ground-state
Zeeman sublevels relative to each other by exactly one vibrational
quantum $\hbar \omega_{\rm osc}$. In this way, the bound states
$|m_F;n_{\rm osc}\rangle$ and $|m_F-1;n_{\rm osc}-1\rangle$ became
degenerate, and Raman transitions between the two states could be
excited with single-frequency light provided by the trapping field
({\em degenerate sideband cooling}) (Deutsch and Jessen, 1998). The
great advantage of degenerate sideband cooling is that works with only
the two lowest-energy atomic ground states being involved, resulting
in the suppression of heating and trap losses caused by inelastic
binary collisions (see Sec.\ \ref{collisions}).

\paragraph*{Evaporative cooling.} 
Evaporative cooling, orginally demonstated with magnetically trapped
hydrogen (Hess {\it et al.}, 1987), has been the key technique to
achieve Bose-Einstein condensation in magnetic traps (Ketterle and van
Druten, 1996). It relies on the selective removal of high-energetic
particles from a trap and subsequent thermalization of the remaining
particles through elastic collisions.  Evaporative cooling requires
high densities to assure fast thermalization rates, and large initial
particle numbers since a large fraction of trapped atoms is removed
from the trap by evaporation. To become effective, the ratio between
inelastic collisions causing losses and heating, and elastic
collisions providing thermalization and evaporation, has to be large.

In dipole traps, inelastic processes can be greatly suppressed when
the particles are prepared in their energetically lowest state.
However, the requirement of large particle numbers and high density
poses a dilemma for the application of evaporative cooling to dipole
traps. In tightly confining dipole traps such as a crossed dipole
trap, high peak densities can be reached, but the trapping volume, and
thus the number of particles transferred into the trap, is small.  On
the other hand, large trapping volumes yielding large numbers of
trapped particles provide only weak confinement yielding small elastic
collision rates. This is why, until now, only one experiment is
reported on evaporative cooling in dipole traps starting with a small
sample of atoms (Adams {\it et al.}, 1995). By precooling large
ensembles in dipole traps with optical methods explained in the
preceeding paragraphs, much better starting conditions for evaporative
cooling are achievable (Engler et al., 1998; Vuletic {\it et al.},
1998; Winoto {\it et al.}, 1998) making evaporative cooling a still
interesting option for future applications.

\paragraph*{Adiabatic expansion.}
When adiabatically expanding a potential without changing its shape,
the temperature of the confined atoms is decreased without increasing
the phase-space density \footnote{Adiabatic changes of the potential
  shape leading to an increase of the phase-space density are
  demonstrated in (Pinkse {\it et al.}, 1997) and (Stamper-Kurn {\it
    et al.}, 1998b).}. In dipole potentials, cooling by adiabatic
expansion was realized by slowly ramping down the trapping light
intensity (Chen et al., 1992; Kastberg {\it et al.}, 1995).  In
far-detuned traps consisting of micropotentials induced by
interference, adiabatic cooling is particularly interesting when the
modulation on the scale of the optical wavelength is slowly reduced
without modifying the large-scale trapping potential, by, e.g.\ 
changing the polarization of the interfering laser beams. Atoms are
initially strongly localized in the micropotentials resulting in high
peak densities. After adiabatic expansion, the temperature of the
sample is reduced, as is the peak density. However, the density
averaged over one period of the interference structure is not
modified.

\subsubsection{Heating mechanisms} \label{heating}

Heating by the trap light is an issue of particular importance for optical 
dipole trapping. 
A fundamental source of heating is the spontaneous 
scattering of trap photons, which due to its random nature causes 
fluctuations of the radiation force\footnote{Under typical conditions
of a dipole trap, the
scattering force that results in traveling-wave configurations 
stays very weak as compared to the dipole force and can thus be neglected.}.
In the far-detuned case considered here, the scattering is completely 
elastic 
(or quasi-elastic if a Raman process changes the atomic ground state). 
This means that the energy of the scattered photon is determined by the 
frequency of the laser and not of the optical transition.

Both absorption and spontaneous re-emission processes show fluctuations and
thus both contribute to the total heating (Minogin and Letokhov, 1987). 
At large detunings, where scattering processes follow Poisson statistics, 
the heating due to fluctuations in absorption 
corresponds to an increase of the thermal energy by exactly 
the recoil energy  
$E_{\rm rec} 
= k_B T_{\rm rec}/2$ per scattering event.
This first contribution occurs in the propagation direction of the light 
field and is thus anisotropic (so-called directional diffusion).
The second contribution is due to the random
direction of the photon recoil in spontaneous emission. 
This heating also increases the thermal energy
by one recoil energy $E_{\rm rec}$ per scattering event, but distributed
over all three dimensions. 
Neglecting the general dependence on the polarization of the spontaneously 
emitted photons, 
one can assume an isotropic distribution for this heating mechanism.
Taking into account both contributions,
the longitudinal motion is heated on an average by $4E_{\rm rec}/3$ per 
scattering process, 
whereas the two transverse dimensions are each heated by $E_{\rm rec}/3$.
The overall heating thus corresponds to an increase of the total thermal 
energy 
by $2 E_{\rm rec}$ 
in a time $\bar{\Gamma}^{-1}_{\rm sc}$.
 
For simplicity, we do not consider the generally anisotropic character 
of heating here; in most cases of interest the trap anyway mixes the 
motional degrees on a time scale faster than or comparable to the heating. 
We can thus use a simple global three-dimensional heating power 
$P_{\rm heat} = \dot{\bar{E}}$ 
corresponding to the increase of the 
mean thermal energy $\bar{E}$ of the atomic motion with time. 
This heating power is directly
related to the average photon scattering rate $\bar{\Gamma}_{\rm sc}$ by
\begin{equation}
P_{\rm heat} = 2 E_{\rm rec} \, \bar{\Gamma}_{\rm sc}
= k_B T_{\rm rec} \, \bar{\Gamma}_{\rm sc} \, .
\label{heatingpower}
\end{equation}

For intense light fields close to resonance, 
in particular in standing-wave configurations, 
it is well known that the induced redistribution of photons 
between different traveling-wave components can lead to dramatic heating 
(Gordon and Ashkin, 1980; Dalibard and Cohen-Tannoudji, 1985). 
In the far off-resonant case, however, this induced heating 
falls off very rapidly with the detuning and is usually completely 
negligible as compared to spontaneous heating.

In addition to the discussed fundamental heating in dipole traps,
it was recently pointed out by Savard {\it et al.}\ (1997),
that technical heating can occur due to intensity fluctuations and pointing 
instabilities in the trapping fields. 
In the first case, fluctuations occuring at twice the
characteristic trap frequencies are
relevant, as they can parametrically drive the oscillatory atomic motion. 
In the second case, a shaking of the
potential at the trap frequencies increases the motional amplitude.
Experimentally, these issues will strongly depend on the particular laser
source and its technical noise spectrum, 
but have not been studied in detail yet. Several experiments have indeed
shown indications for heating in dipole traps by unidentified mechanisms
(Adams {\it et al.}, 1995; Lee {\it et al.}, 1996; Zielonkowski {\it et 
al.}, 1998b; 
Vuletic {\it et al.}, 1998), 
which may have to do with fluctuations of the trapping light.

\subsubsection{Heating rate} \label{equilibrium}

For an ultracold atomic gas in a dipole trap it is often a good assumption
to consider a thermal equilibrium situation, in which the energy 
distribution
is related to a temperature~$T$. The further assumption of a 
power-law potential then allows one to derive very useful expressions for 
the mean photon scattering rate and the corresponding heating rate of the 
ensemble, which also illustrate important differences between traps 
operating
with red and blue detuned light. 

In thermal equilibrium, 
the mean kinetic energy per atom in a three-dimensional trap is 
$\bar{E}_{\rm kin} = 3k_B T/2$. 
Introducing the parameter 
$\kappa \equiv \bar{E}_{\rm pot}/\bar{E}_{\rm kin}$ as the
ratio of potential and kinetic energy, we can express the
mean total energy $\bar{E}$ as
\begin{equation}
\bar{E} = \frac{3}{2}(1 + \kappa) \, k_B T \, .
\label{meanenergy}
\end{equation}

For many real dipole traps as described in Secs.\ \ref{red} and \ref{blue}, 
it is a good approximation to assume a 
separable power-law potential with a constant offset $U_0$
of the form
\begin{equation}
U(x, y, z) = U_0 + a_1 x^{n_1} + a_2 y^{n_2} + a_3 z^{n_3} \, .
\label{potexpansion}
\end{equation}
In such a case, the virial theorem can be used to calculate the ratio
between potential and kinetic energy
\begin{equation}
\kappa 
= \frac{2}{3} \, \left( \frac{1}{n_1} + \frac{1}{n_2} + \frac{1}{n_3} 
\right) \, .
\end{equation}
For a 3D harmonic trap this gives $\kappa=1$, for an ideal 3D box potential
$\kappa = 0$.

The relation between mean energy and temperature 
(Eq.~\ref{meanenergy}) allows us to reexpress the heating power 
resulting from photon scattering (Eq.~\ref{heatingpower}) as a 
heating rate
\begin{equation}
\dot{T} = \frac{2/3}{1+\kappa} \, T_{\rm rec} \,  \bar{\Gamma}_{\rm sc} 
\, ,
\label{heatingrate}
\end{equation}
describing the corresponding increase of temperature with time.

The mean scattering rate $\bar{\Gamma}_{\rm sc}$ can, in turn, be calculated 
from the temperature of the sample, according to the following arguments:
Eq.~\ref{KKrelation} relates the average scattering rate to 
the mean dipole potential $\bar{U}_{\rm dip}$ experienced by the atoms.
In a pure dipole trap\footnote{this excludes hybrid potentials in which
other fields (gravity, magnetic or electric fields) are important for the 
trapping.} described by Eq.~\ref{potexpansion}, 
the mean optical potential is related to the mean potential energy
$\bar{E}_{\rm pot}$, the mean kinetic energy $\bar{E}_{\rm kin}$, and the 
temperature $T$ by
\begin{equation}
\bar{U}_{\rm dip} = U_0 + \bar{E}_{\rm pot} 
= U_0 + \kappa \bar{E}_{\rm kin}
= U_0 + \frac{3 \kappa}{2} \, T 
\, .
\end{equation} 
This relation allows us to express the mean scattering rate as 
\begin{equation}
\bar{\Gamma}_{\rm sc} 
= \frac{\Gamma}{\hbar \Delta} \, ( \, U_0 + \frac{3 \kappa}{2}\, k_B T \,) 
\, .
\label{scatteringT}
\end{equation}

\begin{figure}
\vspace{0.5cm}
\epsfxsize=8.6cm 
\centerline{\epsffile{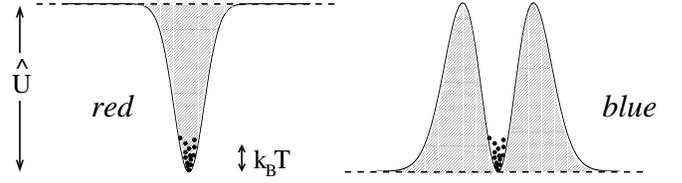}}
\vspace{0.5cm}
\caption{\it Illustration of dipole traps with red and blue detuning.
In the first case, a simple Gaussian laser beam is assumed. In the second
case, a Laguerre-Gaussian LG$_{01}$ ``doughnut'' mode is chosen 
which provides the same 
potential depth and the same curvature in the trap center 
(note that the latter case requires $e^2$ times more laser power or 
smaller detuning).}
\label{heatingfigure}
\end{figure}

Based on this result, let us now discuss two specific situations which are
typical for real experiments as described in 
Secs.~\ref{red} and \ref{blue}; see illustrations in 
Fig.~\ref{heatingfigure}.
In a red-detuned dipole trap ($\Delta<0$), the
atoms are trapped in an intensity maximum with $U_0 < 0$, and the 
trap depth $\hat{U} = |U_0|$ is usually large compared to 
the thermal energy $k_B T$.
In a blue-detuned trap ($\Delta>0$), 
a potential minimum corresponds to an intensity minimum, 
which in an ideal case means zero intensity. In this case,
$U_0 = 0$ and the potential depth $\hat{U}$ 
is determined by the height of the
repulsive walls surrounding the center of the trap.

For red and blue-detuned traps with $\hat{U} \gg k_B T$, 
Eqs.~\ref{heatingrate} and \ref{scatteringT} yield the following 
heating rates: 
\begin{mathletters}
\begin{eqnarray}
\dot{T}_{\rm red} & = & \frac{2/3}{1+\kappa} \,
T_{\rm rec} \, \frac{\Gamma}{\hbar |\Delta|} \, \hat{U} \, ,\\
\label{redheating}
\dot{T}_{\rm blue} & = & \frac{\kappa}{1 + \kappa} \,
T_{\rm rec} \, \frac{\Gamma}{\hbar \Delta} \, k_B T  \, .
\label{blueheating}
\end{eqnarray}
\label{redblueheating}
\end{mathletters}
Obviously, a red-detuned trap shows linear heating (which decreases when
$k_BT$ approaches $\hat{U}$), whereas heating behaves
exponentially in a blue-detuned trap.
Note that in blue-detuned traps a fundamental lower limit
to heating is set by the zero-point energy of the atomic motion, which
we have neglected in our classical consideration.
\pagebreak

Eqs.~\ref{redblueheating} allow for a very illustrative direct comparison 
between a blue and a red-detuned trap: The ratio of heating at the same
magnitude of detuning $|\Delta|$ is given by
\begin{equation}
\frac{\dot{T}_{\rm blue}}{\dot{T}_{\rm red}}
\, = \, \frac{3 \kappa}{2} \, \frac{k_B T}{\hat{U}} \, .
\label{bluered}
\end{equation}
This comparison shows that blue detuning offers
substantial advantages in {\em two experimental situations}: 
\begin{itemize}
\item
$\hat{U} \gg k_BT$, very deep potentials for tight confinement,
\item
$\kappa \ll 1$, box-like potentials with hard repulsive walls.
\end{itemize}
When, in other words, a harmonic potential of moderate depth is
to be realized for a certain experiment, the advantage of blue
detuning is not substantial. The choice of red detuning may be
even more appropriate as the better concentration of the available
laser power in such a trap allows one to use larger detunings 
to create the required potential depth.

\subsection{Experimental techniques} \label{experimental}

\subsubsection{Trap loading} \label{loading}

The standard way to load a dipole trap is to start from a magneto-optical
trap (MOT). This well-known radiation-pressure trap 
operating with near-resonant light
was first demonstrated by Raab {\it et al.}\ in 1987 and has
now become the standard source of ultracold atoms in many laboratories all
over the world. 
A MOT can provide temperatures down to a few 10\,$T_{\rm rec}$,
when its operation is optimized for sub-Doppler cooling 
(see Sec.~\ref{coolingmethods}). This sets a natural scale 
for the minimum depth of a dipole trap as required for efficient loading.
Due to their lower recoil temperatures (Table \ref{alkalitable}), 
heavier alkali atoms require less
trap depth as the lighter ones and thus allow for larger detunings.
For the heavy Cs atoms, for example, dipole traps with depths as low as 
$\sim$\,10\,$\mu$K can be directly loaded from a MOT (Zielonkowski {\it et 
al.},
1998b).
Trap loading at much lower depths can be reached with Bose-Einstein 
condensates (Stamper-Kurn {\it et al.}, 1998).

For dipole-trap loading, a MOT is typically operated in two stages.
First, its frequency detuning is set quite close to resonance 
(detuning of a few natural linewidths) to optimize
capture by the resonant scattering force.
Then, after the loading phase, the MOT parameters are changed to 
optimize sub-Doppler cooling (Drewsen {\it et al.}, 1994; Townsend {\it et 
al.}, 1995). 
Most importantly, the detuning is switched
to much higher values (typically 10\,--\,20 linewidths), and eventually also
the laser intensity is lowered. For the heavier alkali 
atoms\footnote{The lightest alkali atom Li 
behaves in a completely different way: Here optimum
loading is accomplished at larger detunings and optimum cooling 
is obtained relatively close to resonance (Sch\"unemann {\it et al.}, 
1998)}, 
this procedure provides maximum phase-space densities for trap loading.
Another option is to ramp up the magnetic fields of the MOT to spatially 
compress the sample (Petrich {\it et al.}, 1994).

The dipole trap is filled by simply overlapping it with the atomic cloud in 
the 
MOT, before the latter is turned off. In this procedure, it is 
advantageous to switch off the magnetic field of the MOT a few ms before the 
laser fields are extinguished,
because the short resulting optical molasses cooling phase establishes the 
lowest 
possible temperatures and a quasi-thermal distribution in the trap.
For practical reasons, the latter is important because a MOT does not 
necessarily load the atoms into the very center of the dipole trap. 
When MOT position and dipole trap center do not coincide exactly, 
loading results in excess potential energy in the dipole trap. 
When the MOT light is extinguished, it is very important to shield
the dipole trap from any resonant stray light, in particular if very low
scattering rates ($\lesssim$\,1\,s$^{-1}$) are to be reached.

The MOT itself can be loaded in a simple vapor cell (Monroe {\it et al.}, 
1990).
In such a set-up, however, the lifetime of the dipole trap is typically 
limited to less than 1\,s by collsions with atoms in the background gas.
If longer lifetimes are requested for a certain application, 
the loading of the MOT under much better vacuum conditions 
becomes an important issue,  
similar to experiments on Bose-Einstein condensation. 
Loading can then be accomplished from a very dilute vapor 
(Anderson {\it et al.}, 1994),
but more powerful concepts can be realized with a Zeeman-slowed atomic beam 
(Phillips and Metcalf, 1982), 
with a double-MOT set-up (Myatt {\it et al.}, 1997), 
or with slow-atom sources based on modified MOTs 
(Lu {\it et al.}, 1996, Dieckmann {\it et al.}, 1998).

Regarding trap loading, a dipole trap with red detuning can offer an 
important advantage over a blue-detuned trap: 
When MOT and dipole trap are simultaneously turned on, 
the attractive dipole potential leads to a local density increase in the 
MOT, which can substantially enhance the loading process. 
In very deep red-detuned traps, however, the level shifts 
become too large for the cooling light and efficient loading requires rapid 
alternation between cooling and trapping light (Miller {\it et al.}, 1993).

\subsubsection{Diagnostics} \label{diagnostics}

The atomic sample in a dipole trap is characterized by the number of stored 
atoms, the motional temperature of the ensemble (under the assumption of
thermal distribution), and the distribution of population
among the different ground-state sub-levels. Measurements of these 
important quantities can be made in the following ways.

\paragraph*{Number of atoms.}

A very simple and efficient method, which is often used to determine the 
number of atoms in a dipole trap is to recapture them into the MOT and 
to measure the power of the emitted fluorescence light with a calibrated 
photo-diode or CCD camera. In this way, it is quite easy to detect down to 
about one hundred atoms, but even single atoms may be observed in a more
elaborate set-up (Haubrich {\it et al.}, 1996).
This recapture method works particularly well if it is ensured that
the MOT does not capture any other atoms than those released from the dipole
trap. This is hardly possible in a simple vapor-cell set-up, 
but quite easy if an atomic beam equipped with
a mechanical shutter is used for loading the MOT. 

In contrast to the completely destructive recapture, several other
methods may be applied. The trapped atoms can be illuminated with a short
resonant laser pulse of moderate intensity to measure the emitted 
fluorescence
light. This can be done also with spatial resolution by using a CCD camera;
see Fig.~\ref{cat}(b) for an example. If the total number of atoms is not
too low, the detection pulse can be kept weak enough to avoid trap loss 
by heating. Furthermore, absorption imaging can be used
(see Fig.~\ref{fig:trap.salomon}), 
or even more sensitive and less destructive
dark-field or phase-contrast imaging methods as applied to sensitively 
monitor Bose-Einstein condensates (Andrews {\it et al.}, 1996; 
Bradley {\it et al.}, 1997; Andrews {\it et al.}, 1997).

\paragraph*{Temperature.}

In a given trapping potential $U({\bf r})$ the thermal 
density distribution $n({\bf r})$ direct follows from the Boltzmann factor,
\begin{equation}
n({\bf r}) = n_0 \exp\left(-\frac{U({\bf r})}{k_BT}\right) \, .
\label{Boltzmann}
\end{equation}
The temperature can thus be derived from the measured spatial 
density distribution in the trap, which itself can be observed by
various imaging methods (fluorescence, absorptive, and dispersive imaging).
For a 3D harmonic potential 
$U({\bf r}) = \frac{1}{2}m\,(\omega_x^2 x^2 + \omega_y^2 y^2+ \omega_z^2 
z^2)$ 
the resulting distribution is Gaussian in all directions,
\begin{equation}
n({\bf r}) = n_0 \, 
\exp\left(- \frac{x^2}{2 \sigma^2_x}\right) \,
\exp\left(- \frac{y^2}{2 \sigma^2_y}\right) \,
\exp\left(- \frac{z^2}{2 \sigma^2_z}\right) \, ,
\label{gauss}
\end{equation}
with $\sigma_i = \omega^{-1}_i \sqrt{k_B T/m}$.
The temperature is thus related to the spatial extensions of the trapped 
atom
cloud by
\begin{equation}
T = \frac{m}{k_B} \, \sigma^2_i \omega_i^2 \, .
\label{Tsigma}
\end{equation}
Obviously, it is very important to know the exact trap frequencies to 
precisely determine the temperature; a practical example for such a 
measurement is discussed in context with Fig.~\ref{fig:trap.salomon}(c).
This way of measuring the temperature is limited by the 
resolution of the imaging system and therefore becomes difficult for
very tightly confining potentials.

A widely used and quite accurate, 
but completely destructive way to measure temperatures is the time-of-flight 
method. The trap is 
turned off to release the atoms into a free, ballistic flight.
This has to be done in a rapid, completely non-adiabatic way as
otherwise an adiabatic cooling effect (see Sec.~\ref{coolingmethods}) 
would influence the measurement.
After a sufficiently long ballistic expansion phase, 
the resulting spatial distribution, which can again be observed by the 
various
imaging methods, directly reflects the velocity distribution at the time of 
release.
Another method is to detect the Doppler broadening of Raman transitions 
between ground states, using a pair of counterpropagating laser beams
(Kasevich and Chu, 1991), which is not limited by the natural linewidth of
the optical transition.

\paragraph*{Internal distribution.}

The relative population of the two hyperfine ground states
of an alkali atom (see level scheme in Fig.~\ref{alkali}) can be measured 
by application of a probe pulse resonant to the closed sub-transition 
$F=I+1/2 \rightarrow F'=3/2$ in the hyperfine structure, 
which is well resolved for the heavier alkali atoms (see Table 
\ref{alkalitable}). The fluorescence light is then proportional to the
number of atoms in the upper hyperfine state $F=I+1/2$.
If, in contrast, a repumping field 
is present in the probe light (as it is always used in a MOT), 
the fluorescence is proportional to the total number of atoms,
as all atoms are immediately pumped into the closed excitation cycle.
The normalized fluorescence signal thus gives the relative population of 
the upper hyperfine ground state ($F=I+1/2$); such a measurement is 
discussed
in Sec.~\ref{sec:spinrelaxation}.
A very sensitive alternative, which works very well in shallow dipole traps, 
is to blow the total upper-state population out of the trap by the radiation 
pressure of an appropriate resonant light pulse. 
Subsequent recapture into the MOT then shows 
how many atoms have remained trapped in the shelved lower hyperfine ground 
state.

The distribution of population over different magnetic sub-states 
can be analyzed by Stern-Gerlach methods.
When the atomic ensemble is released from the dipole trap and ballistically 
expands in an inhomogeneous magnetic field, then atoms in different magnetic
sub-levels can be well separated in space. Such an analysis has been used 
for 
optically trapped Bose-Einstein condensates (Stamper-Kurn {\it et al.}, 
1998a; 
Stenger, 1998); an example is shown in Fig.~\ref{fig:BEC.ketterle}(b). 
Another possibility, which can be easily 
applied to shallow dipole traps, is to pull atoms out of the trap by the 
state-dependent magnetic force, as has been used 
by Zielonkowski {\it et al.}\ (1998b) for measuring
the depolarizing effect of the trap photon scattering; see discussion in 
Sec.~\ref{sec:startrap}.

\subsection{Collisions} \label{collisions}

It is a well-known experimental fact in the field of 
laser cooling and trapping that collisional processes can lead to 
substantial trap loss. 
Detailed measurements of trap loss under various conditions
provide insight into ultracold collision phenomena, 
which have been subject of extensive research 
(Walker and Feng, 1994; Weiner, 1995).
Here we discuss the particular features of dipole traps
with respect to ultracold collisions.

The decay of the number $N$ of atoms in a trap
can be described by the general loss equation 
\begin{equation}
\dot{N}(t) = - \, \alpha \, N(t) - \beta \int_V n^2({\bf r}, t)\,d^3r 
- \gamma \int_V n^3({\bf r}, t) \, d^3r \, .
\label{lossdgl}
\end{equation}
Here the {\em single-particle} 
loss coefficient $\alpha$ takes into account collisions
with the background gas in the vacuum apparatus.
As a rule of thumb, 
the $1/e$ lifetime $\tau = 1/\alpha$ of a dipole trap 
is $\sim$\,1\,s at a pressure of $3\times10^{-9}$\,mbar.  This is about 
three times lower than the corresponding lifetime in a MOT because of the 
larger cross sections for collisional loss at lower trap depth.

\begin{figure}
\vspace{0.0cm}
\epsfxsize=6.0cm  
\centerline{\epsffile{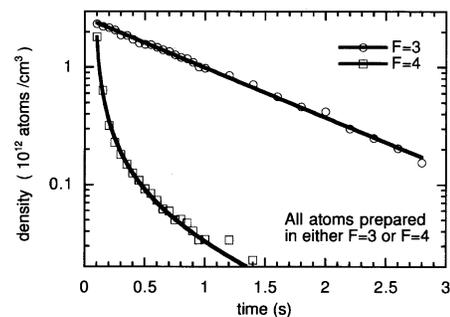}}
\vspace{0.0cm}
\caption{\it Decay of Cs atoms measured in a crossed-beam dipole trap 
(see Sec.~\protect\ref{sec:CBT}) 
realized with the 1064-nm light of a Nd:YAG laser. 
The initial peak density is $2.5\times10^{12}$\,cm$^{-3}$. 
When all atoms are in the lower hyperfine ground state ($F=3$), the purely 
exponential decay ($1/e$-lifetime 1.1\,s) is due to 
collisions with the background gas. When the atoms are in the upper 
hyperfine level ($F=4$), a dramatic loss is observed as a result of
hyperfine-changing collisions (loss coefficient 
$\beta \approx 5\times10^{-11}$\,cm$^3$/s). Unpublished data, 
courtesy of C.\ Salomon.}
\label{hfsloss}
\end{figure}

The {\em two-body loss} coefficient $\beta$ 
describes trap loss due to ultracold binary collisions and reveals
a wide range of interesting physics.
In general, such trap loss becomes important if the colliding
atoms are not in their absolute ground state.
In an inelastic 
process the internal energy can be released into the atomic motion, 
causing escape from the trap.
Due to the shallowness of optical dipole traps, even the collisional release 
of the relatively small amount of energy in the ground-state hyperfine 
structure 
of an alkali atom will always lead to trap loss\footnote{In contrast, a MOT
operated under optimum capture conditions is deep enough to hold atoms after
hyperfine-changing collisions.}. 
{\em Hyperfine-changing collisions}, which occur with large rate 
coefficients
$\beta$ of typically $5\times10^{-11}$\,cm$^3$/s
(Sesko {\it et al.}, 1989; Wallace {\it et al.}, 1992), are thus of 
particular importance 
for dipole trapping. Alkali atoms in the upper hyperfine state ($F=I+1/2$) 
can show very rapid, non-exponential collisional decay, in contrast to a 
sample in the lower ground state ($F=I-1/2$). 
This is impressively demonstrated by the 
measurement in Fig.~\ref{hfsloss}, which shows the decay of a sample
of Cs atoms prepared either in the upper or lower hyperfine ground state.
For the implementaion of laser cooling schemes in dipole traps, 
it is thus a very important issue to keep the atoms predominantly in the
lower hyperfine state; several schemes fulfilling this requirement are
discussed in Secs.~\ref{red} and \ref{blue}.

Trap loss can also occur as a result of light-assisted 
binary collisions involving atoms in the excited state. 
The radiative escape mechanism (Gallagher and Pritchard, 1989)
and excited-state fine-structure changing collisions
strongly affect a MOT, but their influence is negligibly small 
in a dipole trap because of the extremely low optical excitation. 
It can become important, however, if near-resonant cooling light is present.
Another important mechanism for trap loss is photoassociation 
(Lett {\it et al.}, 1995), 
a process in which colliding atoms are excited to bound molecular states, 
which then decay via bound-bound or bound-free transitions. 
Dipole traps indeed represent a powerful tool for photoassociative
spectroscopy (Miller {\it et al.}, 1993b).

{\em Three-body losses}, as described by the coefficient $\gamma$ in 
Eq.~\ref{lossdgl},
become relevant only at extremely high densities (Burt {\it et al.}, 1997;
Stamper-Kurn {\it et al.}, 1998a), far exceeding
the conditions of a MOT. In a collision of three atoms, 
a bound dimer can be formed and the third atom takes up the released
energy, so that all three atoms are lost from the trap.
As a far-detuned dipole trap allows one to completely
suppress binary collision losses 
by putting the atoms into the absolute internal
ground state, it represents an interesting tool for measurements on
three-body collisions; an example is discussed in Sec.~\ref{sec:ketterle}. 

In contrast to inelastic collisions releasing energy,
{\em elastic collisions} lead to a thermalization of the trapped atomic
ensemble. This also produces a few atoms with energies substantially 
exceeding $k_B T$. A loss of these energetic atoms in a shallow trap
leads to evaporative cooling (see Sec.~\ref{coolingmethods}) and 
is thus of great interest for the attainment of Bose-Einstein condensation. 
Regarding the basic physics of elastic collisions, dipole traps are not
different from magnetic traps, but they offer additional experimental 
possibilities. By application of a homogeneous magnetic 
field atomic scattering properties can be tuned
without affecting the trapping itself. Using this advantage of dipole 
trapping, 
Feshbach resonances have been observed with Bose-condensed 
Na atoms (Inouye {\it et al.}, 1998) and thermal Rb atoms (Courteille {\it 
et al.}, 1998).
Moreover, an intriguing possibility is 
to study collisions in an arbitrary mixture of atoms in different 
magnetic sub-states (Stenger {\it et al.}, 1998).

\section{Red-detuned dipole traps} \label{red}

The dipole force points towards increasing intensity if the light
field is tuned below the atomic transition frequency ({\em red
  detuning}).  Therefore, already the focus of a laser beam
constitutes a stable dipole trap for atoms as first proposed by Ashkin
(1978).  The trapping forces generated by intense focused lasers are
rather feeble which was thhe main obstacle for trapping neutral atoms
in dipole traps. Attainable trap depths in a tightly focused beam are
typically in the millikelvin range, orders of magnitude smaller than
the thermal energy of room-temperature atoms. One therefore had to
first develop efficient laser cooling methods for the preparation of
cold atom sources (see Sec.\ \ref{coolingmethods}) to transfer
significant numbers of atoms into a dipole trap.

In their decisive experiment, S.\ Chu and coworkers (1986) succeeded
in holding about 500 sodium atoms for several seconds in the tight
focus of a red-detuned laser beam. Doppler molasses cooling was used
to load atoms into the trap, which was operated at high intensities
and considerable atomic excitation in order to provide a sufficiently
deep trapping potential.  Under these circumstances, the radiation
pressure force still significantly influences the trapping potential
due to a considerable rate of spontaneous emission.  With the
development of sub-Doppler cooling (see Sec.\ \ref{coolingmethods}) 
and the
invention of the magneto-optical trap as a source for dense, cold
atomic samples (see Sec.\ \ref{loading}), dipole trapping of atoms
regained attention with the demonstration of a far-off resonant trap
by Miller {\it et al.}  (1993).  In such a trap, spontaneous emission of
photons is negligible, and the trapping potential is given by from the
equations derived in Sec.\ \ref{background1}.

\begin{figure}
\vspace{0.5cm}
\epsfxsize=86mm
\centerline{\epsfbox{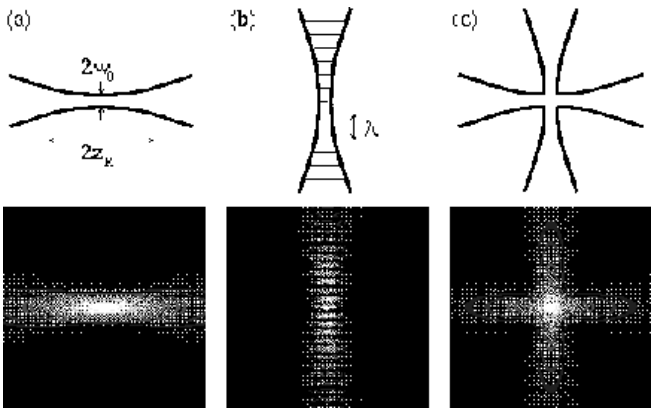}}
\vspace{0.5cm}
\caption{\it Beam configurations used 
  for red-detuned far-off resonance traps. Shown below are the
  corresponding calculated intensity distributions. (a) Horizontal
  focused-beam trap. (b) Vertical standing-wave trap. (c) Crossed-beam
  trap. The waist $w_0$ and the Rayleigh length $z_R$ are indicated.}
\label{fig:traptypes}
\end{figure}

Since then, three major trap types with red-detuned laser beams have
been established, all based on combinations of focused Gaussian beams:
{\it Focused-beam traps} consisting of a single beam, the {\it
  standing-wave traps} where atoms are axially confined in the
antinodes of a standing wave, and {\it crossed-beam traps} created by
of two or more beams intersecting at their foci. The different trap
types are schmetically depicted in Fig.\ \ref{fig:traptypes}.  We
discuss these trap configurations and review applications of the
different trap types for the investigation of interesting physical
questions. Sec.\ \ref{sec:FBT} deals with focused-beam traps, Sec.\ 
\ref{sec:SWT} presents standing-wave traps, and Sec.\ \ref{sec:CBT}
discusses crossed-beam traps.  Far-detuned optical lattices trapping
atoms in micropotentials formed by multiple-beam interference
represent a trap class of their own. Atoms might get trapped in the
antinodes of the interference pattern at red detuning from resonance,
but also in the nodes when the light field is blue-detuned.  Sec.\ 
\ref{sec:lattices} at the end of this chapter is devoted to recent
developments on far-detuned optical lattices.

\subsection{Focused-beam traps}
\label{sec:FBT}

A focused Gaussian laser beam tuned far below the atomic resonance
frequency represents the simplest way to create a dipole trap
providing three-dimensional confinement [see Fig.\ 
\ref{fig:traptypes}(a)]. The spatial intensity distribution of a focused
Gaussian beam (FB) with power $P$ propagating along the $z$-axis is
described by
\begin{equation}
I_{\rm FB}(r,z) =  \frac{2 P}{\pi w^2(z)}\, \exp\left(-
  2 \frac{r^2}{w^2(z)}\right)
\label{eq:GaussInt}
\end{equation}
where $r$ denotes the radial coordinate.  The $1/e^2$ radius $w(z)$
depends on the axial coordinate $z$ via
\begin{equation}
w(z) = w_0 \sqrt{1 + \left(\frac{z}{z_R}\right)^2}\,
\label{eq:GaussRadius}
\end{equation}
where the minimum radius $w_0$ is called the beam waist and $z_R = \pi
w_0^2 / \lambda$ denotes the Rayleigh length. From the intensity
distribution one can derive the optical potential $U(r,z)\propto
I_{\rm FB}(r,z)$ using Eq.\ \ref{UdipTWAgen}, \ref{UdipTWA}, or
\ref{alkaligeneral}. The trap depth $\hat{U}$ is given by $\hat{U} =
\left|U(r=0,z=0)\right|$.

The Rayleigh length $z_R$ is larger than the beam waist by a factor of
$\pi w_0 /\lambda$. Therefore the potential in the radial direction is
much steeper than in the axial direction.  To provide stable trapping
one has to ensure that the gravitational force does not exceed the
confining dipole force. Focused-beam traps are therefore mostly
aligned along the horizontal axis. In this case, the strong radial
force $\sim \hat{U} / w_0$ minimizes the perturbing effects of gravity
\footnote{A new type of trap for the compensation of gravity was
  presented by Lemonde {\it et al.} (1995). It combines a focused-beam
  dipole trap providing radial confinement with an inhomogeneous
  static electric field along the vertical $z$-axis inducing tight
  axial confinement through the dc Stark effekt.}.

If the thermal energy $k_B T$ of an atomic ensemble is much smaller
than the potential depth $\hat{U}$, the extension of the atomic sample
is radially small compared to the beam waist and axially small
compared to the Rayleigh range.  In this case,
the optical potential can be well approximated by a simple
cylindrically symmetric harmonic oscillator
\begin{equation}
U_{FB}(r,z) \simeq -\hat{U} \left[1 - 2 \left(\frac{r}{w_0}\right)^2 -
  \left(\frac{z}{z_R}\right)^2\right]\,. 
\label{eq:HarmonPot}
\end{equation}
The oscillation frequencies of a trapped atom are given by $\omega_r =
(4 \hat{U} / m w_0^2)^{1/2}$ in the radial direction, and $\omega_z =
(2 \hat{U} / m z_R^2)^{1/2}$ in the axial direction.  According to
Eq.\ \ref{potexpansion}, the harmonic potential represents an
important special case of a power law potential for which thermal
equilibrium properties are discussed in Sec.\ \ref{background2}.

\subsubsection{Collisional studies}

In their pioneering work on far-off resonance traps (FORT), Miller et
al.\ (1993a) from the group at University of Texas in Austin have
observed trapping of $^{85}$Rb atoms in the focus of a single,
linearly polarized Gaussian beam with detunings from the $D_1$
resonance of up to $65\,$nm. The laser beam was focused to a waist of
$10\,\mu$m creating trap depths in the mK-range for detunings. Between
$10^3$ and $10^4$ atoms were accumulated in the trap from $10^6$ atoms
provided by a vapor-cell MOT. Small transfer efficiencies are a
general property of traps with tightly focused beams resulting from
the small spatial overlap between the cloud of atoms in the MOT.
Typical temperatures in the trap are below 1\,mK resulting in
densities close to $10^{12}$\,atoms/cm$^3$.  Peak photon scattering
rates were a few hundreds per second leading to negligible loss rates
by photon heating as compared to losses by background gas collisions.
High densities achieved in a tightly-focused beam in combination with
long storage times offer ideal conditions for the investigation of
collisions between trapped atoms.
 
\begin{figure}
\vspace{-0.5cm}
\epsfxsize=60mm 
\centerline{\epsfbox{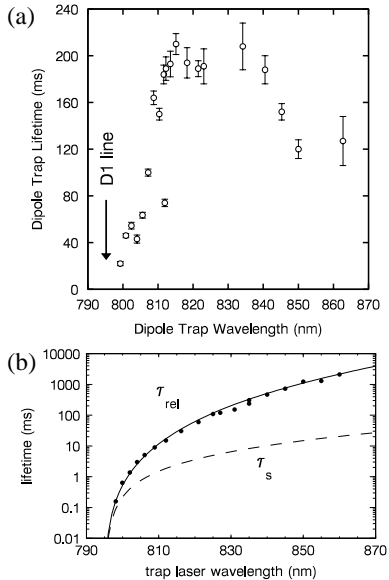}}
\vspace{0.1cm}
\caption{\it Measurement of trap lifetimes and hyperfine relaxation times 
for Rb atoms in a far-detuned
  focused-beam trap. (a) Trap decay time as a function of the trapping
  beam wavelength. From Miller {\it et al.} (1993a).  (b) Time
  constant $\tau_{\rm rel}$ for hyperfine population relaxation versus
  trapping beam wavelength, in comparison to the mean time $\tau_{\rm
    s}$ between two spontaneous scattering events. From Cline {\it et
    al.} (1994 \copyright Optical Society of America).}
\label{fig:FORT.heinzen}
\end{figure}

The trap lifetime without cooling illustrated in Fig.\ 
\ref{fig:FORT.heinzen}(a) showed an increase of the lifetime by about
an order of magnitude for increasing detunings at rather small
detunings.  At larger detunings, the lifetime was found to be
determined by the Rb background pressure of the vapor-cell to a value
of about 200\,ms. The shorter lifetimes at smaller detunings were
explained in a later publication (Miller {\it et al.}, 1993b) by losses
through photoassociation of excited Rb$_2$ dimers which was induced by
the trapping light. This important discovery has inspired a whole
series of experiments on ultracold collisions investigated by the
Austin group with photoassociation spectroscopy in a dipole trap.
Instead of using the trapping light, photoassociation was induced by
additional lasers in later experiments.  The investigations comprise
collisional properties and long-range interaction potentials of ground
state atoms, shape resonances in cold collisions, and the observation
of Feshbach resonances (Cline {\it et al.}, 1994b; Gardner {\it et al.}, 
1995;
Boesten {\it et al.}, 1996; Tsai {\it et al.}, 1997, Courteille {\it et 
al.}, 1998).

\subsubsection{Spin relaxation} \label{sec:spinrelaxation}

If the detuning of the trapping light field is larger than
finestructure splitting of the excited state, photon scattering occurs
almost exclusively into the elastic Rayleigh component. Inelastic
Raman scattering changing the hyperfine ground state is reduced by a
factor $\sim\,1/\Delta^2$ as compared to Rayleigh scattering. This
effect was demonstrated in the Austin group by preparing all trapped
$^{85}$Rb atoms in the lower hyperfine ground state and studying the
temporal evolution of the higher hyperfine state (Cline {\it et al.}, 
1994a).
The relaxation time constant as a function of detuning is plotted in
Fig.\ \ref{fig:FORT.heinzen}(b) in comparison to the calculated
average time between two photon scattering events $\tau_{\rm s} =
\Gamma_{\rm sc}$ . For large detunings, the relaxation time constant
$\tau_{\rm rel}$ is found to exceed $\tau_{\rm s}$ by two orders of
magnitude. This shows the great potential of far-detuned optical
dipole traps for manipulation of internal atomic degrees of freedom
over long time intervals. Using the spin state dependence of the
dipole potential in a circularly polarized light beam (see Sec.\ 
\ref{alkalidiscuss}), Corwin {\it et al.} (1997) from University of Colorado
at Boulder have investigated far-off resonance dipole traps that
selectively hold only one spin state. The small spin relaxation rates
in dipole traps may find useful applications for the search for
beta-decay asymmtries and atomic parity violation.

\subsubsection{Polarization-gradient cooling}

Polarization-gradient cooling in a focused-beam trap has been
investigated by a group at the ENS in Paris (Boiron {\it et al.}, 1998). The
trapped atoms were subjected to blue-detuned molasses cooling in a
near-resonant standing wave (see Sec.\ \ref{coolingmethods}).
Previously, the same group had performed experiments on
polarization-gradient cooling of free atomic samples which were
isotropically distributed. A density-dependent heating mechanism was
found limiting the achievable final temperatures for a given density
(Boiron {\it et al.}, 1996). This heating was attributed to reabsorption of
the scattered cooling light within the dense cloud of cold atoms.  Of
particular interest was the question, whether the anisotropic
geometry of a focused beam influences heating processes through
multiple photon scattering.

Cesium atoms were trapped in the focus ($w_0 = 45\,\mu$m) of a 700\,mW
horizontally propagating Nd:YAG laser beam at 1064\,nm. The trap had a
depth of $50 \,\mu$K, and the radial trapping force exceeded gravity
by roughly one order of magnitude.  From a MOT containing
$3\times10^7$\,atoms, $2\times10^5$ atoms were loaded into the trap
\footnote{The Nd:YAG beam was continuously on also during the MOT
  loading.}.  Polarization-gradient cooling was applied for some tens
of milliseconds yielding temperatures between 1 and 3\,$\mu$K,
depending on the cooling parameters. To avoid trap losses by inelastic
binary collisions involving the upper hyperfine ground state (see
Sec.\ \ref{collisions}), the cooling scheme was chosen in such a way
that the population of the upper hyperfine ground state was kept at a
low level.

\begin{figure} 
\vspace{0.5cm}
\epsfxsize=86mm 
\centerline{\epsfbox{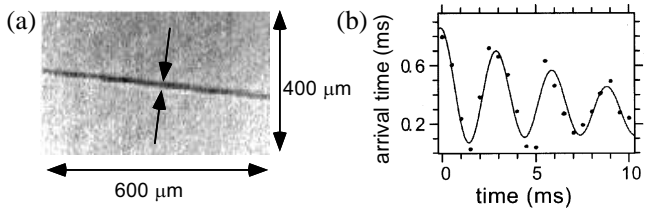}}
\vspace{0.5cm}
\caption{\it Properties of a horizontal focused-beam trap for Cs atoms. 
  (a) Absorption image of the atom distribution. The transverse rms
  radius in the focal plane is 6\,$\mu$m.  (b) Measurement of the
  vertical oscillation period (see text). Adapted from Boiron {\it et
    al.}  (1998).}
\label{fig:trap.salomon} 
\end{figure}

An absorption image of the trapped cesium atoms at $T = 2\,\mu$K is
depicted in Fig.\ \ref{fig:trap.salomon}(a) showing the rod-shaped
cloud of atoms. The distribution of atoms had a radial extension
$\sigma_r = 6\,\mu$m and a axial size of $\sigma_z = 300\,\mu$m
corresponding to a peak density of $\sim
1\times10^{12}$\,atoms/cm$^3$. The picture was taken 30\,ms after the
cooling had been turned off. The time interval is large compared to
the radial oscillation period ($\omega_r/2\pi=330$\,Hz), but short
compared to the axial oscillation period ($\omega_z/2\pi=1.8$\,Hz). In
this transient regime, the axial distribution had not yet reached its
thermal equilibrium extension of $\sigma_z = 950\,\mu$m which leads to
a decrease of the density.  For the radial extension, the measured
value coincided with the expectation for thermal equilibrium as
determined by the measured temperature and oscillation frequency (see
Eq.\ \ref{Tsigma}).

To measure the transverse oscillation frequency, the Nd:YAG beam was
interrupted for a $\sim 1\,$ms time interval, during which the cold
atoms were accelerated by gravity to a mean velocity about 1\,cm/s.
After the trapping laser had been turned on again, the atoms
vertically oscillate in the trap.  The oscillation in vertial velocity
was detected by measuring the mean arrival time of the atoms at a
probe laser beam 12\,cm below the trap [ordinate in Fig.\ 
\ref{fig:trap.salomon}(b)] as a function of the trapping time
intervals [abscissa in Fig.\ \ref{fig:trap.salomon}(b)]. The measured
vertical oscillation frequency $\omega_r = 330\,$Hz is consistent with
the value $\omega_r = 390\,$Hz derived from the trap depth and the
beam waist.

The temperatures measured in the dipole trap were about 30 times lower
than one would expect on the basis of the heating rates found for a
isotropic free-space sample at the corresponding densities (Boiron et
al., 1995).  No evidence was found for heating of the trapped sample.
The reduction of density-dependent heating is a benefit from the
strongly anisotropic trapping geometry of a focused-beam trap. Due to
the much smaller volume to surface ratio of an atomic cloud in the
focused-beam trap as compared to a spherical distribution, photons
emitted during cooling have a higher chance to escape without
reabsorption from the trap sample which causes less heating through
reabsorption.

\subsubsection{Bose-Einstein condensates} \label{sec:ketterle}

Optical confinement of Bose-Einstein condensates was demonstrated for
the first time by a group at MIT in Cambridge, USA (Stamper-Kurn et
al., 1998a).  Bose condensates represent the ultimately cold state of
an atomic sample and are therefore captured by extremely shallow
optical dipole traps.  High transfer efficiencies can be reached in
very far-detuned traps, and the photon scattering rate acquires
negligibly small values (see Eq.\ \ref{KKrelation}). Various specific
features of dipole traps can fruitfully be applied to the
investigation of many aspects of Bose-Einstein condensation which were
not accessible formerly in magnetic traps.

Sodium atoms were first evaporatively cooled in a magnetic trap to
create Bose condensates containing $5-10\times10^6$\,atoms in the
$3S_{1/2}(F=1, m_F=1)$ state (Mewes {\it et al.}, 1996). Subsequently, the
atoms were adiabatically transferred into the dipole trap by slowly
ramping up the trapping laser power, and then suddenly switching off
the magnetic trap. The optical trap was formed by a laser beam at
985\,nm (396\,nm detuning from resonance) focused to a waist of about
$6\,\mu$m. A laser power of 4\,mW created a trap depth of about
$4\,\mu$K which was sufficient to transfer 85\% of the Bose condensed
atoms into the dipole trap and to provide tight confinement. Peak
densities up to $3 \times 10^{15}\,$atoms/cm$^3$ were reported
representing unprecendented high values for optically trapped atomic
samples.

Condensates were observed in the dipole trap even without initially
having a condensate in the magnetic trap. This strange effect could be
explained by an adiabatic increase of the local phase-space density
through changes in the potential shape (Pinkse {\it et al.}, 1997). The slow
increase of the trapping laser intensity during loading leads to a
deformation of the trapping potential created by the combination of
magnetic and laser fields. The trapping volume of the magnetic trap
was much larger than the volume of the dipole trap.  Therefore,
phase-space density was increased during deformation, while entropy
remained constant through collisional equilibration. Using the
adiabatic deformation of the trapping potential, a 50-fold increase of
the phase-space density was observed in a later experiment
(Stamper-Kurn et al, 1998b).

\begin{figure}
\vspace{0.5cm}
\epsfxsize=86mm 
\centerline{\epsfbox{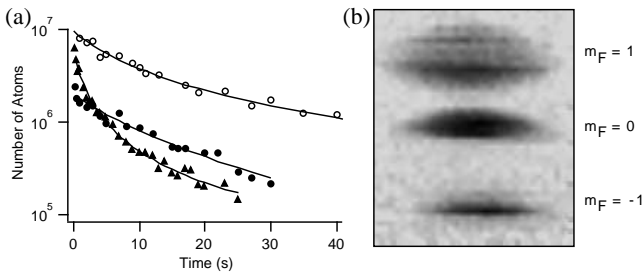}}
\vspace{0.5cm}
\caption{\it Bose-Einstein condensates of Na atoms in focused-beam traps. 
(a)
  Trapping lifetime in optical traps and magnetic traps. The number of
  condensed atoms versus trapping time is shown. Closed symbols
  represent the data for optical traps with best transfer efficiency
  (triangles) and slowest decay (circles). Open circles represent data
  for a magnetic trap with optimized lifetime. The lines are fits
  based on single-particle losses and three-body decay. (b) Optical
  trapping of condensates in all hyperfine spin states of the $F=1$
  ground state. An absorption image after 340\,ms of optical
  confinement and subsequent release from the trap is shown. Hyperfine
  states were separated by a pulsed magnetic field gradient during
  time of flight. The field view of the image is 1.6 by 1.8\,mm.
  Adapted from Stamper-Kurn {\it et al.} (1998a).}
\label{fig:BEC.ketterle}
\end{figure}

The lifetime of atoms in the dipole trap is shown in Fig.\ 
\ref{fig:BEC.ketterle}(a) in comparison to the results obtained in a
magnetic trap. In the case of tight confinement and high densities
(triangles in Fig.\ \ref{fig:BEC.ketterle}), atoms quickly escape
through inelastic intra-trap collisions. Long lifetimes (circles in
Fig.\ \ref{fig:BEC.ketterle}) were achieved in the dipole trap and the
magnetic trap, respectively, when the trap depth was low enough for
collisionally heated atoms to escape the trap.  Trap loss was found to
be dominated by three-body decay with no identifiable contributions
from two-body dipolar relaxation. The lifetime measurements delivered
the three-body loss rate constant $\gamma=1.1(3)\times10^{-31}$\,cm$^6$/s
(see Eq.\ \ref{lossdgl}) for collisions among condensed sodium atoms.

Fig.\ \ref{fig:BEC.ketterle}(b) demonstrates simultaneous confinement
of a Bose condensate in different Zeeman substates $m_F = 0,\pm1$ of
the $F=1$ ground state. To populate the substates, the atoms were
exposed to an rf field sweep (Mewes {\it et al.}, 1997). The distribution
over the Zeeman states was analyzed through Stern-Gerlach separation
by pulsing on a magnetic field gradient of a few G/cm after turning
off the dipole trap. It was verified that all $F=1$ substates were
stored stably for several seconds.

After the first demonstration of optical trapping, a spectacular
series of experiments with Bose-Einstein condensates in a dipole trap
was performed by the MIT group.  By adiabatically changing the
phase-space density in the combined magnetic and optical dipole trap
(see above), Stamper-Kurn {\it et al.}  (1998b) were able to reversibly
cross the transition to BEC. Using this technique, the temporal
formation of Bose-Einstein condensates could extensively be studied
(Miesner et al, 1998a).  The possibility of freely manipulating the
spin of trapped atoms without affecting the trapping potential has led
to the observation of Feshbach resonances in ultracold elastic
collisions (Inouye {\it et al.}, 1998) and the investigation of spin domains
and metastable states in spinor condensates (Stenger {\it et al.}, 1998,
Miesner {\it et al.}, 1998b).

\subsubsection{Quasi-electrostatic traps}
\label{sec:QUEST}

The very interesting case of quasi-electrostatic dipole trapping has
so far not been considered in this review. When the frequency of the
trapping light is much smaller than the resonance frequency of the
first excited state $\omega \ll \omega_0$, the light field can be
regarded as a quasi-static electric field polarizing the atom.
Quasi-electrostatic traps (QUEST) were first proposed (Takekoshi et
al., 1995) and realized (Takekoshi and Knize, 1996) by a group at
University of Southern California in Los Angeles.  In the
quasi-electrostatic approximation $\omega \ll \omega_0$, one can write
the dipole potential as
\begin{equation}
U_{\rm dip}({\bf r}) = - \alpha_{\rm stat} \frac{I({\bf r})}{2
  \varepsilon_0 c}
\label{Ustatic}
\end{equation}
with $\alpha_{\rm stat}$ denoting the static polarizability ($\omega =
0$).  The light-shift potential of the excited states is also
attractive contrary to far-off resonant interaction discussed before.
Atoms can therefore be trapped in all internal states by the same
light field. Since the trap depth in Eq.\ \ref{Ustatic} does not
depend on the detuning from a specific resonance line as in the case
of a FORT, different atomic species or even molecules may be trapped
in the same trapping volume.

For the ground state of alkali atoms, Eq.\ \ref{Ustatic} is
well approximated by applying the quasi-static approximation to Eq.\ 
\ref{UdipTWAgen} which gives
\begin{equation}
U_{\rm dip}({\bf r}) =  - \frac{3 \pi c^2}{\omega_0^3} \,
 \frac{\Gamma}{\omega_0} \, I({\bf r})\,.
\label{Uquest}
\end{equation}
Compared to a FORT at a detuning $\Delta$, the potential depth for
ground state atoms in a QUEST is smaller by a factor
$2\Delta/\omega_0$. Therefore, high power lasers in the far-infrared
spectral range have to be employed to create sufficiently deep traps.
The CO$_2$ laser at 10.6\,$\mu$m which is commercially available with
cw powers up to some kilowatts is particulary well suited for the
realization of a QUEST (Takekoshi {\it et al.}, 1995).

An important feature of the QUEST is the practical absence of photon
scattering.  The relation between the photon scattering rate and the
trap potential can be derived from Eqs.\ \ref{UdipTWAgen} and
\ref{GscTWAgen} in the quasi-electrostatic approximation:
\begin{equation}
\hbar \Gamma_{\rm sc}({\bf r}) =  2 \,
\left( \frac{\omega}{\omega_0} \right)^3 \,
\frac{\Gamma}{\omega_0} \,U_{\rm dip}({\bf r}). 
\label{eq:Gquest}
\end{equation}
When compared to the corresponding relation for a FORT given by Eq.
\ref{KKrelation}, the dramatic decrease of the photon scattering rate
in a QUEST becomes obvious. Typical scattering rates are below
$10^{-3}$s$^{-1}$ showing that the QUEST represents an ideal
realization of a purely conservative trap.

Takekoshi and Knize (1996) have realized trapping of cesium atoms in a
QUEST by focussing a 20\,W CO$_2$ laser to a waist of $100\,\mu$m
resulting in a trap depth of $115\,\mu$K. Around $10^6$ atoms prepared
in the $F=3$ state were loaded into the trap from a standard MOT. The
atom loss rates of $\sim 1\,$s${-1}$ were consistent with pure losses
through background gas collisions. Hyperfine relaxation times were
found to exceed 10\,s.

\subsection{Standing-wave traps} \label{sec:SWT}

A standing wave (SW) trap provides extremely tight confinement in
axial dimension as can be seen from Fig.\ \ref{fig:traptypes}(b). The
trap can be realized by simply retroreflecting the beam while
conserving the curvature of the wave fronts and the polarization.
Assuming small extensions of the atomic cloud, one can write the
potential in the form
\begin{equation}
U_{SW}({\bf r}) \simeq -\hat{U} \cos^2\left(kz\right)
 \left[1 - 2 \left(\frac{r}{w_0}\right)^2 -
  \left(\frac{z}{z_R}\right)^2\right]
\label{eq:StandingPot}
\end{equation}
with the standing wave oriented along the $z$-axis. The potential
depth is four times as large than the corresponding trap depth for a
single focused beam discussed in Sec.\ \ref{sec:FBT}. As for a single
focused beam, 
radial confinement is provided by
a restoring force $\sim \hat{U} / w_0$. The axial trapping potential
is spatially modulated with a period of $\lambda/2$.  Atoms are
strongly confined in the antinodes of the standing wave (restoring
force $\sim \hat{U} k$) resulting in a regular one-dimensional lattice
of pancake-like atomic subensembles. When aligned vertically, the
axial confinement greatly exceeds the gravitational force $m g$.  One
can therefore use rather shallow trap, just sufficiently deep to trap
a pre-cooled ensemble, which results in small photon scattering rates
(see Eq.\ \ref{KKrelation}) and large loading efficiencies.

The tight confinement along the axial direction leads to large
oscillation frequencies $\omega_z = \hbar k (2\hat{U}/m)^{1/2}$ at the
centre of the trap. The oscillation frequency decreases when moving
along the $z$-axis due to the decreasing light intensity. At low
temperatures, the energy of the axial zero-point motion $\frac{1}{2}
\hbar \omega_z$ in the centre of the trap can become of the same order
of magnitude as the thermal energy $\frac{1}{2}k_B T$ even for
moderate trap depths \footnote{Using the recoil temperature $T_{\rm
    rec}$ for the cooling transition at the wavelength $\lambda_0$
  introduced in Sec.\ \ref{coolingmethods}, one can write the
  zero-point energy as $(\lambda_0/\lambda) (k_B T_{\rm rec}
  \hat{U}/2)^{1/2}$.}.  In this regime, the axial atomic motion can no
longer be described classically but has to be quantized, and the
vibrational ground state of the axial motion is substantially
populated. The axial spread of the wavepacket is much smaller than the
wavelength of an optical transition (Lamb-Dicke regime) giving rise to
spectral line-narrowing phenomena.  One might even enter a regime
where the wavepacket extension comes close to the s-wave scattering
length leading to dramatic changes in the collisional properties of
the trapped gas.

\subsubsection{Optical cooling to high phase-space densities}
\label{sec:sideband}

Well-resolved vibrational levels and Lamb-Dicke narrowing, as realized
along the axis of a standing-wave dipole trap, are necessary
requirements for the application of optical sideband cooling as
explained in Sec.\ \ref{coolingmethods}.  By employing
degenerate-sideband Raman cooling, high phase-space densities of an
ensemble containing large particle numbers have been achieved by
Vuletic {\it et al.} (1998) from a group in Stanford. In a vertical
far-detuned standing wave, peak phase- space densities around 1/180
have been obtained with $10^7$ cesium atoms, corresponding to a mean
temperature of $2.8\,\mu$K and a peak spatial density of $1.4 \times
10^{13}$ atoms/cm$^3$.

The trap was generated by a Nd:YAG laser with 17\,W single-mode power
at $\lambda = 1064$\,nm. The large beam waist of $260\,\mu$m created a
trap depth of $160\,\mu$K which resulted in high loading efficiencies
($\approx 30\%$ from a blue-detuned molasses released from a MOT) and
small photon scattering rates ($\approx 2\,s^{-1}$).  Atoms oscillated
at $\omega_r/2\pi = 120$\,Hz in the radial (horizontal) direction, and
at $\omega_z/2\pi = 130$\,kHz in the axial (vertical) direction. A
dramatic dependence of the trap lifetime on the hyperfine state of the
cesium atoms was observed similar to Fig.\ \ref{hfsloss} in Sec.\ 
\ref{collisions}.  Around $1 \times 10^7$ atoms were contained in the
trap populating 4700 vertical potential wells. Degenerate Raman
sideband cooling was applied between vibrational states of a pair of
Zeeman shifted magnetic sublevels in the lowest hyperfine ground state
as explained in Sec.\ \ref{coolingmethods}. Suppressed collisional
losses through inelastic binary collisions at high densities are
greatly suppressed since population in the upper hyperfine state is
kept extremely low in this cooling scheme. The Raman coupling was
provided by the lattice field itself (Deutsch and Jessen, 1998).

\begin{figure}
\vspace{0.2cm}
\epsfxsize=60mm
\centerline{\epsfbox{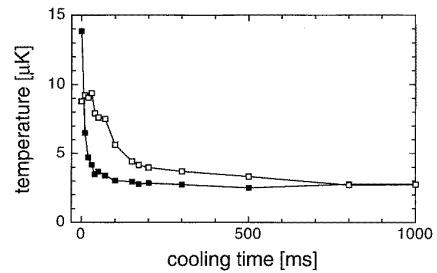}}
\vspace{0.1cm}
\caption{\it Resolved-sideband cooling of Cs atoms in a vertical
  standing-wave trap.  Evolution of vertical (solid squares) and
  horizontal (open squares) temperatures. Cooling is applied only
  along the tightly confining vertical axis, the horizontal degrees of
  freedom are indirectly cooled through collisional thermalization.
  From Vuletic {\it et al.} (1998).}
\label{fig:sideband.vladan}
\end{figure}

Axial and radial temperatures evolved differently during cooling as
indicated in Fig.\ \ref{fig:sideband.vladan}.  The axial direction was
directly cooled causing the axial temperature to quickly drop to $T_z
= 2.5\,\mu$K. The radial temperature followed by collisional
thermalization with a time constant of 150\,ms. After sideband cooling
was turned off, the temperature increased at a rate of $4\,\mu$K/s
limiting the achievable final temperatures. This heating rate was much
larger than the rate estimated on the basis of photon scattering which
might indicate additional heating sources such as laser noise.  The
achieved high thermalization rates in combination with large particle
numbers would provide excellent starting conditions for subsequent
evaporative cooling.

A different approach to optical cooling of large particle numbers to
high phase-space densities was followed by a group at Berkeley (Winoto
{\it et al.}, 1998), who have recently applied polarization-gradient
cooling to cesium atoms in a vertical standing-wave trap.  The
standing wave was linearly polarized resulting in equal light shifts
for all magnetic sublevels of the atomic ground state (see Sec.\ 
\ref{alkalidiscuss}).  Therefore, polarization-gradient cooling, which
relies on optical pumping between the ground-state sublevels (see
Sec.\ \ref{coolingmethods}), could be applied to the trapped sample in
a very efficient way. A phase-space density around $10^{-4}$ with the
large number of $10^8$ atoms was reached.

\subsubsection{Quantum interference}

The regular arrangement of atoms in a standing wave-dipole trap has
amazing consequences when a Bose condensate is loaded into such a
trap.  Macroscopic interference of Bose condensed $^{87}$Rb atoms
tunneling from an extremely shallow 1D lattice under the influence of
gravity has recently been observed by Anderson and Kasevich (1998) at
Yale University.  The dipole potential was created by a vertical
standing wave at 850\,nm (detuning of 65\,nm from the $D_1$ line) with
a waist of $80\,\mu$m, an order of magnitude larger than the
transverse radius of the condensate. The condensate of $10^4$ atoms
was coherently distributed among $\approx 30$ very shallow potential
wells. The wells supported only one bound energy band below the
potential edge which is equivalent to $\hat{U} \simeq k_B T_{\rm rec}$
where $T_{\rm rec}$ is the recoil temperature (see Eq.\ \ref{Trec}) at
the wavelength of the trapping field.

The gravitational field induces an offset between adjacent wells. For
weak potential gradients, the external field can be treated as a
perturbation th the band structure associated with the lattice. In
this limit, wavepackets remain confined in a single band. The external
field drives coherent oscillations at the Bloch frequency as has been
demonstrated with ultracold atoms confined in an accelerated
far-detuned 1D standing wave by Ben Dahan {\it et al.}  (1996) at ENS
in Paris and by Wilkinson {\it et al.}  (1996) at University of Texas
in Austin.

The shallow potential wells allow for tunneling of particles into
unbound continuum states.  In the Yale experiment, the lifetime of the
atoms confined in the lattice was purely determined by the tunneling
losses.  For a lattice of depth $\hat{U} = 1.1 k_B T_{\rm rec}$, the
observed lifetime was $\sim 50$\,ms.  Each lattice site can be seen as
a point emitter of a deBroglie wave.  Interference between the
different emitter ouputs lead to the formation of atom pulses falling
out of the standing wave, quite similar to the output of a mode-locked
pulsed laser. The pulse repetition frequency $\omega_{\rm rep} = m g
\lambda / 2 \hbar$ was determined by the gravitational increment
between two adjacent walls. The repetition frequency can be
interpreted as the difference in chemical potential divided by
$\hbar$. This indicates the close relation of the observed effect of
coherent atomic deBroglie waves to the ac Josephson effect resulting
from quantum interference of two superconducting reservoirs.

\subsubsection{Spin manipulation} \label{sec:startrap}

Zielonkowski {\it et al.} (1998b) from the MPI f\"ur Kernphysik in
Heidelberg have used a vertical standing wave trap for the
manipulation of spin-polarized atoms. By using a large-volume, shallow
red-detuned standing wave trap, depolarizing effects of photon
scattering and atomic interactions could be kept at a low level while
allowing for large numbers of stored atoms.

\begin{figure}
\vspace{0.5cm}
\epsfxsize=96mm 
\centerline{\epsfbox{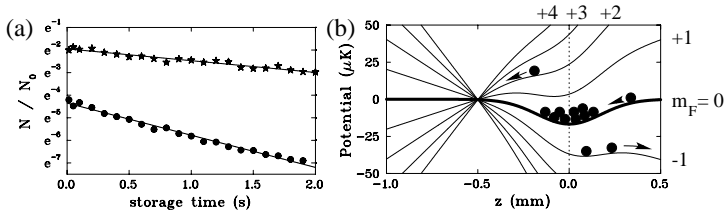}}
\vspace{0.5cm}
\caption{\it Spin manipulation of Cs in a vertical standing-wave trap. (a)
  Storage time with all ground-state sublevels populated (stars), and
  for the Stern-Gerlach selected $m_F=0$ state (circles). Shown is the
  trapped particle number $N$ relative to the number of atoms trapped
  in the MOT before transfer $N_0$.  The line represents an
  exponential fit to the data yielding a time constant of 1.9\,s for
  the unpolarized sample and 0.7\,s for the $m_F = 0$ polarized atoms.
  (b) Trap potentials for the combined magnetic quadrupole and optical
  trap used for Stern-Gerlach selection of the $m_F=0$ state from the
  $F=4$ hyperfine ground state. The zero-point of the magnetic
  quadrupole has been horizontally shifted with respect to the centre
  of the standing-wave trap. Atoms with $m_F \neq 0$ are expelled from
  the dipole trap by the magnetic field gradient.  Adapted from
  Zielonkowski {\it et al.} (1998b).}
\label{fig:STAR.rudi}
\end{figure}

Cesium atoms were trapped in a retroreflected 220-mW beam at a
wavelength of $\lambda = 859$\,nm (detuning of 6.1\,nm above the $D_2$
line). The waist of the beam in the interaction region was 0.50\,mm.
The maximum potential depth amounted to only $17\,\mu$K which was
sufficiently deep to directly load atoms from a MOT at sub-Doppler
temperatures.  Despite the shallow potential depth, the confining
force exceeded the gravitational force by about three orders of
magnitude.  The transfer efficiency from the MOT into the shallow trap
was about 14\% resulting in $\sim 10^5$ trapped atoms at a liftime of
$\tau = 1.9\,$s as shown by the stars in Fig.\ \ref{fig:STAR.rudi}(a).

In a first experiment, the spin state $m_F = 0$ was selected by a
Stern-Gerlach (SG) force. The force was created by horizontally
shifting the zero of the MOT quadrupole field with respect to the
position of the dipole trap (see Fig.\ \ref{fig:STAR.rudi}(b)).  In
this way, only atoms with $m_F = 0$ are trapped by the dipole trap,
all other magnetic substates are pulled out of the trap by the
magnetic dipole force. The depolarizing effect of the trap light was
determined by measuring the lifetime $\tau_{SG}$ of the $m_F = 0$
atoms (see circles in Fig.\ \ref{fig:STAR.rudi}(a)) and comparing this
value with the trap lifetime without Stern-Gerlach selection. The
depolarization rate $\Gamma_{depol} = 1/\tau_{SG} - 1/\tau =
0.9$\,s$^{-1}$ allowed a determination of the photon scattering rate
from calculated values of the $m_F$-state branching ratios for
spontaneous scattering. The resulting scattering rate agrees well with
the expectation based on Eq.\ \ref{Gsclin}.

In a second experiment, spin precession in a ficitious magnetic field
(Cohen-Tannoudji and Dupont-Roc, 1972, Zielonkowski {\it et al.},
1998a) was demonstrated. The field was induced by an additional
off-resonant circularly polarized laser beam which induced a light
shift scaling linear with $m_F$ as discussed in Sec.
\ref{alkalidiscuss}.  The resulting splitting corresponds to a
fictitious magnetic field of 50\,mG.  The $m_F=0$ state was selected
by a short Stern-Gerlach pulse resulting in a macroscopic
magnetization of the sample. The population of the same state was
analyzed after a 150\,ms delay.  Between preparation and analysis, the
atoms interacted with a pulse of the fictitious field laser in
combination with a holding magnetic field.  The $m_F=0$ population
oscillates with the duration of the laser pulses which can be directly
interpreted as the Larmor precession of the spin in the superposition
of fictitous and holding magnetic field.  A $2\pi$ and a $4\pi$
rotation of the magnetization were observed.

\subsubsection{Quasi-electrostatic lattices} \label{sec:QUEL}

A standing wave created by the light at $10.6\,\mu$m from a CO$_2$
laser creates a QUEST (see Sec.\ \ref{sec:QUEST}) with a spacing
between the axial potential wells which is large compared to the
transition wavelength of the trapped atoms. In a group at the MPI
f\"ur Quantenoptik in Garching, such a lattice of mesoscopic potential
wells was recently realized with rubidium atoms (Friebel {\it et al.},
1998a; 1998b). A 5\,W CO$_2$ laser beam was focused to a waist of
50\,$\mu$m creating a trap depth of about $360\,\mu$K.  Up to $3
\times 10^5$\,atoms could be loaded into the horizontal standing wave
trap which had a lifetime of $1.8\,$s limited by background gas
collisions. Temperatures around 10\,$\mu$K were achieved by
polarization gradient cooling in the trap.

The vibrational frequencies of atoms inside the potential wells were
measured by modulating the laser intensity in order to drive
parametric excitation of the atomic oscillations. When the modulation
frequency equals twice the vibrational frequency $2\omega_{\rm osc}$
(or subharmonics $2 \omega_{\rm osc}/n$), atoms are heated out of the
trap leading to a reduced lifetime. This effect was demonstrated by
varying the modulation frequency and measuring the number of atoms
\begin{figure}
\vspace{0.5cm}
\epsfxsize=86mm
\centerline{\epsfbox{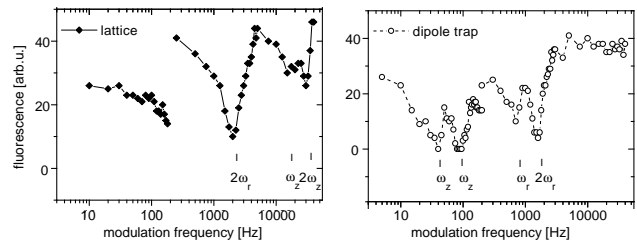}}
\vspace{1cm}
\caption{\it Excitation spectra of Rb atoms trapped with a CO$_2$ laser at 
  10.6\,$\mu$m creating quasi-electrostatic traps. The spectrum of a
  standing-wave trap (left graph) is compared to the spectrum from a
  single-beam trap (right graph).  The horizontal axis is the
  fluorescence of atoms which are illuminated after 0.6\,s trapping
  time. The vertical axis gives the modulation frequency of the trap
  light intensity. Parametric resonances at the oscillation
  frequencies and twice their value are observed. The jump at
  100\,Hz in both graphs is an artefact resulting from a change of the
  modulation depth and modulation time.  Adapted from Friebel {\it et
    al.} (1998a).}
\label{fig:CO2.weitz}
\end{figure}
\noindent
that were left in the trap after a fixed trapping time of 600\,ms.
The remaining atoms were detected by switching on a resonant light
field and recording the fluorescence.  The left graph in Fig.\ 
\ref{fig:CO2.weitz} shows the excitation spectrum for the
standing-wave trap. Parametric resonances at 2\,kHz and at 32\,kHz can
be identified which are attributed to excitations of the radial and
axial vibrations, respectively. A subharmonic resonance at 16\,kHz is
also observed. In the left graph of Fig.\ \ref{fig:CO2.weitz}, the
excitation spectrum of a single focused-beam dipole trap is presented.
The trap was realized by interrupting the retroreflected laser beam
which had formed the standing wave. The radial resonance shifts to
1.6\,kHz (subharmonic at 0.8\,kHz) because of the four times lower
trap depth.  The reduction of the axial vibrational frequency is more
dramatic since it scales as $1/(2 k z_R)$ relative to the
standing-wave trap (including the factor 4 in trap depth).  The axial
resonance is now found at 80\,Hz (subharmonic at 40\,Hz).

\subsection{Crossed-beam traps} \label{sec:CBT}

A single focused beam creates a highly anisotropic trap with
relatively weak confinement along the propagation axis and tight
confinement in the perpendicular direction [see Figs.\ 
\ref{fig:traptypes}(a) and \ref{fig:trap.salomon}(a)]. In a
standing-wave trap, the anisotropic atomic distribution of the single
beam trap is split into anisotropic subensembles with extremely tight
confinement along the axial direction. Crossing two beams with
orthogonal polarization and equal waist under an angle of about
$90^\circ$ as indicated in Fig.\ \ref{fig:traptypes}(c) represents an
obvious way to create nearly isotropic atomic ensembles with tight
confinement in all dimensions.  In this case, the dipole potential for
small extensions of the atomic cloud can be approximated as
\begin{equation}
U_{CB}(x,y,z) \simeq -\hat{U} \left(1 - \frac{x^2 + y^2 + 2
    z^2}{w_0^2}\right)\,.
\label{eq:CrossedPot90}
\end{equation}
It should be noted that the effective potential depth is only
$\hat{U}/2$ as atoms with larger energy leave the steep trap along one
of the beams.

\subsubsection{Evaporative cooling}

Crossed-beam dipole traps provide a good compromise between decent
trapping volumes and tight confinement, and are therefore suited
for the application of evaporative cooling as explained in Sec.\ 
\ref{coolingmethods}.  Adams {\it et al.} (1995) at Stanford used a
crossed-beam configuration oriented in the horizontal plane for
evaporative cooling of sodium. A single-mode Nd:YAG
laser generated two beams of 4\,W each, focused to a waist of
$15\,\mu$m and crossing under $90^\circ$. The trap depth was close to
1\,mK. The polarizations of the two beams were chosen orthogonal which
results in a spin-independent trapping potential for the ground states
because of the large detuning of the 1064\,nm light from the two
fine-structure lines of sodium (see Table \ref{alkalitable}).

\begin{figure}
\vspace{0.5cm}
\epsfxsize=86mm
\centerline{\epsfbox{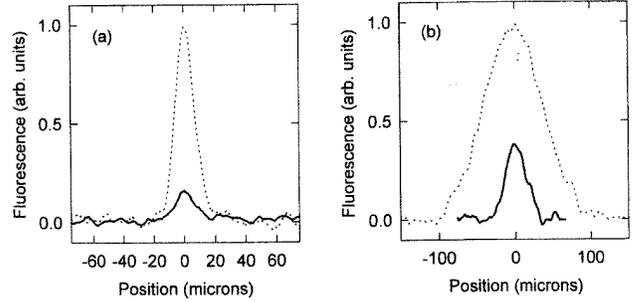}}
\vspace{0.5cm}
\caption{\it Evaporative cooling of Na atoms in a crossed-dipole trap.
  (a) Atom density distribution before (dotted line) and after (solid
  line) evaporative cooling of Na in a crossed dipole trap. The
  density decreased by a factor of about 7. (b) Time-of-flight
  measurement of the temperature before (dotted line) and after (solid
  line) evaporative cooling. The temperature decreased from
  140\,$\mu$K to 4\,$\mu$K. From Adams {\it et al.}  (1995).}
\label{fig:evapcool.chu}
\end{figure}

After transfer from a MOT, evaporative cooling was started with
$\sim5000\,$atoms at a temperature of $140\,\mu$K and a peak density
of $4\times10^{12}$\,atoms/cm$^3$ as indicated by the dotted lines in
Fig.\ \ref{fig:evapcool.chu}.  To force evaporation of high-energetic
particles, the Nd:YAG power was exponentially ramped down from 8\,W
(trap depth $\sim\,900\,\mu$K) to 0.4\,W (trap depth $\sim\,45\,\mu$K)
within 2\,s . After evaporation, $\sim\,500\,$atoms were left in the
trap. Temperature was reduced by a factor of 35 to $4\,\mu$K. Yet,
density decreased by about an order of magnitude since the restoring
force in the trap is reduced when ramping down the intensity. To keep
the density at a high value for efficient evaporation, one would have
to further compress the cloud.  However, compared to the initial
conditions, phase-space density was increased by a factor of 28 in the
experiment indicating the potential of evaporative cooling for the
enhancement of phase-space density in dipole traps.

\subsubsection{Interference effects}

Let us now consider the more general case of two beams of equal waist
$w_0$ propagating in the $xy$-plane and intersecting at the focal
points under an arbitrary angle.  The angle between the beams and the
$x$-axis is $\pm\phi$. This configuration includes as special cases
the single focussed-beam trap ($\phi = 0$), discussed in Sec.
\ref{sec:FBT}, and the standing wave trap ($\phi = 90^\circ$),
discussed Sec.\ \ref{sec:SWT}.  In the harmonic approximation, the
trapping potential can be written as
\begin{equation}
U_{CB}(x,y,z) \simeq -\hat{U}(y) \left(1 - 2 \frac{x^2}{w_x^2}
- 2 \frac{y^2}{w_y^2} -2 \frac{z^2}{w_0^2}\right)
\label{eq:CrossedPot}
\end{equation}
The potential radii $w_{x,y}$ are given by $w_x^2 = \left(
  \cos^2\phi/w_0^2 + \sin^2\phi/2z_R^2\right)^{-1}$ and $w_y^2 =
\left(\sin^2\phi/w_0^2 + \cos^2\phi/2z_R^2\right)^{-1}$.  The
$y$-dependence of the trap depth $\hat{U}$ reflects interference
effects between the two waves which lead to a modulation of the trap
depth on the scale of an optical wavelength.

If both beams are linearly polarized along the $y$-direction
(lin$\parallel$lin), interference results in a pure intensity
modulation with period $D = \lambda/(2\sin\phi)$ yielding $\hat{U}(y)
= \hat{U}_{\rm max}\cos^2(\pi y/D)$. Trapped atoms are therefore bound
in the antinodes of the interference pattern forming a one-dimensional
lattice along the $y$-direction with lattice constant $D$. In the case
of orthogonal polarization of the two beams (lin$\perp$lin), the
intensity exhibits no interference effects, but the polarization is
spatially modulated between linear and circular with the same period
$D$. As discussed in Sec.\ \ref{alkalidiscuss}, the light shift
depends on the spin state of the atoms giving rise to a spatial
modulation of the potential. For detunings $\Delta$ large compared to
the fine-structure splitting $\Delta'_{\rm FS}$, the potential depth
is modulated with a relative amplitude $\frac{1}{3}g_F m_F
\Delta'_{\rm FS}/\Delta$ (see Eq.\ \ref{alkaliexpand}).

Making use of these interference effects, the ENS group has
investigated Raman cooling (Kuhn {\it et al.}, 1996) and sideband
cooling (Perrin {\it et al.}, 1998; Bouchoule {\it et al.}, 1998) in a
crossed-beam trap for cesium atoms.  The trap consisted of two Nd:YAG
laser beams ($\lambda = 1064\,$nm) propagating along a vertical
$xy$-plane with a power of about 5\,W in each beam. The beams crossed
at their common waists ($w_0 \approx 100\,\mu$m) under an angle $\phi
= \pm53^\circ$ with the horizontal $x$-axis. The hyperfine splitting
of ground and excited state were small compared to the detuning of the
laser from the $D_1$ and $D_2$ lines of cesium at 894\,nm and 852\,nm,
respectively. For the lin$\perp$lin case, the comparatively large
fine-structure splitting of cesium ($\Delta'_{\rm FS}/\Delta \approx
5$) lead to spatially modulated potentials with a small, yet
significant modulation amplitude of $\approx \hat{U}/15$ for the
stretched Zeeman states $|m_F| = I + 1/2$.

The trap was loaded from $\sim 10^7$\,cesium atoms in a MOT. The
loading efficiency and the shape of the atomic cloud strongly depended
on the laser polarizations of the dipole trap (Kuhn {\it et al.}, 1996).
The potential wells formed by interference of the laser beams were
used for resolved-sideband cooling with Raman transitions between the
$F=3$ and $F=4$ hyperfine ground states. Great differences in the
performance of sideband cooling were found for the lin$\parallel$lin
and the lin$\perp$lin case due to the different character of the
potential wells being weakly or strongly modulated, respectively
(Perrin {\it et al.}  1998). By optimizing the sideband cooling in the
lin$\parallel$lin configuration, single vibrational states (motional
Fock states) could be prepared in the 1D standing wave. Atoms were
first sideband-cooled into the lowest vibrational state $|n_{\rm osc}
= 0\rangle$ from where they could be transferred into other pure
$|n_{\rm osc}\rangle$ states by Raman transitions at multiples of the
resolved vibrational sidebands (Bouchoule {\it et al.}, 1998).

\subsection{Lattices} \label{sec:lattices}

When adding more laser beams, one can design a whole variety of
interference patterns to create two- and three-dimensional lattices
confining the atoms in micropotentials of submicron extension (Deutsch
and Jessen, 1998).  When combined with efficient cooling methods,
significant population of the vibrational ground state can be
achieved.  Many important aspects of these optical lattices have
extensively been studied for near-resonant trapping fields which
simultaneously provide tight confinement and dissipation (Jessen and
Deutsch, 1996; Hemmerich {\it et al.}, 1996; Grynberg and Triche,
1996).

In this chapter, we have so far concentrated on red-detuned traps
because of the conceptual differences in the practical realization of
dipole traps as compared to blue-detuned traps discussed in the next
chapter.  In the case of three-dimensional far-detuned lattices, this
distinction becomes faint since both lattice types are realized
through appropriate interference patterns of multiple beams. The atoms
are trapped either in the antinodes (red detuning) or nodes (blue
detuning) of the interference pattern. The main difference for red
and blue detuning lies in the photon scattering rates as
discussed in Sec.\ \ref{equilibrium}.  Here, we shortly present novel
developments for both types of far-off resonance lattices recently
realized by several groups (Anderson {\it et al.}, 1996;
M\"uller-Seydlitz {\it et al.}, 1997; Hamann {\it et al.}, 1998;
Boiron {\it et al.}, 1998; DePue {\it et al.}, 1998).  Localized
wavepackets oscillating in conservative microtraps are promising
systems in which to study fundamental questions related to
quantum-state preparation, coherent control and decoherence of
macroscopic superposition states.  Furthermore, optical lattices can
serve as prototype systems for the study of condensed matter models
based on periodic arrangements of weakly interacting particles.

Anderson {\it {\it et al.}} (1996) have confined lithium atoms in
three-dimensional optical lattices formed in the intersection of four
laser beams.  For lithium, the fine-structure splitting is only
10\,GHz leading to ground-state optical potentials which are
independent of the light polarization and the atomic spin state. A
face-centered cubic lattice with a nearest-neighbor spacing of
$1.13\lambda$ was realized by a four-beam configuration.
Three-dimensional lattices with periodicity much larger than $\lambda$
could be created by reducing the angles subtended by each possible
pair of the four lattice beams \footnote{A different approach to
  create structures with large periodicities has been used by Boiron
  {\it et al.} (1998) by using interference from multiple beams emerging
  from a holographic phase grating.}. Up to $10^5$ atoms were trapped
in the lattice. By reducing the intensity of the trapping light so
that only the coldest atoms from a MOT were confined in the lattice, a
trapped ensemble with an rms velocity spread corresponding to
1.8\,$\mu$K was prepared.  By adiabatically expanding deep potential
wells (see Sec.\ \ref{coolingmethods}), cooling at the expense of
smaller spatial density could be achieved.

\begin{figure}
\vspace{0.5cm}
\epsfxsize=60mm
\centerline{\epsfbox{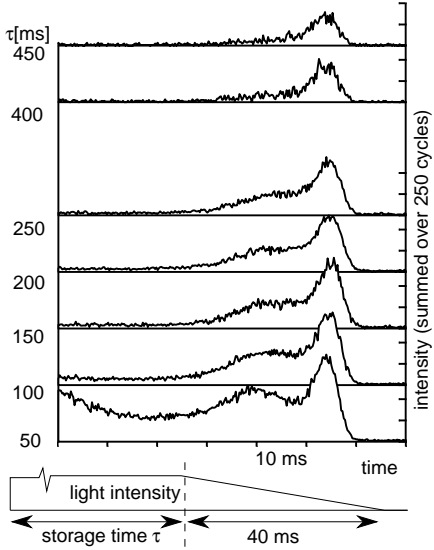}}
\vspace{0.5cm}
\caption{\it Time-of-flight spectra of metastable argon atoms trapped in
  the nodes of a three-dimensional far-off resonance optical lattice.
  The spectra were taken after various storage times $\tau$. Atoms
were released from the trap by slowly ramping down the trapping light
intensity. Higher-energetic bound states arrive first, deeper confined
states later. After long storage times, only atoms in the lowest bound
state are found. From M\"uller-Seydlitz {\it et al.} (1998).}
\label{fig:bluelattice}
\end{figure}

The temporal evolution of metastable argon atoms stored in the
intensity nodes of a blue-detuned optical lattice was investigated by
M\"uller-Seydlitz {\it et al.} (1997) at Konstanz University. The
lattice of simple-cubic symmetry was formed by three mutually
orthogonal standing waves with othogonal linear polarizations
resulting in isotropic potential wells with a lattice constant of
$\lambda/2 = 397\,$nm.  About $10^4$ atoms are initially captured in
the lattice. Fig.\ \ref{fig:bluelattice} shows time-of-flight spectra
for variable trapping times.  After a certain storage time,
\noindent
the light
intensity is ramped down releasing one bound state after the other.
For increasing storage times, the population of excited bands in the
potential wells decreases faster than the population of the
vibrational ground-state band leaving about 50 atoms populating the
motional ground state after storage times of about 450\,ms. Two
processes could be identified for this state selective loss of
particles: Firstly, atoms in higher excited bands have a higher
probability of leaving the finite extension of the trapping field
($w_0 = 0.55\,$mm) by tunneling.  Secondly, atoms interacting with the
far-off resonance trapping light (detuning 2\,nm from resonance for
the metastables) are optically pumped into the electronic ground state
of argon and are therefore lost from the trap. The probability for an
optical pumping process depends on the spatial overlap between the
optical lattice field and the atomic wavepacket confined to the
potential well. The smaller the spatial extension of the wavepackets,
the smaller is the excitation rate being smallest for the motional
ground state. The reduced photon scattering probability of deeper
bound states is an important specific property of a blue-detuned
optical lattice.

\begin{figure}
\vspace{0.5cm}
\epsfxsize=75mm
\centerline{\epsfbox{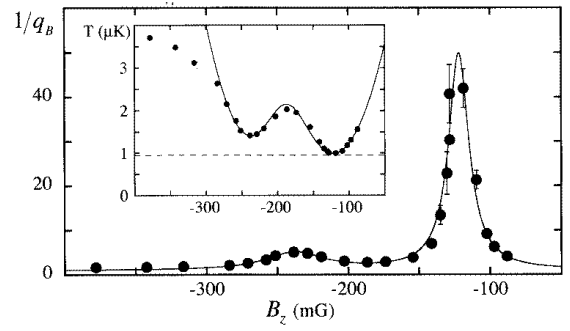}}
\vspace{0.5cm}
\caption{\it Resolved-sideband Raman cooling of Cs atoms in a two-
dimensional
  far-off resonance lattice. Shown is the inverse Boltzmann factor
  $1/q_B$ as a function of the applied magnetic field. When the
  magnetic field tunes the lattice-field induced Raman coupling to the
  first-order (at $B_z \approx 0.12\,G)$) or the second-order (at $B_z
  \approx 0.24\,G)$) motional sidebands, Raman cooling is most
  efficient resulting in the peaks of $1/q_B$.  Solid circles are data
  points, the solid line is a fit to the sum of two Lorentzians.
  Inset: Corresponding kinetic temperatures measured to determine the
  Boltzmann factor. The dashed lines indicates the kinetic temperature
  of the vibrational ground state. From Hamann {\it et al.}
  (1998).}
\label{fig:sideband.jessen}
\end{figure}

The first demonstration of resolved-sideband Raman cooling in a dipole
trap explained in Sec.\ \ref{coolingmethods} was performed with cesium
in a two-dimensional lattice by a group at University of Arizona in
Tucson (Hamann {\it {\it et al.}}, 1998). The 2D lattice consisted of
three coplanar laser beams with polarizations in the lattice plane.  A
magnetic field perpendicular to the lattice plane was applied to
Zeeman-shift the motional states $|m_F=4; n_{\rm osc}\rangle$ and
$|m_F=3; n_{\rm osc}-1\rangle$ of the upper hyperfine ground state
$F=4$ into degeneracy (see Sec.\ \ref{coolingmethods}). By adding a
small polarization component orthogonal to the lattice plane, Raman
coupling between magnetic sublevels with $\Delta m_F = \pm1$ was
introduced. In Fig.\ \ref{fig:sideband.jessen}, the inverse Boltzmann
factor $1/q_B=\exp(\hbar \omega_{\rm osc}/k_B T)$ is plotted versus
the magnetic field.  The inverse Boltzmann factor reaches a maximum
when the two motional states are shifted into degeneracy by the
magnetic field, so that sideband cooling by the lattice field becomes
effective. The second, smaller peak shows well-resolved cooling on the
second-order Raman sideband ($\Delta n_{\rm osc} = -2$). As shown by
the inset in Fig.\ \ref{fig:sideband.jessen}, the thermal energy of
the atoms closely approaches the zero-point energy in the potential
wells indicating a population $>95\%$ of the vibrational ground state.

Unity occupation of sites in a three-dimensional far-off resonance
optical lattice was very recently realized by the Berkeley group
(DePue {\it et al.}, 1998). In the regime of unity occupation,
interactions between highly localized atoms have dramatic effects, and
studies of collisional properties of tightly bound wavepackets become
possible.  The necessary high densities were achieved by applying
polarization gradient cooling to the 3D lattice filled with $10^8$
cesium atoms, and subsequent adiabatic toggling between the 3D lattice
and a 1D standing wave. The 3D lattice was formed by three mutual
orthogonal standing waves with controlled time-phase differences.
After the atoms were cooled in the lattice, the horizontal beams of
the lattice are adiabatically turned off leaving the atoms in a
vertical standing wave trap. Due to their low temperature (700\,nK),
the atoms were essentially at rest at their respecive positions.
Under the action of the transverse trapping potential, the atoms
radially collapsed towards the trap centre.  All atoms arrived
simultaneously at the centre after about a quarter radial oscillation
period. Thereby, the density in the trap centre was transiently
enhanced by a factor of ten reaching $6\times10^{12}$\,atoms/cm$^3$.
At the moment of peak density, the horizontal lattice beams were
adiabatically turned on again. A substantial fraction of lattice sites
was then multiply occupied and underwent fast inelastic collisions.
After multiply occupied sites had decayed, 44\% of the lattice site
were occupied by a single atom cooled close near its vibrational
ground state.

\section{Blue-detuned dipole traps} \label{blue}

Laser light acts repulsively on the atoms when its frequency is higher 
than the transition frequency (``blue'' detuning). 
The basic idea of a blue-detuned dipole trap is 
thus to surround a spatial region with repulsive laser light. 
Such a trap offers the great advantage of atom storage
in a ``dark'' place with low influence of the trapping light, which 
minimizes unwanted effects like
photon scattering, light shifts of the
atomic levels, and light-assisted collisional losses.
According to the discussion following Eq.~\ref{bluered}, this advantage
becomes substantial in the case of hard repulsive optical walls 
($\kappa \ll 1$ in Eq.~\ref{bluered}) or large potential depth
for tight confinement ($\hat{U} \gg k_B T$).

Experimentally, it is not quite as simple and straightforward 
to realize a blue-detuned trap as it is in the red-detuned case, 
where already a single tightly focused 
laser beam constitutes an interesting dipole trap.
Therefore, the development of appropriate methods to produce the 
required repulsive ``optical walls'' has played a central role in
experiments with blue-detuned traps.
Three main methods have been applied for this purpose: 
{\em Light sheets}, produced by strong elliptical focusing of a laser beam, 
can be used as nearly flat optical walls
(Davidson {\it et al.}, 1995; Lee {\it et al.}, 1996). 
{\em Hollow laser beams} can provide spatial confinement in at least 
two dimensions (Yang {\it et al.}, 1986).
{\em Evanescent waves}, formed by total internal reflection on the surface 
of
a dielectric medium, represent nearly ideal mirrors to reflect atoms 
(Cook and Hill, 1982; Dowling and Gea-Banacloche, 1996).
In most blue-detuned traps, gravity is used to close the confining
potential from above. Such traps are referred to as {gravito-optical 
traps}.
As a further experimental possibility, which we have already discussed
in Sec.\ \ref{sec:lattices}, atoms can be trapped in the micropotentials of
far blue-detuned optical lattices (M\"uller-Seydlitz {\it et al.}, 
1997). 

In this Chapter, we discuss various blue-detuned traps 
that have been realized experimentally and their particular features; 
an overview is given in Table \ref{bluetable}.
In the following, these traps are classified according to the
main method applied for producing the optical walls: 
Light-sheet traps (Sec.\ \ref{lstraps}), 
hollow-beam traps (Sec.\ \ref{hbtraps}), 
and evanescent-wave traps (Sec.\ \ref{ewtraps}).

\subsection{Light-sheet traps} \label{lstraps}

In experiments performed at Stanford University, 
Davidson {\it et al.}\ (1995) and Lee {\it et al.}\ (1996, 1998)
have realized light-sheet traps of various configurations and
applied them for rf spectroscopy on trapped atoms and
for optical cooling to high phase-space densities.
The light sheets were derived from the two strongest lines 
of an all-line argon-ion laser.
A laser power of up to 10\,W at 514\,nm and up to 6\,W at 488\,nm 
was focused with cylindrical lenses to cross sections of 
typically 15\,$\mu$m\,$\times$\,1\,mm. 
For the Na atoms used in the experiments 
(resonance line at 589\,nm), this leads to maximum light-sheet 
potentials in the order of 100\,$\mu$K.

When two light sheets are combined and overlap in space, there are
two ways to avoid 
perturbing interference effects, which could open escape channels in the 
optical potential.
First, if the two light sheets have different frequencies then the
relevant potential is determined by the time average over the rapid beat 
node, in which the interference averages out.
Second, if the light sheets have same frequencies but orthogonal 
polarizations then interference leads to a spatial modulation of the
polarization. In the case of large detunings greatly exceeding
the fine-structure splitting, as it was very well fulfilled in the Stanford
experiments, the polarization modulation 
has negligible effect on the dipole potential 
(see 
Eq.~\ref{alkaliexpand}).

\begin{figure}
\vspace{0.5cm}
\epsfxsize=8.6cm 
\centerline{\epsffile{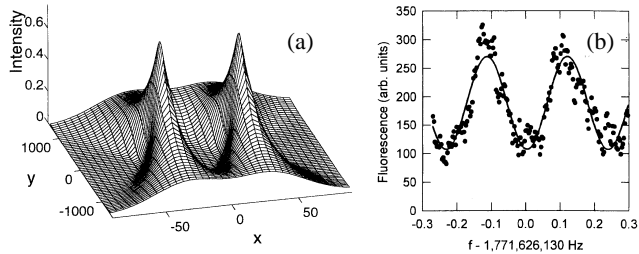}}
\vspace{0.5cm}
\caption{\it Light-sheet trap used for rf spectroscopy. 
(a) Laser intensity produced in a horizontal plane 30\,$\mu$m above the 
intersection of the two focused sheets of light. The $x,y$ dimensions 
are in microns, and the intensity is normalized to the peak laser 
intensity. (b) The central Ramsey fringes observed for rf-induced
hyperfine transitions for a measurement time of 4\,s.
From Davidson {\it et al.}\ (1995).}
\label{davidson}
\end{figure}

In the first experiment (Davidson {\it et al.}, 1995), 
two horizontally propagating light sheets were combined
to form a vertical ``V'' cross section. This configuration
already provides 3D confinement, as the trapping potential is closed
along the propagation direction due to the divergence of the 
tightly focused light, see Fig.\ \ref{davidson}(a).
Vertically, the atoms are kept in the trap by gravity. 
The gravito-optical trap thus has the form of a boat with a length of about 
2\,mm (for trap parameters see Table \ref{bluetable}).

Using this light-sheet trap, 
Davidson {\it et al.}\ have stored about 3000 Na atoms and 
impressively demonstrated the advantages of blue-detuned
dipole traps for spectroscopic applications.
By using the method of separated oscillatory fields, 
they have measured Ramsey fringes of the
$F=1, m_{F}=0 \rightarrow F=2, m_{F}=0$ hyperfine transition
of the Na ground state.
For the excitation of this transition an rf travelling wave with a frequency 
of 1.77\,GHz was used.
The Ramsey fringes were measured by applying
two $\pi/2$ rf pulses separated by a time delay of up to 4\,s.
Initially, all trapped atoms 
were optically pumped into the lower hyperfine state $(F=1)$. 
After applying of the two rf pulses, the number of atoms
transfered into the 
upper state $(F=2)$ was measured by applying
a short pulse of light resonant with the cycling 
$F=2 \rightarrow F'=3$
transition and detecting the induced fluorescence.
The two central Ramsey fringes observed  
by varying the rf frequency for a pulse delay of 4\,s
are shown in Fig.\ \ref{davidson}(b).
By analyzing the dependence of the fringe contrast on the
delay between the two rf pulses,
a $1/e$ coherence decay time of $4.4$\,s was obtained.

In the same experiments, the Stanford group has also measured the mean 
residual light shift (ac Stark shift)
of the hyperfine transition frequency as caused by the trapping light.
From the frequency of the central Ramsey fringe, 
a corresponding shift of 270\,mHz was observed. 
The absolute light shifts of the two hyperfine sublevels are larger
by the ratio of the optical detuning ($\sim$\,90\,THz) 
to the hyperfine splitting (1.77\,GHz) and 
thus amounted to $\sim$\,14\,kHz. This number directly gives the average
dipole potential $\bar{U}_{\rm dip} \approx h \times 14\,$kHz$ 
\approx k_B \times 0.7\,\mu$K experienced by the atom and also allows 
one to determine an average photon scattering rate of 0.01\,s$^{-1}$ 
according to equation \ref{KKrelation}.

The authors also mention similar experiments performed in 
a red-detuned dipole trap realized with a Nd:YAG laser. In this case,
the longest observed coherence times were 
$\sim$300 times lower, which highlights the advantage of the 
blue-detuned geometry for in-trap spectroscopy.

\begin{figure}
\vspace{0.5cm}
\epsfxsize=8.6cm 
\centerline{\epsffile{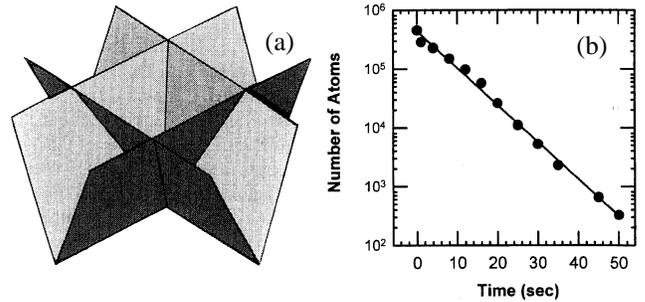}}
\vspace{0.5cm}
\caption{\it Inverted-pyramid trap used for 
Raman cooling. (a) Schematic of the trap geometry, and (b) lifetime
measurement performed after Raman cooling at a background pressure
of $\sim$\,$3\times 10^{-11}$\,mbar. From Lee {\it et al.}\ (1996).}
\label{invpyramid}
\end{figure}

In a later experiment (Lee {\it et al.}, 1995), the  
light-sheet trapping was improved by a new trap geometry 
in the form of an {\em inverted pyramid}. 
As illustrated in Fig.\ \ref{invpyramid}(a),
this trap was produced by four sheets of light.
Due to the much larger trapping volume provided by the pyramidal geometry
this trap could be loaded with $4.5\times10^5$ atoms, which constitutes an
improvement over the previous configuration by more than a factor of 100
(see also Table \ref{bluetable}). Lee {\it et al.}\ have also tested a
similar, tetrahedral box trap, the performance of
which was inferior to the inverted pyramid.

The particular motivation of the experiment by Lee {\it et al.}\ was to 
obtain high phase-space densities of the trapped atomic ensemble 
by optical cooling. The authors therefore applied 1D Raman cooling 
(Kasevich and Chu, 1992) as a sub-recoil cooling method, 
see also Sec.\ \ref{coolingmethods}.
The geometry of the inverted-pyramid trap is very advantageous for getting
dense atomic samples, because of a strong spatial compression with
decreasing temperature. 
In an ideal inverted pyramid the density 
$n$ would scale as $T^{-3}$ in contrast to a 3D harmonic oscillator, where
$n \propto T^{-3/2}$. Moreover, this trap configuration provides very
fast motional coupling between the different degrees of freedom, which
is particularly interesting in case of a 1D cooling scheme.

About $4.5 \times 10^{5}$ atoms were loaded at a temperature 
of 7.7\,$\mu$K and a peak density of $2 \times 10^{10}$\,cm$^{-3}$. 
By applying a sequence of Raman cooling pulses
during the following 180\,ms, the temperature of the atoms was reduced to 
1\,$\mu$K and, keeping practically all atoms in the trap, 
a peak density of 4$\times 10^{11}$ cm$^{-3}$ was reached. 
This density increase by a factor of 20 resulting from a temperature 
reduction by a factor of about 7 indicates that the trap was already 
in a regime where the confining potential behaves rather harmonically
than like in an ideal pyramid. This may be explained by the 
finite transverse decay length of the light-sheets potential walls.
As a result of the Raman cooling, the atomic phase-space density 
was increased by a factor 
of 320 over the initial one and reached a value which was about a 
factor of $400$ from Bose-Einstein condensation.

After an initial loss of atoms observed in the first second after the Raman 
cooling process, an exponential decay of the trapped atom number
was measured with a time constant of 7\,s, 
see Fig.\ \ref{invpyramid}(b). 
This loss did not significantly depend on the background pressure below 
$10^{-10}$\,mbar, which points to
the presence of an additional single-particle loss mechanism. 
This loss was consistent with an observed
heating process in the trap that 30 times exceeded 
the calculated heating by photon scattering. This heating of unknown origin
was identified as the main obstacle to implement evaporative cooling in the
inverted pyramid trap. 
The experiments moreover showed evidence  
that ground-state hyperfine-changing collisions, ejecting atoms 
out of the trap, were limiting the maximum density achievable with Raman
cooling to $\sim$10$^{12}$\,cm$^{-3}$. 

In a later experiment, Lee and Chu (1998) used the inverted-pyramid trap 
for preparing a Raman-cooled sample with spin polarization in any of the
three magnetic sub-levels of the lower hyperfine ground state ($F=1$). 
A small bias magnetic field was used to lift the degeneracy of the ground 
state and appropriate polarization was chosen for the Raman cooling light. 
The attained temperature and phase-space density was similar to the 
unpolarized case. 
This experiment can be seen as  
a nice illustration of the general advantage of dipole trapping to leave the 
full ground-state manifold available for experiments.

\subsection{Hollow-beam traps} \label{hbtraps}

Hollow blue-detuned laser beams, which provide radial confinement,
are particularly interesting and versatile 
tools for the construction of dipole traps.
Hollow laser beams may be divided into two classes:
Beams in pure higher-order Laguerre-Gaussian (LG) modes 
and other hollow beams which cannot
be represented as single eigenmodes of an optical resonator.
The main difference is that LG beams preserve their 
transverse profile with propagation 
(only converging or diverging in width), which is of particular interest
for atom guiding (Dholakia, 1998; Schiffer {\it et al.}, 1998). 
Other, non-LG hollow beams
can change their profile substantially, thus offering additional 
features of interest for atom trapping.

A Laguerre-Gaussian mode LG$_{p\,l}$ is characterized by a radial
index $p$ and an azimuthal index $l$. A hollow beam with a
``doughnut'' profile is obtained for $p=0$ and $l\ne0$ with an 
intensity distribution given by
\begin{equation}
I_{0\,l}(r)=P\frac{2^{l+1}r^{2l}}{\pi l! \, w_{0}^{2(l+1)}}
\exp(-2r^{2}/w_{0}^{2}) \, ;
	\label{lg}
\end{equation}
here $P$ is the power and $w_{0}$ is the waist of the beam.
For $l=0$ this equation gives the transverse profile of a usual Gaussian 
beam 
(TEM$_{00}$ mode) 
as described by Eq.\ \ref{eq:GaussInt}. The higher the mode index $l$, 
the larger is the ratio between the beam radius and the width of the ring, 
i.e.\ the harder is the repulsive optical wall radially confining the
atoms and the weaker is heating by photon scattering.

\subsubsection{Plugged doughnut-beam trap}

\begin{figure}
\vspace{0.0cm}
\epsfxsize= 8.6cm  
\centerline{\epsffile{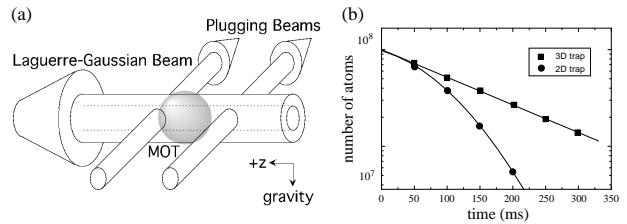}}
\vspace{0.5cm}
\caption{\it (a) Plugged doughnut beam trap, and 
(b) measurement of the storage time (filled squares). 
The filled circles refer to a 2D trap formed by the
doughnut beam alone without plugging beams (no axial confinement).
Adapted from Kuga {\it et al.}\ (1997).}
\label{kuga}
\end{figure}

The dougnut-beam trap shown in Fig.\ \ref{kuga}(a) was recently realized by 
Kuga {\it et al.}\ (1997) at the University of Tokyo; see also Table 
\ref{bluetable}. 
The LG$_{03}$ doughnut beam ($w_{0}$=0.6\,mm) was derived from a laser which 
was forced to 
oscillate in a Hermite-Gaussian HG$_{03}$ mode by insertion of a thin wire 
into the cavity. 
An astigmatic mode convertor (Beijersbergen {\it et al.}, 1993) then 
transformed the beam into the LG$_{03}$ mode. As a LG beam does not provide 
axial confinement, the
hollow beam was plugged by two additional laser beams (diameter 0.7\,mm),
which were separated by 2\,mm and perpendicularly intersected the hollow 
beam. The plugging beams were derived from the recycled doughnut beam.

An exponential decay of the stored atom number
was observed with a time constant of about 150\,ms 
[see Fig.\ \ref{kuga}(b)], which was 
explained by heating out of the trap by photon scattering.
The authors also observed that trapping in a LG$_{01}$ mode showed 
inferior performance with very 
short lifetimes of a few milliseconds only, 
which highlights the benefit of ``hard'' optical walls for
reducing heating by photon scattering.

In subsequent experiments (Torii {\it et al.}, 1998), the storage time of 
the trap was improved by application of a pulsed optical molasses 
cooling scheme to a value of 1.5\,s, which then was dominated by  
collisions with the background gas. 
This non-continuous molasses cooling scheme allows one to 
cool the atoms down to nearly the same temperatures as in a continuous
cooling scheme, but suppresses trap losses by 
light-assisted collisions 
and the interference of light shifts from the molasses light with 
the trapping potential. Moreover, potentially severe 
losses by hyperfine-changing collisions
(see Sec.~\ref{collisions})
can be strongly reduced by keeping the atoms in the lower hyperfine ground
states in the off-times of the molasses.
In such a pulsed cooling scheme, the off-time 
can be as long as heating by photon scattering (see Sec.~\ref{heating})
does not significantly degrade the trap performance. 
The on-time can be as short as the typical cooling
time in the molasses. 

\subsubsection{Single-beam traps}

A blue-detuned trap based on a single laser beam was 
recently demonstrated by Ozeri {\it et al.}\ (1998) at the Weizman Institute 
in
Israel; see also Table \ref{bluetable}. 
The trapping beam was produced in an experimentally very simple way by 
passing a single Gaussian beam through
a phase plate of appropriate size, which shifted the center 
of the beam by a phase angle of $\pi$. In the focus of such a beam, 
destructive interference leads to a reduced or even vanishing light 
intensity. By choosing the proper ratio between the diameter of the 
phase plate and the laser beam diameter, Ozeri {\it et al.}\  obtained a
darkness ratio (central intensity normalized to maximum intensity) of
1/750.
Close to the center of this single-beam trap, the intensity increases 
radially with the fourth power of the distance (like in the case of a 
LG$_{02}$ beam) and axially the dependence is quadratic. 

In order to measure the average light intensity experienced by the atoms
in the trap, Ozeri {\it et al.}\  have studied the relaxation of hyperfine 
population 
caused by photon scattering from the trapping light in a similar way
as done by Cline {\it et al.}\ (1994) 
in a red-detuned far-off-resonance trap; 
see also Sec.~\ref{sec:spinrelaxation}.
From measurements performed at a detuning of 0.5\,nm it was concluded 
that the average intensity experienced by a trapped atom was as
low as $\sim$1/700 of the maximum intensity. In another series of 
measurements they observed that, at constant temperature, the photon 
scattering rate scaled linearly with the inverse detuning.
This observed behavior represents a nice confirmation of 
Eq.~\ref{scatteringT}, which
for a power-law potential directly relates the average scattering rate 
to the temperature with an inverse proportionality to the detuning.

In a recent proposal, Zemanek and Foot (1998) considered a 
blue-detuned dipole trap formed by two counterpropagating laser beams 
of equal central intensities, but different diameters.
Along the axis of such a standing-wave configuration, completely destructive 
interference 
would lead to minima of the dipole potential with zero intensity.
These traps would be radially closed because of the incomplete destructive
interference at off-axis positions. 
This resulting linear array of three-dimensional dipole traps 
could combine the interesting features of standing-wave 
trapping schemes (see Sec.~\ref{sec:SWT}) with the advantages of blue 
detuning. Experimentally, it seems straightforward to realize such a 
trap by retroreflection and appropriate attenuation of a 
{\em single} slightly converging laser beam.

\subsubsection{Conical atom trap}

\begin{figure}
\vspace{0.5cm}
\epsfxsize=6.5cm
\centerline{\epsffile{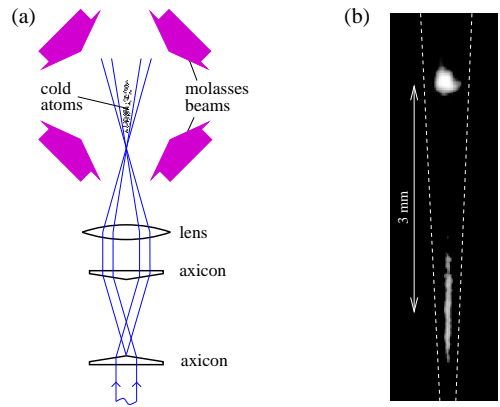}}
\vspace{0.5cm}
\caption{\it (a) Illustration of the conical atom trap (CAT).
(b) Fluorescence image of atoms in the CAT (lower, elongated blob) 
combined with an image of atoms 
in the MOT (upper spot). The dashed lines indicate the conical trapping
field. Adapted from Ovchinnikov {\it et al.}\ (1998).}
\label{cat}
\end{figure}

A single-beam gravito-optical trap was recently demonstrated by 
Ovchinnikov {\it et al.}\ (1998) at the MPI f\"ur Kernphysik in Heidelberg.
The ``conical atom trap'' (CAT), illustrated in Fig.\ \ref{cat}(a),
is based on a conical hollow beam and
combines experimental simplicity with several features of
interest for the trapping of a large number of atoms at high 
densities: high loading efficiency, tight confinement, low collisional
losses, and efficient cooling.

In the experiment (parameters see Table \ref{bluetable})
the upward directed conical trapping beam was generated
by using an arrangement of two axicons and one spherical lens. 
In the focal plane, the beam profile was roughly Gaussian with a 
diameter of about 100\,$\mu$m.
Within a few millimeters distance from the focus, 
the beam evolved into a ring-shaped profile
with a dark central region resembling a higher-order Laguerre-Gaussian 
mode. The opening angle of the conical beam was about 150\,mrad. 

The trap was operated relatively close to resonance
as compared to
the other blue-detuned traps discussed so far.  
With an optical detuning
of 3\,GHz (12\,GHz) with respect to the lower (upper) hyperfine 
ground state of Cs 
a large potential depth was realized, which together with the
funnel-like geometry facilitated the transfer of as much as 80\%
of all atoms from the MOT into the CAT.
The relatively small detuning
requires efficient cooling for removing the heat resulting from photon 
scattering from the trapping light.
For this purpose, an optical molasses was applied continuously to cool 
atoms in the upper hyperfine ground state. 
This cooling, however, takes place with an inherently reduced duty cycle
and is thus similar to the pulsed molasses cooling scheme of 
Torii {\it et al.}\ (1998), which was discussed before. 
In the molasses scheme applied in the CAT, the short phases of cooling in 
the upper hyperfine ground state (typically 50\,$\mu$s) are self-terminating 
by optical pumping into the lower hyperfine level. There the atoms
stay for a much longer time (typically $\gtrsim$\,1\,ms) 
until they are repumped by the light of the intense conical trapping beam. 

This inherent duty cycle for the molasses provides 
cooling phases which are long enough to efficiently remove heat, 
while the average population of the upper hyperfine state is kept very 
low. This efficiently suppresses trap loss due to hyperfine-changing 
collisions (see Sec.\ \ref{collisions}). At an atomic
number density of $\sim\,$$10^{11}\,$cm$^{-3}$ lifetime measurements of
the trapped atoms ($1/e$ lifetime of 7.8\,s due to background gas 
collisions) 
showed no significant loss due to ultracold collisions, which 
indirectly confirmed the predominant population of the lower hyperfine 
ground state.

The CAT was also operated in a pure ``reflection cooling'' mode without
any molasses cooling, similar to the situation that was theoretially 
considered by Morsch and Meacher (1998). 
Reflection cooling (Ovchinnikov {\it et al.}, 1995a and b) is based on
the inelastic reflection of an atom from a blue-detuned light field,
as discussed in more detail in the following section in context with
evanescent waves. In a Sisyphus-like
process the atom is pumped from the strongly repulsive lower to the
weakly repulsive upper hyperfine state. A closed cooling cycle requires
a weak repumping beam to bring the atom back to the lower state.
Such a beam was applied in the CAT experiment from above.
The detuning of the conical beam was increased to a few ten GHz to optimize 
reflection cooling. Because of the lower potential depth the loading 
was less efficient ($\sim$10\% transfer from the MOT instead of 80\% reached 
before), but stable background-gas limited trapping was achieved. 
Without the repumping beam, atoms 
were rapidly heated out of the trap. These observations clearly 
demonstrated reflection cooling in a blue-detuned trap made of 
free-propagating light fields. However, optimum reflection cooling
requires very steep optical walls so that only evanescent waves
are suited to reach temperatures close to the recoil limit by this
mechanism, as will be discussed in Sec.~\ref{gosttrap}.

\subsection{Evanescent-wave traps} \label{ewtraps}

A hard repulsive optical wall with nearly ideal properties can be realized 
by a blue-detuned evanescent wave (EW), produced by total internal 
reflection of a laser beam from a dielectric-vacuum interface. 
In the vacuum the EW intensity falls off exponentially 
within a typical distance of 
$\lambda/2\pi$ from the surface and thus provides a very large gradient. 
The use of an EW as a mirror for neutral atoms was suggested by Cook and 
Hill (1982), 
who also proposed to trap atoms in a box realized with EW mirrors.
The first experimental demonstration of an atom mirror was
made by Balykin {\it et al.}\ (1987) by grazing-incidence reflection of a 
thermal
atomic beam. A few years later, Kasevich {\it et al.}\ (1990) observed 
reflection of cold atoms at normal incidence. 
Since then many experiments have been conducted with EW atom mirrors.
An extensive review on EW atom mirrors and related trapping schemes 
has already been given by Dowling and Gea-Banacloche (1996). 
We thus concentrate on the essential issues and some interesting, new 
developments.

\subsubsection{Evanescent-wave atom mirror} 

The exponential shape of the repulsive optical potential of a
far-detuned EW leads to simple expressions for the basic 
properties of such an atom mirror. 
The EW intensity as a function of the distance $z$ from the surface 
is given by 
\begin{equation}
I(z) = I_0 \exp(- 2 z/ \Lambda)\, ,
\end{equation}
where the $1/e^2$ decay length 
$\Lambda = \lambda / (2\pi \sqrt{n^2\sin^2\theta - 1})$
depends on the
angle of incidence $\theta$ and the refractive index $n$ of the dielectric. 
The maximum EW intensity $I_0$ is related to the
incident light intensity $I_1$ by
$I_0/I_1 = 4 n \cos^2\theta / (n^2-1)$ \,.

The repulsive dipole potential of an atom mirror is independent of the 
magnetic substate if the detuning is large compared to the excited-state
hyperfine splitting and the EW is linearly polarized (see 
Sec.~\ref{alkali}); 
the latter is obtained
for an incoming linear polarization perpendicular to 
the plane of incidence (TE polarization). 
In the interesting case of low saturation,  
the EW dipole potential can then be calculated according to 
Eq.~\ref{alkaligeneral}.

A very important quantity to characterize the EW reflection process is its
probability to take place coherently, i.e.\  without spontaneous photon
scattering. 
The probability for an (in)coherent reflection can be calculated
by integrating the intensity-dependent scattering rate $\Gamma_{\rm sc}$
(Eq.~\ref{Gsclin}) 
over the classical trajectory of the atom in the repulsive potential. 
For large enough laser detuning (still close to one of the $D$ lines), 
the resulting small probability $p_{\rm sp} \ll 1$ for 
an incoherent reflection process is given by
\begin{equation}
p_{\rm sp}= \frac{m \Lambda}{\hbar} \frac{\Gamma}{\Delta} \, v_{\perp} \, 
,	
	\label{psp}
\end{equation}
where $v_{\perp}$ is the velocity component perpendicular to the surface.

The light shift of the ground-state sublevel
integrated over time in a single reflection process  
corresponds to a phase shift 
\begin{equation}
\Phi_{\rm ls} = \frac{m \Lambda}{\hbar} \, v_{\perp} \, 
\label{phils}
\end{equation}
experienced by the atom.
The simple relation $p_{\rm sp} = (\Gamma/\Delta) \, \Phi_{\rm ls}$ 
is a result of the fundamental connection between the
absorptive and dispersive effect of the interaction with the light field
(see also Eq~\ref{KKrelation}).
As an interesting consequence of the exponential shape of the EW potential,
both equations Eq.~\ref{psp} and Eq.~\ref{phils} do not depend on the
intensity of the light field, as long as 
the potential barrier is high enough. For higher/lower intensities the 
reflection just takes place at larger/smaller distances from the surface.

Very close to the surface ($z \lesssim 100\,$nm), the 
van-der-Waals attraction becomes significant. Its main effect is to 
reduce the maximum potential barrier provided by the EW mirror
(Landragin {\it et al.}, 1996b). For small kinetic energies, the reflected
atoms do not penetrate deeply into the EW and thus do not feel the
surface attraction. We have thus neglected the effect of the van-der-Waals
force in Eqs.~\ref{psp} and \ref{phils}, and we will do so in the following.

\subsubsection{Gravito-optical cavities} \label{gocavities}
 
A great deal of interest in EW atom mirrors has been
stimulated by the intriguing prospect to
build resonators and cavities for 
atomic de-Broglie waves (Balykin and Letokhov, 1989; Wallis {\it et al.}, 
1992).
The simplest way to realize such a scheme is a single atom mirror on
which the atoms classically bounce like on a trampoline. 
Such a trapping scheme is referred to as gravito-optical cavity.

The time $t_{\rm b}$ between 
two bounces in a gravito-optical cavity is related to 
the maximum height $h$ and the maximum velocity $v_{\perp}$ 
of the atoms by 
$t_{\rm b}=2 \sqrt {2h/g}=2v_{\perp}/g$, where $g$ is the gravitational
acceleration.
Using Eq.~\ref{psp} for the probability for photon scattering per bounce, 
the average photon scattering rate can be expressed as
\begin{equation}
\bar{\Gamma}_{\rm sc}
= \frac{p_{\rm sp}}{t_{\rm b}} = \frac{mg\Lambda}{2 \hbar} \, 
\frac{\Gamma}{\Delta} \, .
	\label{nsp}
\end{equation}
Analogously using Eq.~\ref{phils}, the mean light shift 
$\delta \bar{\omega}_{\rm ls} = \bar{U}_{\rm dip}/\hbar$
experienced by the bouncing atom is obtained as
\begin{equation}
\delta\bar{\omega}_{\rm ls}
= \frac{\Phi_{\rm ls}}{t_{\rm b}}
= \frac{mg \Lambda}{2\hbar} \, .
	\label{shift}
\end{equation}
It is a remarkable consequence of the exponential shape of the 
EW potential that $\bar{\Gamma}_{\rm sc}$ and $\delta\bar{\omega}_{\rm ls}$
do not depend on the energy of the atom. 
For an atom with less energy in a gravito-optical
cavity, the decrease in scattering probability and light shift is
exactly compensated by the higher bounce rate.

The average scattering
rate $\bar{\Gamma}_{\rm sc}$ can be interpreted as the decoherence rate
of gravito-optical resonator due to photon scattering. 
It also determines the heating power
according to Eq.~\ref{heatingpower}. 
The mean light shift is of interest for
possible spectroscopic applications of gravito-optical cavities.

The eigenenergies of the vertical modes in a gravito-optical 
cavity can be approximately calculated by idealizing the EW potential 
as a hard wall (Wallis {\it et al.}, 1992). 
In this case the vertical potential has the shape of 
a wedge and the energy of the $n$-th vertical mode is given by
\begin{equation}
	E_{n}=\hbar\omega_{\rm v}\left( n-\frac{1}{4} \right) ^{2/3} \, ,
	\label{En}
\end{equation}
where $\omega_{\rm v}=(9\pi^{2}mg^{2}/8\hbar)^{1/3}$ is a
characteristic frequency.
For example, for cesium atoms $\omega_{\rm v}/2\pi$=2080$\ $Hz,
which corresponds to a temperature of 
$\hbar\omega_{\rm v}/k_{B}\simeq 95\ $nK. Consequently, the population of 
a single vertical mode with many atoms, a challenging issue for
future experiments, requires cooling below this temperature.

As an important step for EW trapping, the ENS group in Paris 
observed the bouncing of atoms in a stable 
gravito-optical cavity (Aminoff {\it et al.}, 1993). 
The atom mirror used in the experiment
(diameter $\sim$\,1.5\,mm) was produced on a 
{\em concave} spherical substrate (radius of curvature 2\,cm)
to obtain additional transverse confinement of the atomic motion
(Wallis {\it et al.}, 1992); the resulting transverse trap depth was
about 5\,$\mu$K.

\begin{figure}
\vspace{0.5cm}
\epsfxsize=8.6cm 
\centerline{\epsffile{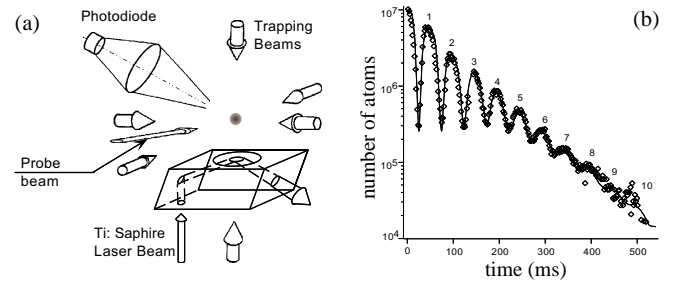}}
\vspace{0.5cm}
\caption{\it Observation of atoms bouncing in a gravito-optical cavity. 
(a) Experimental set-up, and (b) number of atoms detected
in the probe beam for different times after their release (points). The 
solid curve is the result of a corresponding Monte-Carlo simulation.
Adapted from Aminoff {\it et al.}\ (1993).}
\label{parisbounces}
\end{figure}

In the set-up sketched in Fig.\ref{parisbounces}(a), about
$10^{7}$ cold atoms were dropped onto the mirror
from a MOT located at a height of 3\,mm above the prism 
(corresponding bounce period $t_{\rm b} = 50\,$ms).
The bouncing atoms were detected by measuring the fluorescence in a 
resonant probe beam, which was applied to the atoms after a variable
time delay. 
The experimental results displayed in Fig.\ \ref{parisbounces}(b) show
up to ten resolved bounces, after which the
number of atoms dropped below the detection limit.
The measured loss per bounce of $\sim$\,40\% was  
attributed to photon scattering during the EW reflection process 
($\sim$\,5\% according to Eq.~\ref{psp}),  photon
scattering from stray light from the mirror ($\sim$\,10\%), collisions
with the background gas ($\sim$\,10\%), 
and another, not identified source of loss ($\sim$\,20\%). 
An explanation for the latter may be the diffusive 
reflection from an EW mirror, as observed later 
by Landragin {\it et al.}\ (1996a).
The ENS gravito-optical cavity represents the first
atom trap realized with evanescent waves.

\subsubsection{Gravito-optical surface trap} \label{gosttrap}

\begin{figure}
\vspace{0.1cm}
\epsfxsize=8.6cm 
\centerline{\epsffile{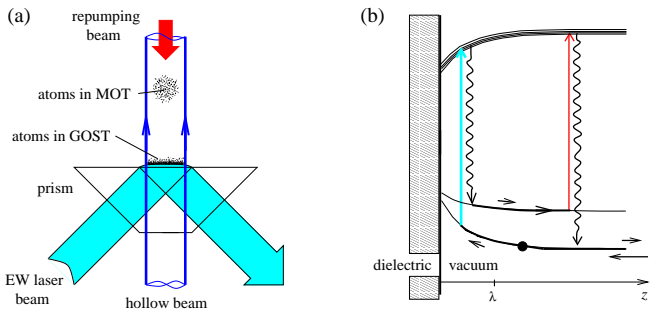}}
\vspace{0.5cm}
\caption{\it (a) Schematic of the gravito-optical surface trap (GOST). 
(b) Illustration of the evanescent-wave cooling cycle. 
The dot indicates an atom approaching the dielectric surface 
in the lower hyperfine state, scatters an EW photon, leaves the EW in
the upper, less repulsive ground state, and is finally pumped back into
the lower state.}
\label{gost}
\end{figure}

A new step for gravito-optical EW traps 
was the introduction of a {\em dissipative} mechanism, 
following suggestions by 
Ovchinnikov {\it et al.}\ (1995a) and S\"oding {\it et al.}\ (1995) 
to cool atoms by inelastic reflections.
In a corresponding
experiment at the MPI f\"ur Kernphysik in Heidelberg, 
Ovchinnikov {\it et al.}\ (1997) have realized the gravito-optical surface 
trap (GOST) schematically shown in Fig.\ \ref{gost}(a); see also Table
\ref{bluetable}.
This trap facilitates storage and efficient cooling of a dense atomic 
gas closely above an EW atom mirror.

A {\em flat} atom mirror was used in the GOST and 
horizontal confinement was achieved by
a hollow, cylindrical laser beam, far blue-detuned from the atomic 
resonance.
This beam with a 
ring-shaped transverse intensity profile providing very steep optical
walls was generated using an axicon (Manek {\it et al.}, 1998). 
A Laguerre-Gaussian beam of similar performance
would require extremly high order (about LG$_{0,\,100}$).
The optical potentials of the GOST thus come very close to ideal hard walls,
which leads to a strong reduction of photon scattering from the trapping
light even relatively close to resonance.

Cooling by inelastic EW reflections (evanescent-wave Sisyphus cooling)
is the key to stable trapping in the GOST. The basic mechanism, which
was experimentally studied before in grazing-incidence reflection of 
an atomic beam (Ovchinnikov {\it et al.}\ 1995b; Laryushin {\it et al.}, 
1996) 
and normal-incidence reflection 
of cold atoms (Desbiolles {\it et al.}, 1996), is  
based on the
splitting of the $^2$S$_{1/2}$ ground state of an alkali atom 
into two hyperfine sublevels. In the case of linear EW polarization,
the atom can be modeled as a three-level scheme (S\"oding {\it et al.}, 
1995;
see also Sec.~\ref{alkalidiscuss})
with two ground states separated by $\Delta_{\rm HFS}$ 
and one excited state, for which the hyperfine splitting
can be neglected.

An inelastic reflection takes place when the atom enters the EW
in the lower ground state and, by scattering an EW photon during the 
reflection
process, is pumped into the less repulsive upper state;
see Fig.\ \ref{gost}(b).
The cooling cycle is then closed by pumping the atom back into the lower 
hyperfine state with a weak resonant repumping laser (coming from above
in Fig.~\ref{gost}(a)).
The mean energy loss $\Delta E_{\perp}$
from the motion perpendicular to the surface per inelastic reflection
is given by
$\Delta E_{\perp}/E_{\perp}
= - \frac{2}{3} \Delta_{\rm HFS}/ (\Delta +\Delta_{\rm HFS})$,
where $E_{\perp} = m v_{\perp}^2/2$ is the
kinetic energy of the incoming atom and $\Delta$ 
is the relevant detuning with respect to the lower hyperfine state.
With the probability $p_{\rm sp}$ for an incoherent reflection
according to Eq.~\ref{psp}, the branching ratio $q$ for spontaneous
scattering into the upper ground state, 
and the bounce rate $t^{-1}_{\rm b} = g/2 v_{\perp}$ 
EW Sisyphus cooling can be characterized by a simple cooling rate
\begin{equation}
\beta = 
 \frac{q}{3} \, \frac{\Delta_{\rm HFS}}{\Delta + \Delta_{\rm HFS}} \, 
\frac{m g \Lambda}{\hbar} \, \frac{\Gamma}{\Delta} \, .
\label{coolingrate}
\end{equation}
The vertical motion 
is damped exponentially because the average photon scattering rate is
independent of the kinetic energy (see Eq.~\ref{nsp}).
The final attainable temperature is
recoil-limited to a value of $\sim$\,$10\,T_{\rm rec}$, similar to a 
polarization-gradient optical 
molasses (see Sec.~\ref{coolingmethods}).

The GOST was experimentally realized for Cs atoms, 
the high mass of which is very favorable for gravito-optical trapping: 
In the gravitational potential, 
the recoil temperature $T_{\rm rec}$  corresponds to a height 
$h_{\rm rec} = k_B T_{\rm rec}/ (m g)$, which for the heavy Cs atoms 
is particularly low ($h_{\rm rec} = 1.3\,\mu$m, 
see also Table \ref{alkalitable}).  
As a consequence, Cs atoms cooled close to the recoil limit
can be accumulated very close to the dielectric surface.

In the experiment, trapped atoms were observed for storage times up to 
25\,s, 
corresponding to more than 10.000 (unresolved) bounces. The observed
exponential loss with a $1/e$-lifetime of 6\,s 
was completely consistent with collisions with the background gas. 
By time-of-flight diagnostics, a temperature of 3.0\,$\mu$K 
was measured, which is quite close to the theoretical cooling limit. 
In thermal equilibrium, the number density as function of the distance
from the surface follows from Eq.~\ref{Boltzmann} (in this case, 
equivalent to the {\em barometric equation}), 
which for the measured temperature of 3.0\,$\mu$K
corresponds to an exponential decay with a $1/e$ height as low as 
19\,$\mu$m.

\begin{figure}
\vspace{0.2cm}
\epsfxsize=6.0cm 
\centerline{\epsffile{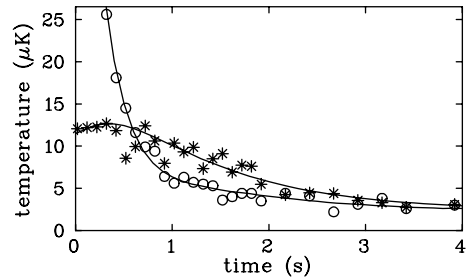}}
\vspace{0.5cm}
\caption{\it Cooling dynamics in the GOST. The vertical ($\circ$) and 
horizontal
($\bullet$) 
temperatures measured for about 10$^5$ trapped Cs atoms are plotted as a 
function of the storage time. The solid lines are theoretical fits based
on Eq.~\protect\ref{coolingrate} 
and the assumption of an EW reflection with a 
small diffusive, non-specular component. From Ovchinnikov {\it et al.}\ 
(1997).}
\label{gostresults}
\end{figure}

The cooling dynamics observed in the experiments is shown in 
Fig.~\ref{gostresults}. The initial horizontal temperature is determined by
the temperature of the MOT, whereas the much higher initial vertical
temperature results from the release of the atoms at a height of 
$\sim$\,800$\mu$m. With increasing storage time, 
the vertical temperature follows 
a nearly exponential decay with a time constant of $400\,$ms, in good 
agreement with the cooling rate according to Eq.~\ref{coolingrate}
(with $\Delta/2\pi$ = 1\,GHz, $\Delta_{\rm HFS}/2\pi$ = 9.2\,GHz, $q = 
0.25$). The temperature of the horizontal motion, which is not cooled 
directly, follows the vertical one
with a clearly visible time lag and approaches the same final value of
about 3\,$\mu$K. This apparent motional coupling can be fully explained 
by a small diffusive component of the reflection from the EW atom mirror,
as observed before by Landragin {\it et al.}\ (1996a).

A very important feature of the GOST and the evanescent-wave cooling 
mechanism is the predominant population of the absolute internal atomic
ground state: The lower hyperfine level is populated by more than 99.99\% 
of the atoms. A unique feature of EW cooling is that the trapping and 
cooling light does not penetrate the atomic sample.
Moreover, a possible reabsorption of scattered trap photons is strongly 
reduced because of the large surface area for photons to escape.
Due to these facts trap loss by ultracold collisions 
(see Sec.~\ref{collisions}) and other 
density-limiting mechanisms are strongly suppressed as compared to a MOT.
One can therefore expect EW Sisyphus cooling to work very well up to 
number densities of 10$^{13}$\,cm$^{-3}$, or even higher (S\"oding {\it et 
al.},
1995). In the first GOST experiments, this interesting 
regime was out of reach because the trap could be loaded with only 
$10^5$ atoms, leading to peak densities of $2 \times 10^{10}$\,cm$^{-3}$.
In present experiments with the GOST, 
being performed with a substantially improved loading scheme,
the high-density regime of evanescent-wave cooling is being explored. 

The GOST offers several interesting options for future experiments 
on dense atomic gases (Engler {\it et al.}, 1998). 
By detuning the EW and the hollow beam 
very far from resonance, a situation can be reached in which 
the photon scattering rate is far below 1\,s$^{-1}$. 
For a sufficiently dense gas, one can then
expect very good starting conditions for evaporative cooling, which
may be implemented by ramping down the EW potential. This appears to be
a promising route to quantum-degeneracy of Cs, which because of anomalously 
fast dipolar relaxation seems not attainable in a magnetic trap 
(S\"oding {\it et al.}, 1998; Gu\'ery-Odelin {\it et al.}, 1998).

Other interesting applications of the GOST, arising from its particular
geometry, are related to the possible formation of a two-dimensional 
quantum gas. 
The vertical motion is much more strongly confined than the horizontal one, 
so that the corresponding level spacings of the quantized atomic motion 
differ 
by nearly six orders of magnitude.
In such a highly anisotropic situation, Bose-Einstein
condensation is predicted to occur in two distinct steps
(van Druten and Ketterle, 1997). First, the vertical
motion condenses into its ground state (Wallis, 1996).
Then, in a second step occuring at much lower temperatures, 
the system condenses to its absolute ground state. 
The highly anisotropic nature of the trapping potential may be further
enhanced by addition of a second, attractive evanescent-wave 
(Ovchinnikov {\it et al.}, 1991). This can create a wavelength-sized 
potential
well close to the surface, which could be efficiently loaded by elastic
collisions. The situation would then resemble the situation of atomic 
hydrogen 
trapped on liquid helium, for which evidence of a Kosterlitz-Thouless
phase transition was reported very recently (Safonov {\it et al.}, 1998).
The prospect to realize such a system with alkali atoms\footnote{Other
experimental approaches to two-dimensional systems based on dipole trapping
make use of the particular properties of the optical transition structure 
in metastable noble gas atoms  
(Schneble {\it et al.}, 1998; Gauck {\it et al.}, 1998).}
is of particular interest to study effects of 
quantum-degeneracy in a 2D system.

\section{Concluding remarks}

In this review, we have discussed the basic physics of dipole trapping in 
far-detuned light, the typical experimental techniques and procedures, 
and the different presently available 
trap types along with their specific features.  
In the discussed experiments, optical dipole traps have already shown their 
great potential for a variety of different applications.

The particular advantages of dipole trapping 
can be summarized in the following three main points:
\vspace{-2mm}

\begin{itemize}

\item
The ground-state trapping potential can be designed to be either 
independent of the particular sub-level, or dependent in a well-defined 
way. In the first case,  
the internal ground-state dynamics 
under the influence of additional fields behaves in 
the same way as in the case of a free atom. 

\item
Photon scattering from the trap light can take place on an extremely
long time scale exceeding many seconds.
The trap then comes
close to the ideal case of a conservative, non-dissipative trapping 
potential, allowing for 
long coherence times of the internal and external dynamics 
of the stored atoms.

\item
Light fields allow one to realize a great variety of different trap
geometries, e.g., highly anisotropic traps, multi-well potentials,
mesoscopic and microscopic traps, and potentials for low-dimensional 
systems.

\end{itemize}

\noindent
Regarding these features, 
the main research lines for future experiments in optical dipole trapping
may be seen in the following fields:

\paragraph*{Ultracold collisions and quantum gases.} The behavior of
ultracold, potentially quantum-degenerate atomic gases  
is governed by their specific collisional properties, which strongly 
depend on the particular species, the internal
states of the colliding atoms, and possible external fields.
In this respect, dipole traps offer unique possibilities as they allow
to store atoms in any sub-state or combination of sub-states 
for the study of
collisional properties and collective behavior. Experiments along
these lines have already been performed (Gardner {\it et al.}, 1995; 
Tsai {\it et al.}, 1997; Stenger {\it et al.}, 1998; 
Miesner {\it et al.}, 1998b), but it seems that this has just 
opened the door to many other studies involving other atomic species, 
or even mixtures of different species (Engler {\it et al.}, 1998).

Of particular importance is the trapping of atoms in the 
absolute internal ground state, which cannot be trapped magnetically.
In this state, inelastic binary collision are completely suppressed for
energetical reasons. In this respect,
an ultracold cesium gas represents a particularly interesting situation, 
as Bose-Einstein condensation seems only attainable for the 
absolute ground state (S\"oding {\it et al.}, 1998; 
Gu\'ery-Odelin {\it et al.}, 1998). 
As a consequence, an optical trap may be the only
way to realize a quantum-degenerate gas of Cs atoms. 

Tuning of scattering properties by external fields is another very 
interesting subject. In this respect, Feshbach resonances at particular
values of magnetic fields play a very important role. In optical traps
one is completely free to choose any magnetic field without changing the
trap itself. Using this advantage, Feshbach resonances have been observed
for sodium and rubidium (Inouye {\it et al.}, 1998; Courteille {\it et al.}, 
1998), and future experimental work will certainly explore corresponding 
collisional properties
of other atomic species.

Highly anisotropic dipole traps offer a unique 
environment for the realization of low-dimensional quantum gases.
In this respect, interesting  trapping configurations are standing-wave traps 
(Sec.~\ref{sec:SWT}), optical lattices (Sec.~\ref{sec:lattices}), 
evanescent-wave surface traps (Sec.\ \ref{ewtraps}), and combinations
of such trapping fields (Gauck {\it et al.}, 1998). 
New phenomena could be related to a step-wise Bose-Einstein 
condensation (van Druten and Ketterle, 1997) and modifications of 
scattering properties in cases, in which the atomic
motion is restricted to a spatial scale on the order of the $s$-wave 
scattering length. 

Another fascinating issue would be to study collisional properties of 
ultracold fermions (e.g., the alkali atoms $^6$Li and $^{40}$K) 
with the challenging goal to produce a quantum-degenerate Fermi gas. 
Direct optical cooling in a standing-wave dipole trap by 
degenerate sideband cooling, similar to the method of Vuletic et al.\
(1998), seems to be a particularly promising route.

\paragraph*{Spin physics and magnetic resonance.}

As dipole traps allow for confinement with negligible effect on the
atomic ground-state spin, experiments related to the 
coherent ground-state dynamics in
external fields can be performed in essentially 
the same way as in the case of free atoms. In such experiments, 
the trap would provide much longer observation times as attainable in atomic 
beams or vapor cells.
As a further advantage, trapped atoms stay at the same place which keeps 
inhomogeneties of external fields very low. 
First demonstrations of along this line are the experiments  
by Davidson {\it et al.}\ (1995) and Zielonkowski {\it et al.}\ (1998b), as 
discussed in Sec.\ \ref{lstraps} and \ref{sec:startrap}, respectively.
In principle, a dipole trap constitutes an appropriate environment 
to perform any kind of magnetic resonance experiment with optically and 
magnetically manipulated ground-state spins (Suter, 1997). 
This could be of interest, e.g., for storing and processing quantum 
information in the spin degrees of freedom. 

A possible, very fundamental application would be the measurement of 
a permanent electric dipole moment of a heavy paramagnetic atom like cesium 
(Khriplovich and Lamoreaux, 1997). 
Such an experiment, testing time-reversal symmetry,
could greatly benefit from extremely long spin lifetimes and
coherence times of spin polarization, which seem to be attainable in 
dipole traps.


\paragraph*{Trapping of other atomic species and molecules.} 

Optical dipole traps do neither rely on a resonant interaction with laser
light nor on spontaneous photon scattering. Therefore
any polarizable particle can be trapped in powerful, 
sufficiently far-detuned light. For this purpose, the quasi-electrostatic
trapping with far-infrared laser sources, 
like CO$_2$ lasers (see Secs.\ \ref{sec:QUEST} and \ref{sec:QUEL}),
appears to be very attractive.  
The trapping mechanism could be applied to many other atomic species 
or molecules, which are not accessible to direct laser cooling. 
The problem is not the trapping itself, 
but the cooling to the very low temperatures required for trap loading. 
There may be several ways to overcome this problem.

A possible way to load a dipole trap 
could be based on buffer-gas loading of atoms or
molecules into magnetic traps (Kim {\it et al.}, 1997; 
Weinstein {\it et al.}, 1998) and subsequent evaporative cooling.
The effectiveness of 
cryogenic loading and subsequent evaporative cooling 
is demonstrated by the recent attainment of Bose-Einstein condensation 
of atomic hydrogen confined in a magnetic trap (Fried {\it et al.}, 1998),
without any optical cooling involved. 
Similar strategies based on buffer-gas loading
could open ways for filling optical dipole traps with many more atomic species 
than presently available or even with molecules.

Another possible way could be the production of 
ultracold molecules by photoassociation of laser-cooled atoms
(Fioretti {\it et al.}, 1998). The translational energies of these molecules 
can be low enough for loading into far-infrared dipole traps. 
A very recent experiment by 
Takekoshi {\it et al.} (1998) indeed reports evidence of dipole trapping
of a few Cs$_2$ molecules in a CO$_2$-laser beam, which may just be the
beginning of a new class of experiments in dipole trapping. 

\medskip
Optical dipole traps can be seen as storage devices at the low 
end of the presently explorable energy scale.
We are convinced that future experiments exploiting the particular
advantages of these traps will reveal interesting 
new phenomena and show many surprises.

\section*{Acknowledgments}
We would like to thank 
Hans Engler, Markus Hammes, Moritz Nill, Ulf Moslener, M. Zielonkowski, 
and in particular Inka Manek 
from the Heidelberg cooling and trapping group 
for many useful discussions and assistance in preparing the manuscript.
Further useful discussions are acknowledged with Steven Chu, Lev Khaykovich, 
Takahiro Kuga, Heun Jin Lee, Roee Ozeri, Christophe Salomon, and Vladan Vuletic.
We are grateful to Dirk Schwalm for supporting the work on laser cooling and
trapping at the MPI f\"ur Kernphysik. 
One of us (Yu.B.O.) acknowledges a fellowship by the 
Alexander von Humboldt-Stiftung. 
Our work on dipole trapping is generously supported 
by the Deutsche Forschungsgemeinschaft in the frame of the 
Gerhard-Hess-Programm.

\section*{References}

\parindent=0mm
\parskip=1.7mm

Adams, C.S., Sigel, M., and Mlynek, J. (1994).
{\it Phys. Rep.} {\bf 240}, 143.

Adams, C.S., Lee, H.J., Davidson, N., Kasevich, M., and Chu, S. (1995).
{\it Phys. Rev. Lett.} {\bf 74}, 3577.

Adams, C.S., and Riis, E. (1997). 
{\it Prog. Quant. Electron.} {\bf 21}, 1.

Allen, L., and Eberly, J.H. (1975).
{\it Optical resonance and two-level atoms} (Wiley, New York).

Aminoff, C.G., Steane, A., Bouyer, P., Desbiolles, P., Dalibard, J., and
Cohen-Tannoudji, C. (1993).
{\it Phys. Rev. Lett.} {\bf 71}, 3083.

Anderson, B.P., Gustavson, T.I., and Kasevich, M.A. (1996).
{\it Phys. Rev. A} {\bf 53}, R3727.

Anderson, M.H., Ensher, J.R., Matthews, M.R., Wieman, C.E., 
and Cornell, E. (1995).
{\it Science} {\bf 269}, 198.

Anderson, B.P., and Kasevich, M.A. (1998). 
{\it Science}, in press.

Andrews, M.R., Mewes, M.-O., van Druten, N.J., Durfee, D.S., Kurn, D.M., and
Ketterle, W. (1996).
{\it Science} {\bf 273}, 84.

Andrews, M.R., Kurn, D.M., Miesner, H.-J., Durfee, D.S., Townsend, C.G., 
Inouye, S., and Ketterle, W. (1997).
{\it Phys. Rev. Lett.} {\bf 79}, 553.

Arimondo, E., Phillips, W.D., and Strumia, F., eds. (1992).
{\it Laser Manipulation of Atoms and Ions}
Proceedings of the International School of Physics ``Enrico Fermi'', 
Varenna, 9 - 19 July 1991,
(North Holland, Amsterdam).

Ashkin, A (1970).
{\it Phys. Rev. Lett.} {\bf 24}, 156.

Ashkin, A. (1978).
{\it Phys. Rev. Lett.} {\bf 40}, 729.

Askar'yan, G.A. (1962).
{\it Sov. Phys. JETP} {\bf 15}, 1088.

Aspect, A., Arimondo, E., Kaiser, R., Vansteenkiste, N., and 
Cohen-Tannoudji, C. (1988). 
{\it Phys. Rev. Lett.} {\bf 61}, 826.

Balykin, V.I., and Letokhov, V.S. (1989).
{\it Appl. Phys. B} {\bf 48}, 517. 

Balykin, V.I., Letokhov, V.S., Ovchinnikov, Yu.B., and Sidorov, A.I. (1987). 
{\it JETP Lett.} {\bf 45}, 353; {\it Phys. Rev. Lett.} {\bf 60}, 2137 
(1988).

Beijersbergen, M.W., Allen, L., van der Veen, H.E.L.O., and Woerdman, J.P.,
{\it Opt. Commun.} {\bf 96}, 123 (1993).

Ben Dahan, M., Peik, E., Reichel, J., Castin, Y., and Salomon, C. (1996). 
{\it Phys. Rev. Lett.} {\bf 76}, 4508. 

Bergeman, T., Erez, G., and Metcalf, H.J. (1987).
{\it Phys. Rev. A} {\bf 35}, 1535.

Bergstr\"om, I., Carlberg, C., and Schuch, R., eds.\ (1995).
{\it Trapped charged particles and fundamental applications}
(World Scientific, Singapore); also published as 
{\it Physica Scripta} {\bf T59}.

Bjorkholm, J.E., Freeman, R.R., Ashkin, A., and Pearson, D.B. (1978).
{\it Phys. Rev. Lett.} {\bf 41}, 1361.

Boesten, H.M.J.M., Tsai, C.C., Verhaar, B.J., and Heinzen, D.J. (1997).
{\it Phys. Rev. Lett.} {\bf 77}, 5194.

Boiron, D., Trich\'{e}, C., Meacher, D.R., Verkerk, P., and Grynberg,
G. (1995). 
{\it Phys. Rev. A} {\bf 52}, R3425.

Boiron, D., Michaud, A., Lemonde, P., Castin, Y., Salomon, C., 
Weyers, S., Szymaniec, K., Cognet, L., and Clairon, A. (1996).
{\it Phys. Rev. A} {\bf 53}, R3734.

Boiron, D., Michaud, A., Fournier, J.M., Simard, L., Sprenger, M., 
Grynberg, G., and Salomon, C. (1998).
{\it Phys. Rev. A} {\bf 57}, R4106.

Bouchoule, I., Perrin, H., Kuhn, A., Morinaga, M., and Salomon, C. (1998). 
{\it Phys. Rev. A}, in press.

Bradley, C.C., Sackett, C.A., Tollet, J.J.,  and Hulet, R.G. (1995).
{\it Phys. Rev. Lett.} {\bf 75}, 1687.

Bradley, C.C., Sackett, C.A., and Hulet, R.G. (1997).
{\it Phys. Rev. Lett.} {\bf 78}, 985.



Chen, J., Story, J.G., Tollett, J.J., and Hulet, R.G. (1992). 
{\it Phys. Rev. Lett.} {\bf 69}, 1344.

Cho, D. (1997).
{\it J. Kor. Phys. Soc.} {\bf 30}, 373.

Chu, S., Bjorkholm, J.E., Ashkin, A., and Cable, A. (1986).
{\it Phys. Rev. Lett.} {\bf 57}, 314.

Chu, S. (1998).
{\it Rev. Mod. Phys.} {\bf 70}, 686.

Cline, R.A., Miller, J.D., Matthews, M.R., and Heinzen, D.J. (1994a).
{\it Opt. Lett.} {\bf 19}, 207.

Cline, R.A., Miller, J.D., and Heinzen, D.J. (1994b).  
{\it Phys. Rev. Lett.} {\bf 73}, 632.

Cohen-Tannoudji, C., and Dupont-Roc, J. (1972).
{\it Phys. Rev. A} {\bf 5}, 968.

Cohen-Tannoudji, C., Dupont-Roc, J., and Grynberg, G. (1992).
{\it Atom-Photon Interactions: Basic Processes and Applications}
(Wiley, New York).

Cohen-Tannoudji, C. (1998).
{\it Rev. Mod. Phys.} {\bf 70}, 707.

Cook, R.J. (1979).
{\it Phys. Rev. A} {\bf 20}, 224;
{\it ibid.} {\bf 22}, 1078 (1980).

Cook, R.J., and Hill, R.K. (1982). 
{\it Opt. Commun.} {\bf 43}, 258.

Corwin, K.L., Lu, Z.-T., Claussen, N., , Wieman, C., and Cho, D. (1997). 
{\it Bull. Am. Phys. Soc.} {\bf 42}, 1057.  

Courteille, P., Freeland, R.S., Heinzen, D.J., van Abeelen, F.A., and
Verhaar, B.J. (1998).
{\it Phys. Rev. Lett.} {\bf 81}, 69.

Dalibard, J., and Cohen-Tannoudji, C. (1985).
{\it J. Opt. Soc. Am. B} {\bf 2}, 1707.

Dalibard, J., and Cohen-Tannoudji, C. (1989).
{\it J. Opt. Soc. Am. B} {\bf 6}, 2023.

Davidson, N., Lee, H.J., Kasevich, M., and Chu, S. (1994).  
{\it Phys. Rev. Lett.} {\bf 72}, 3158.

Davidson, N., Lee, H.J., Adams, C.S., Kasevich, M., and Chu, S. (1995).
{\it Phys. Rev. Lett.} {\bf 74}, 1311.

Davis, K.B., Mewes, M.-O., Andrews, M.R., van Druten, N.J., Durfee, D.S, 
Kurn, D.M, and Ketterle, W. (1995).
{\it Phys. Rev. Lett.} {\bf 75}, 3969. 

DePu, M.T., McCormick, C., Winoto, S.L., Oliver, S., and Weiss, D.S.
(1998). ``Unity Occupation of Sites in a 3D Optical Lattice'', preprint.

Desbiolles, P., Arndt, M., Szriftgiser, P., and Dalibard, J. (1996a).
{\it Phys. Rev. A} {\bf 54}, 4292.


Deutsch, I.H., and Jessen, P. (1997).
{\it Phys. Rev. A} {\bf 57}, 1972.

Dholakia, K (1998).
{\it Contemp. Phys.} {\bf 39}, 351.

Dieckmann, K., Spreeuw, R.J.C., Weidem\"uller, M., and Walraven, J. (1998).
{\it Phys. Rev. A} {\bf 58}, 3891.


Dowling, J.P., and Gea-Banacloche, J. (1996). 
{\it Adv. At. Mol. Opt. Phys.} {\bf 37}, 1.

Drewsen, M., Laurent, P., Nadir, A., Santarelli, G., Clairon, A., 
Castin, Y., Grison, D., and Salomon, C. (1994).
{\it Appl. Phys. B} {\bf 59}, 283.

Engler, H., Manek, I., Moslener, U., Nill, M., Ovchinnikov, Yu.B.,
Schl\"oder, U., Sch\"unemann, U., Zielonkowski, M., Weidem\"uller, M.,
and Grimm, R. (1998).
{\it Appl. Phys. B} {\bf 67}, 709.

Fioretti, A., Comparat, D., Crubellier, A., Dulieu, O., Masnou-Seewes, F.,
and Pillet, P. (1998).
{\it Phys. Rev. Lett.} {\bf 81}, 4402.

Foot, C.J. (1991). 
{\it Comtemp. Phys.} {\bf 32} 369.

Friebel, S., D'Andrea, C., Walz, J., Weitz, M., and H\"ansch,
T.W. (1998a). 
{\it Phys. Rev. A} {\bf 57}, R20.

Friebel, S., Scheunemann, R.,, Walz, J., H\"ansch,T.W., and Weitz, M.
(1998b). 
{\it Appl. Phys. B} {\bf 67}, 699.

Fried, D.G., Killian, T.C., Willmann, L., Landhuis, D., Moss, S.C.,
Kleppner, D., Greytak, T.J. (1998).
{\it Phys. Rev. Lett.} {\bf 81}, 3811.

Gallagher, A., and Pritchard, D.E. (1989).
{\it Phys. Rev. Lett.} {\bf 63}, 957.

Gardner, J.R., Cline, R.A., Miller, J.D., Heinzen, D.J., Boesten, 
H.M.J.M., and Verhaar, B.J. (1995).  
{\it Phys. Rev. Lett.} {\bf 74}, 3764.

Gauck, H., Hartl, M., Schneble, D., Schnitzler, H., Pfau, T., 
and Mlynek, J. (1998).
{\it Phys. Rev. Lett.} {\bf 81}, 5298.

Ghosh, P.K. (1995). 
{\it Ion Traps} (Clarendon Press, Oxford).

Gordon, J.P., Ashkin, A. (1980).
{\it Phys. Rev. A} {\bf 21}, 1606.

Gu\'ery-Odelin, D., S\"oding, J., Desbiolles, P., and Dalibard, J. (1998a).
{\it Opt. Express} {\bf 2}, 323.

Grynberg, G., and Trich\'{e}, C., (1996), 
in {\it Coherent and Collective Interactions of Particles and Radiation 
Beams},
Proceedings of the International School of Physics ``Enrico Fermi'', 
Varenna, 11 - 21 July 1995,
edited by Aspect, A., Barletta, W. and Bonifacio, R. (IOS Press, Amsterdam),
p. 95.

Gu\'ery-Odelin, D., S\"oding, J., Desbiolles, P., and Dalibard, J. (1998b).
{\it Europhys. Lett.}, submitted.

Hamann, S.E., Haycock, D.L., Klose, G., Pax, P.H., Deutsch, I.H., and
Jessen, P.S. (1998).  
{\it Phys. Rev.  Lett.} {\bf 80}, 4149.

Haubrich, D., Schadwinkel, H., Strauch, F., Ueberholz, B., Wynands, R., and
Meschede, D. (1996).
{\it Europhys. Lett.} {\bf 34}, 663. 

Hemmerich, A., Weidem\"uller, M., Esslinger, T., Zimmermann, C., and
H\"ansch, T.W. (1995). 
{\it Phys. Rev. Lett.} {\bf 75}, 37. 

Hemmerich, A., Weidem\"uller, M., and H\"ansch, T.W. (1996), 
in {\it Coherent and Collective Interactions of Particles and Radiation
Beams}, 
Proceedings of the International School of Physics ``Enrico Fermi'', 
Varenna, 11 - 21 July 1995,
edited by Aspect, A., Barletta, W. and Bonifacio, R. (IOS Press, Amsterdam),
p. 503.

Henkel, C., Molmer, K., Kaiser, R., Vansteenkiste, N., Westbrook, C.I., 
and Aspect, A., 
{\it Phys. Rev. A} {\bf 55}, 1160 (1997).

Hess, H.F., Kochanski, G.P., Doyle, J.M., Masuhara, N., Kleppner, D., 
and Greytak, T.J. (1987). 
{\it Phys. Rev. Lett.} {\bf 59}, 672.

Inouye, S., Andrews, M.R., Stenger, J., Miesner, H.-J., Stamper-Kurn, D.M., 
and Ketterle, W. (1998). 
{\it Nature} {\bf 392}, 151.

Jackson, J.D. (1962).
{\it Classical electrodynamics} (Wiley, New York).

Jessen, P.S., and Deutsch, I.H. (1996). 
{\it Adv. At. Mol. Opt. Phys.} {\bf 37}, 95.

Kasevich, M.A., Weiss, D.S., and Chu, S. (1990).
{\it Opt. Lett.} {\bf 15}, 607.

Kasevich, M., and Chu, S. (1991). 
{\it Phys. Rev. Lett.} {\bf 67}, 181.

Kasevich, M.A., and Chu, S., (1992).
{\it Phys. Rev. Lett.} {\bf 69}, 1741.

Kastberg, A., Phillips, W.D., Rolston, S.L., Spreeuw, R.J.C., and 
Jessen, P.S. (1995). 
{\it Phys. Rev.  Lett.} {\bf 74}, 1542.

Kazantsev, A.P. (1973).
{\it Sov. Phys. JETP} {\bf 36}, 861; 
{\it ibid.} {\bf 39}, 784 (1974).

Ketterle, W., and van Druten, N.J. (1996).
{\it Adv. At. Mol. Opt. Phys.} {\bf 37}, 181.

Khripolovich, I.B., and Lamoreaux, S.K. (1997).
{\it CP Violation Without Strangeness: Electric Dipole Moments of 
Particles, Atoms, and Molecules}
(Springer, Berlin).

Kim, J., Friedrich, B., Katz, D.P., Patterson, D., Weinstein, J.D., 
DeCarvalho, R., and Doyle, J.M. (1997).
{\it Phys. Rev. Lett.} {\bf 78}, 3665.

Kuga, T., Torii, Y., Shiokawa, N., and Hirano, T. (1997).
{\it Phys. Rev. Lett.} {\bf 78}, 4713.

Kuhn, A., Perrin, H., H\"ansel, W., and Salomon, C. (1996). 
{\it OSA Trends in Optics and Photonics} {\bf 7}, 58.

Landragin, A., Labeyrie, G., Henkel, C., Kaiser, R., Vansteenkiste, N., 
Westbrook, C.I., and Aspect, A. (1996a).
{\it Opt. Lett.} {\bf 21}, 1591.

Landragin, A., Courtois, J.-Y., Labeyrie, G., Vansteenkiste, N., 
Westbrook, C.I., and Aspect, A. (1996b).
{\it Phys. Rev. Lett.} {\bf 77}, 1464.

Laryushin, D.V., Ovchinnikov, Yu.B., Balykin, V.I., and Letokhov, V.S. 
(1997). 
{\it Opt. Commun.} {\bf 135}, 138.

Lee, H.J., Adams, C., Davidson, N., Young, B., Weitz, M., Kasevich,
M., and Chu, S. (1994). 
In {\it Atomic Physics 14} edt. by Wineland, D.J., Wieman, C.E., 
and Smith, S.J. (AIP, New York).

Lee, H.J., Adams, C.S., Kasevich, M., and Chu, S. (1996).
{\it Phys. Rev. Lett.} {\bf 76}, 2658.

Lee, H.J., and Chu, S. (1998).
{\it Phys. Rev. A} {\bf 57}, 2905.

Lemonde, P., Morice, O., Peik, E., Reichel, J., Perrin, H.,
H\"ansel, W, and Salomon, C. (1995).
{\it Europhys. Lett.} {\bf 32}, 555.

Letokhov, V.S. (1968).
{\it JETP Lett.} {\bf 7}, 272.

Lett, P.D., Watts, R.N., Westbrook, C.I., Phillips, W.D., Gould, P.L., 
and Metcalf, H.J. (1988). 
{\it Phys. Rev. Lett.} {\bf 61}, 169.

Lett, P.D., Phillips, W.D., Rolston, S.L., Tanner, C.E., Watts, R.N., 
and Westbrook, C. (1989).
{\it J. Opt. Soc. Am. B} {\bf 6}, 2084 (1989).

Lett, P.D., Julienne, P.S., and Phillips, W.D. (1995).
{\it Annu. Rev. Phys. Chem.} {\bf 46}, 423.

Lu, Z.T., Corwin, K.L., Renn, M.J., Anderson, M.H., Cornell, E.A., and
Wieman, C.E. (1996).
{\it Phys. Rev. Lett.}{\bf 77}, 3331.

Manek, I., Ovchinnikov, Yu.B., and Grimm, R. (1998).
{\it Opt. Commun.} {\bf 147}, 67.

Metcalf, H., and van der Straten, P. (1994).
{\it Phys. Rep.} {\bf 244}, 203.

Mewes, M.-O., Andrews, M.R., van Druten, N.J., Kurn, D.M., Durfee, D.S., 
and Ketterle, W. (1996). 
{\it Phys. Rev. Lett.}  {\bf 77}, 416.

Mewes, M.-O., Andrews, M.R., Kurn, D.M., Durfee, D.S., Townsend, and
Ketterle, W. (1997). 
{\it Phys. Rev. Lett.}  {\bf 78}, 582.

Miesner, H.-J., Stamper-Kurn, D.M., Andrews, M.R., Durfee, D.S.,
Inouye, S., and Ketterle, W. (1998a). 
{\it Science} {\bf 279}, 1005.

Miesner, H.-J., Stamper-Kurn, D.M., Stenger, J., Inouye, S.,
Chikkatur, A.P., and Ketterle, W. (1998b). 
Preprint cond-mat/9811161.

Migdall, A.L, Prodan, J.V., Phillips, W.D., Bergemann, T.H., 
and Metcalf, H.J. (1985).
{\it Phys. Rev. Lett.} {\bf 54}, 2596.

Miller, J.D., Cline, R.A., and Heinzen, D.J. (1993a).
{\it Phys. Rev. A} {\bf 47}, R4567.

Miller, J.D., Cline, R.A., and Heinzen, D.J. (1993b).
{\it Phys. Rev. Lett.} {\bf 71}, 2204.

Minogin, V.G., and Letokhov, V.S. (1987).
{\it Laser light pressure on atoms}
(Gordon and Breach, New York).

Monroe, C., Swann, W., Robinson, H., and Wieman, C. (1990).
{\it Phys. Rev. Lett.} {\bf 65}, 1571

Morsch O., and Meacher, D.R., 
{\it Opt. Commun. {\bf 148}, 49 (1998).}

M\"uller-Seydlitz, T., Hartl, M., Brezger, B., H\"ansel, H., Keller, C.,
Schnetz, A., Spreeuw, R.J.C., Pfau, T., and Mlynek, J. (1997).
{\it Phys. Rev. Lett.}{\bf 78}, 1038.

Myatt, C.J., Burt, E.A., ghrist, R.W., Cornell, E.A., and Wieman, C.E. 
(1997).
{\it Phys. Rev. Lett.}{\bf 78}, 586.

Ovchinnikov, Yu.B., Shul'ga, S.V., and Balykin, V.I. (1991). 
{\it J. Phys. B: At. Mol. Opt. Phys.} {\bf 24}, 3173. 

Ovchinnikov, Yu.B., S\"oding, J., and Grimm, R. (1995a). 
{\it JETP Lett.} {\bf 61}, 21.

Ovchinnikov, Yu.B., Laryushin, D.V., Balykin, V.I., and Letokhov, V.S. 
(1995b). 
{\it JETP Lett.} {\bf 62}, 113.

Ovchinnikov, Yu.B., Manek, I., and Grimm, R. (1997).
{\it Phys. Rev. Lett.} {\bf 79}, 2225.

Ozeri, R., Khaykovich, L., and Davidson, N. (1998). 
``Long spin relaxation times in a single-beam blue-detuned optical
trap'', preprint.

Perrin, H., Kuhn, A., Bouchoule, I., and Salomon, C. (1998). 
{\it Europhys. Lett.} {\bf 42}, 395.

Petrich, W., Anderson, M.H., Ensher, J.R., and Cornell, E.A. (1994).
{\it J. Opt. Soc. Am. B} {\bf 11}, 1332.

Phillips, W., and Metcalf, H. (1982).
{\it Phys. Rev. Lett.} {\bf 48}, 596.

Phillips, W.D. (1998).
{\it Rev. Mod. Phys.} {\bf 70}, 721.

Pinkse,  P.W.H., Mosk, A., Weidem\"uller, M., Reynolds, M.W.,
Hijmans, T.W., and Walraven, J.T.M., 
{\it Phys. Rev. Lett.} {\bf 78}, 990 (1997).

Pritchard, D.E., Raab, E.L., Bagnato, V.S., Wieman, C.E., 
and Watts, R.N. (1986).
{\it Phys. Rev. Lett.} {\bf 57}, 310.

Raab, E.L., Prentiss, M., Cable, A., Chu, S., and Pritchard, D. (1987).
{\it Phys. Rev. Lett.} {\bf 59}, 2631.

Reichel, J., Bardou, F., Ben Dahan, M., Peik, E., Rand, S., Salomon,
C., and Cohen-Tannoudji, C. (1995).  
{\it Phys. Rev. Lett.} {\bf 75}, 4575.

Safonov, A.I., Vasilyev, S.A., Yashnikov, I.S., Lukashevich, I.I., and 
Jaakola, S. (1998).
{\it Phys. Rev. Lett.} {\bf 81}, 4545.

Salomon, C., Dalibard, J., Phillips, W.D., and Guellatti, S. (1990).
{\it Europhys. Lett.} {\bf 12}, 683.


Savard, T.A., O`Hara, K.M., and Thomas, J.E. (1997).
{\it Phys. Rev. A} {\bf 56}, R1095.

Schiffer, M., Rauner, M., Kuppens, S., Zinner, M., Sengstock, K., and
Ertmer, W. (1998).
{\it Appl. Phys. B} {\bf 67}, 705 (1998).


Schneble, D., Gauck, H., Hartl, M., Pfau, T., and Mlynek, J. (1998).
in {\it Bose-Einstein condensation of atomic gases},
Proceedings of the International School of Physics ``Enrico Fermi'', 
Varenna, 7 - 17 July 1998, 
edited by Inguscio, M., Stringari, S., and Wieman, C., in press.

Sch\"unemann, U., Engler, H., Zielonkowski, M., Weidem\"uller, M., and
Grimm, R. (1998).
{\it Opt. Commun.}, in press.

Sengstock, K., and Ertmer, W., (1995). 
{\it Adv. At. Mol. Opt. Phys.} {\bf 35}, 1.

Sesko, D., Walker, T., Monroe, C., Gallagher, A., and Wieman, C. (1989).
{\it Phys. Rev. Lett.} {\bf 63}, 961.

Sobelman, I.I. (1979).
{\it Atomic Spectra and Radiative Transitions} (Springer, Berlin).

S\"oding, J., Grimm, R., and Ovchinnikov, Yu.B. (1995).
{\it Opt. Commun.} {\bf 119}, 652.

S\"oding, J., Gu\'ery-Odelin, D., Desbiolles, P., Ferrari, G., 
and Dalibard, J. (1998).
{\it Phys. Rev. Lett.} {\bf 80}, 1869.

Stamper-Kurn, D.M., Andrews, M.R., Chikkatur, A.P., Inouye, S., Miesner, H.-
J.,
Stenger, J., and Ketterle, W. (1998a).
{\it Phys. Rev. Lett.} {\bf 80}, 2027.

Stamper-Kurn, D.M., Miesner, H.-J., Chikkatur, A.P., Inouye, S.
Stenger, J., and Ketterle, W. (1998b). 
{\it Phys. Rev. Lett.} {\bf  81}, 2194.

Steane, A.M., and Foot, C.J. (1991).  
{\it Europhys. Lett.} {\bf 14}, 231.

Stenger, J., Inouye, S., Stamper-Kurn, D.M., Miesner, H.-J.,
Chikkatur, A.P., and Ketterle, W. (1998). 
{\it Nature}, in press.

Stenholm, S. (1986).
{\it Rev. Mod. Phys.} {\bf 58}, 699.

Suter, D. (1997).
{\it The Physics of Laser-Atom Interactions}
(University Press, Cambridge).

Takekoshi, T., Yeh, J.R., and Knize, R.J. (1995). 
{\it Opt. Comm.} {\bf 114}, 421.

Takekoshi, T., and Knize, R.J. (1996). 
{\it Opt. Lett.} {\bf 21}, 77.

Takekoshi, T., Patterson, B.M., and Knize, R.J. (1998).
{\it Phys. Rev. Lett.} {\bf 81}, 5105.

Torii, Y., Shiokawa, N., Hirano, T., Kuga, T., Shimizu, Y., 
and Sasada, H.,
{\it Eur. Phys. J. D} {\bf 1}, 239 (1998).

Townsend, C.G., Edwards, N.H., Cooper, C.J., Zetie, K.P., Foot, C.J., 
Steane, A.M., Szriftgiser, P., Perrin, H., and Dalibard, J. (1995).
{\it Phys. Rev. A} {\bf 52}, 1423.

Tsai, C.C., Freeland, R.S., Vogels, J.M., Boesten, H.M.J.M., Verhaar, B.J., 
and Heinzen, D.J. (1997). 
{\it Phys. Rev. Lett.} {\bf 79}, 1245.

van Druten, N.J., and Ketterle, W., 
{\it Phys. Rev. Lett.} {\bf 79}, 549 (1997).

Vuletic, V., Chin, C., Kerman, A.J., and Chu, S., (1998). 
{\it Phys. Rev. Lett.}, in press.

Walker, T. and Feng, P. (1994).
{\it Adv. At. Mol. Opt. Phys.} {\bf 34}, 125.
 
Wallace, C.D., Dinneen, T.P., Tan, K.-Y.N., Grove, T.T., and 
Gould, P.L. (1992).
{\it Phys. Rev. Lett.} {\bf 69}, 897.

Wallis, H., Dalibard, J., and Cohen-Tannoudji, C. (1992).
{\it Appl. Phys. B} {\bf 54}, 407.

Wallis, H. (1996).
{\it Quantum Semiclass. Opt.} {\bf 8}, 727 (1996).

Walther, H. (1993).
Adv. At. Mol. Opt. Phys. {\bf 31}, 137.

Weiner, J. (1995). 
{\it Adv. At. Mol. Opt. Phys.} {\bf 35}, 45.

Weinstein, J.D., deCarvalho, R., Guillet, T., Friedrich, B., and
Doyle, J.M. (1998).
{\it Nature} {\bf 395}, 148.

Wilkinson, S.R., Bharucha, C.F., Madison, K.W., Niu, Q., 
and Raizen, M.G. (1996). 
{\it Phys. Rev. Lett.} {\bf 76}, 4512.

Winoto, S.L., DePu, M.T., Bramall, N.E., and Weiss, D.S. (1998).
``Laser Cooling Large Numbers of Atoms at High Densities'', preprint.

Wineland, D.J., Bergquist, J.C., Bollinger, J.J., and Itano, W.M. (1995).
in {\it Trapped charged particles and fundamental applications}
edt. by Bergstr\"om, I., Carlberg, C., and Schuch, R.
(World Scientific, Singapore); also published as 
{\it Physica Scripta} {\bf T59}, 286. 

Zemanek, P., and Foot, C.J. (1998).
{\it Opt. Commun.} {\bf 146}, 119.

Zielonkowski, M., Steiger, J., Sch\"unemann, U., DeKieviet, M., and
Grimm, R. (1998a).
{\it Phys. Rev. A} {\bf 58}, in press.

Zielonkowski, M., Manek, I., Moslener, U., Rosenbusch, P., and
Grimm, R. (1998b).
{\it Europhys. Lett.} {\bf 44}, 700.

\widetext

\begin{table}
\makebox[4.5cm][l]{}
\vspace{4.5cm}
\caption{\it Properties of the alkali atoms of relevance for optical dipole 
trapping. Transition wavelengths $\lambda_{D_2}$ and $\lambda_{D_1}$, fine-
structure splitting $\Delta'_{\rm FS}$, nuclear spin $I$, 
ground-state hyperfine splitting  $\Delta_{\rm HFS}$, 
excited-state hyperfine splitting  $\Delta'_{\rm HFS}$,
natural linewidth $\Gamma$, recoil temperature $T_{\rm rec}$ 
(for $D_2$ line), and corresponding height 
$h_{\rm rec} = k_B T_{\rm rec}/(m g)$ in the field of gravity.}
\vspace{5mm}
\epsfxsize=18cm
\centerline{\epsffile{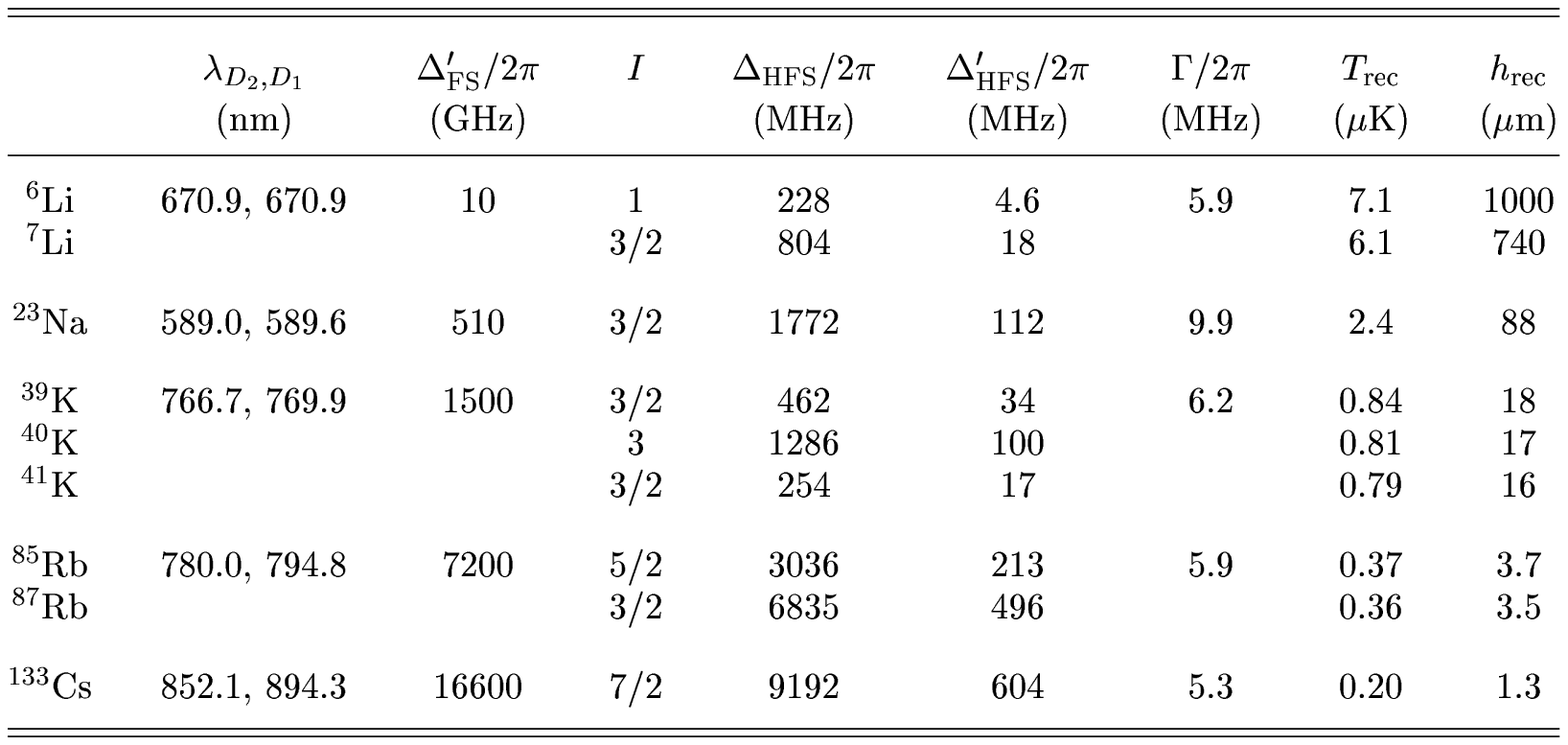}}
\vspace{1cm}
\label{alkalitable}
\end{table}

\pagebreak 
\makebox[1cm][l]

\pagebreak 

\begin{table}
\caption{\it Parameters of various experimentally realized
blue-detuned dipole traps.}
\vspace{5mm}
\epsfxsize=18cm
\centerline{\epsffile{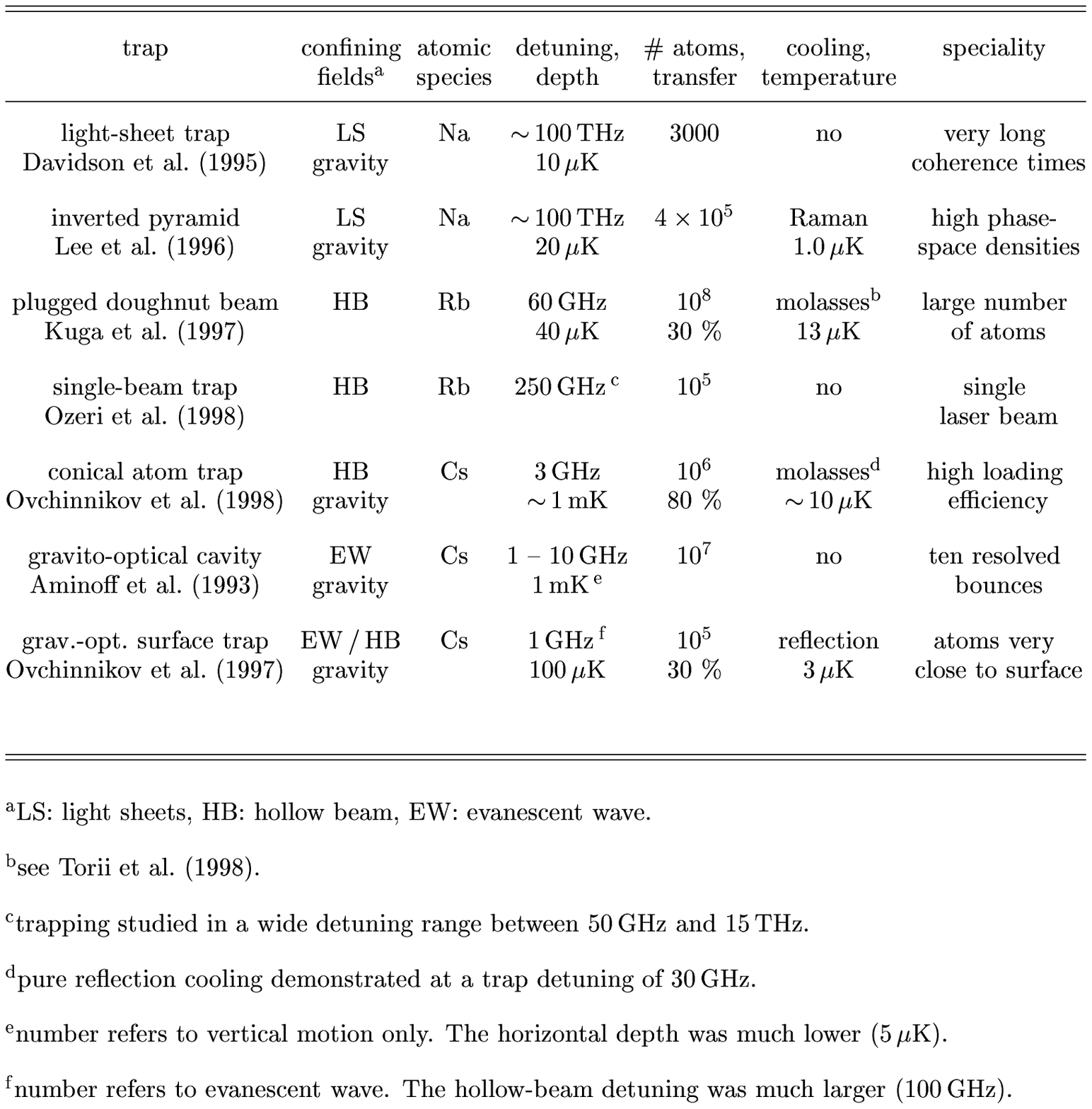}}
\vspace{1cm}
\label{bluetable}
\end{table}

\end{document}